\documentclass[fleqn,usenatbib, onecolumn]{mnras}
\usepackage[utf8]{inputenc}
\usepackage{amsmath,amssymb, graphicx, multicol, array, upgreek, natbib, booktabs, multirow, bm, dsfont, bbold} 
\usepackage[table,xcdraw]{xcolor}
\usepackage{lineno}
\usepackage{simplewick}
\usepackage{scalerel}
\usepackage[mathscr]{euscript}
\usepackage{extarrows}
\usepackage{caption}
\usepackage{subcaption}
\usepackage{multirow}

\usepackage[thinc]{esdiff}
\definecolor{color26}{rgb}{0.9, 0.9, 0.9}
\definecolor{color27}{rgb}{0.2, 0.6, 0.8}
\definecolor{color38}{rgb}{0.9,.17,0.31}

\definecolor{myblue}{rgb}{0.1, 0.4, 0.5}
\definecolor{color1}{rgb}{0.5, 0.5, 0.5}

\usepackage{tikz}
\usetikzlibrary{tikzmark}

\def\begeqar{\begin{eqnarray}}
\def\endeqar{\end{eqnarray}}
\def\begeq{\begin{equation}}
\def\endeq{\end{equation}}
\def\non{\nonumber\\}

\DeclareMathOperator{\parity}{\mathbb{\hat{P}}}

\def\patchy{{\textsc{Patchy }}}

\def\mpch{{h^{-1}\rm{Mpc}}}

\renewcommand{\vec}[1]{\boldsymbol{#1}}

\def\bfk{{\bf k}}

\def\bfkhat{{\hat\bfk}}

\def\bfr{{\bf r}}

\def\bfrhat{{\hat\bfr}}

\def\calC{{\cal C}}
\def\calD{{\cal D}}
\def\calE{{\cal E}}

\def\calN{{\cal N}}

\def\calP{{\cal P}}
\def\calR{{\cal R}}

\def\bfx{{\bf x}}
\def\bfxhat{{\hat\bfx}}

\newcommand{\av}[1]{\left\langle{#1}\right\rangle} 
\newcommand{\vd}{\vec d}

\newcommand{\vs}{\vec s}

\newcommand{\bfs}{\mathbf s}

\newcommand{\vn}{\vec n}

\newcommand{\bfnhat}{\hat{\bf n}}

\newcommand{\bfyhat}{\hat{\bf y}}
\newcommand{\bfzhat}{\hat{\bf z}}

\newcommand{\dprod}[2]{\bfrhat_{#1}\cdot\bfrhat_{#2}}

\def\triple{{\bfrhat_1\cdot(\bfrhat_2\times\bfrhat_3)}}
\newcommand{\six}[6]{\left(\begin{array}{ccc}
									{#1}& {#2}& {#3}\\
									{#4}& {#5}& {#6} \\
\end{array}\right)}

\newcommand{\nine}[9]{\left\{\begin{array}{ccc}
									{#1}& {#2}& {#3}\\
									{#4}& {#5}& {#6} \\
							    	{#7}& {#8}& {#9} \\
\end{array}\right\}}

\begin{document}
\title[Parity-Odd 4PCF Detection]{Measurement of Parity-Odd Modes in the Large-Scale 4-Point Correlation Function of SDSS BOSS DR12 CMASS and LOWZ Galaxies}
\author[Hou, Slepian \& Cahn]{Jiamin Hou,$^{1,2}$\thanks{E-mail: jiamin.hou@ufl.edu (JH)}
Zachary Slepian,$^{1,3}$\thanks{E-mail: zslepian@ufl.edu (ZS)} \& Robert N. Cahn$^{3}$
\\
$^{1}$Department of Astronomy, University of Florida, Gainesville, FL 32611, USA\\
$^{2}$ Max-Planck-Institut für Extraterrestische Physik, Postfach 1312, Giessenbachstr., 85748 Garching, Germany\\
$^{3}${Lawrence Berkeley National Laboratory, Berkeley, CA 94720, USA}}
\date{\vspace{-5ex}}

\maketitle

\label{firstpage}

\begin{abstract}
A tetrahedron is the simplest shape that cannot be rotated into its mirror image in 3D. The 4-Point Correlation Function (4PCF), which quantifies excess clustering of quartets of galaxies over random, is the lowest-order statistic sensitive to parity violation. Each galaxy defines one vertex of the tetrahedron. Parity-odd modes of the 4PCF probe an imbalance between tetrahedra and their mirror images. We measure these modes from the largest currently available spectroscopic samples, the 280,067 Luminous Red Galaxies (LRGs) of {\color{black} the Baryon Oscillation Spectroscopic Survey (BOSS)} DR12 LOWZ ($\bar{z} = 0.32$) and the 803,112  LRGs of BOSS DR12 CMASS ($\bar{z} = 0.57$). In LOWZ we find $3.1\sigma$ evidence for a non-zero parity-odd 4PCF, and in CMASS we detect a parity-odd 4PCF at $7.1\sigma$. Gravitational evolution alone does not produce this effect; parity-breaking in LSS, if cosmological in origin, must stem from the epoch of inflation. We have explored many sources of systematic error and found none that can produce a spurious parity-odd \textit{signal} sufficient to explain our result. Underestimation of the \textit{noise} could also lead to a spurious detection. Our reported significances presume that the mock catalogs used to calculate the covariance sufficiently capture the covariance of the true data. We have performed numerous tests to explore this issue. The odd-parity 4PCF opens a new avenue for probing new forces during the epoch of inflation with 3D {\color{black} large-scale structure}; such exploration is timely given large upcoming spectroscopic samples such as DESI and Euclid.
\end{abstract}


\noindent Submission to MNRAS: June 08 2022, Received: June 11 2022, Accepted: April 01 2023

\section{Introduction}
The laws of nature respect certain symmetries; the physical processes governed by them are invariant under the corresponding transformations. Parity transformation (P), which reverses the sign of each coordinate axis, had been thought to be such a symmetry. Indeed, the electromagnetic and strong interactions are invariant under P. However, this symmetry is broken in the weak interaction~\citep{Lee1956, Wu1957}. \citet{Sakharov1967} showed that the matter-antimatter asymmetry of the Universe requires that the combination CP of P and charge-conjugation (C) symmetry be broken. The currently-known CP violation is inadequate to explain the observed matter-antimatter asymmetry. Whatever additional CP violation is responsible may involve pure P violation as well.

Most of the cosmological studies of parity invariance to date have focused on Cosmic Microwave Background (CMB) polarization~\citep{Lue1999PVcmb,Kamionkowski2011CMBpv,Shiraishi2011CMBbispectraTensor,Minami2020} or on gravitational waves~\citep{Saito2007PGW,Yunes2010GWpv,Jeong2012ClusFossil,Wang2013PVpgw,Zhu2013PVpgw,Nishizawa2018GWpv,Orlando2021sgwg}. A recent CMB study of parity violation reported $2.4\sigma$ evidence for cosmic birefringence (where  the two polarization states of a wave propagate differently)~\citep{Minami2020}. 
{\color{black} \citet{Eskilt2022birefringence} refined this analysis and found $3.6\sigma$ evidence.}




{\color{black} A number of mechanisms producing parity violation at cosmological scales have been presented in the literature. For instance, one can add a Chern-Simons coupling to the standard cosmological paradigm at early or at late times. {\color{black} This term typically describes an interaction between a pseudo-scalar field and a spin-1 field~\citep{Sorbo2011PVpseudoscalar,Barnaby2011PsdudoscalatInflaton,Ozsoy2021PVaxion} or a spin-2 field~\citep{Jackiw2003CSGR,Alexander2009PVgravity,Soda2011PVgraviton,Dyda2012CSgravity}. The pseudo-scalar field can be axion-like and if present at late times, can play the role of dark matter or dark energy. In this case, the Chern-Simons coupling can rotate the polarizations of initially linearly-polarized CMB photons.\footnote{Helical primordial gravitational waves could {\color{black} in principle} leave an imprint on the cross-spectrum between the $E$ and $B$ modes of CMB polarization, or between $B$ modes and temperature fluctuations. However, these observables are suppressed by the two-dimensional nature of the CMB \citep{Masui2017}} In contrast, if the axion-like field plays the role of the inflaton, the coupling} can give rise to non-vanishing parity-odd polyspectra of the primordial curvature perturbations~\citep{Bartolo2015pv,Shiraishi2016PVtrispectrum}. Since the curvature perturbations seed the subsequent formation of large-scale structure, the primordial parity-odd polyspectra would produce the same in the late-time distribution of galaxies.}

\begin{figure}
    \centering
    \includegraphics[width=.6\textwidth]{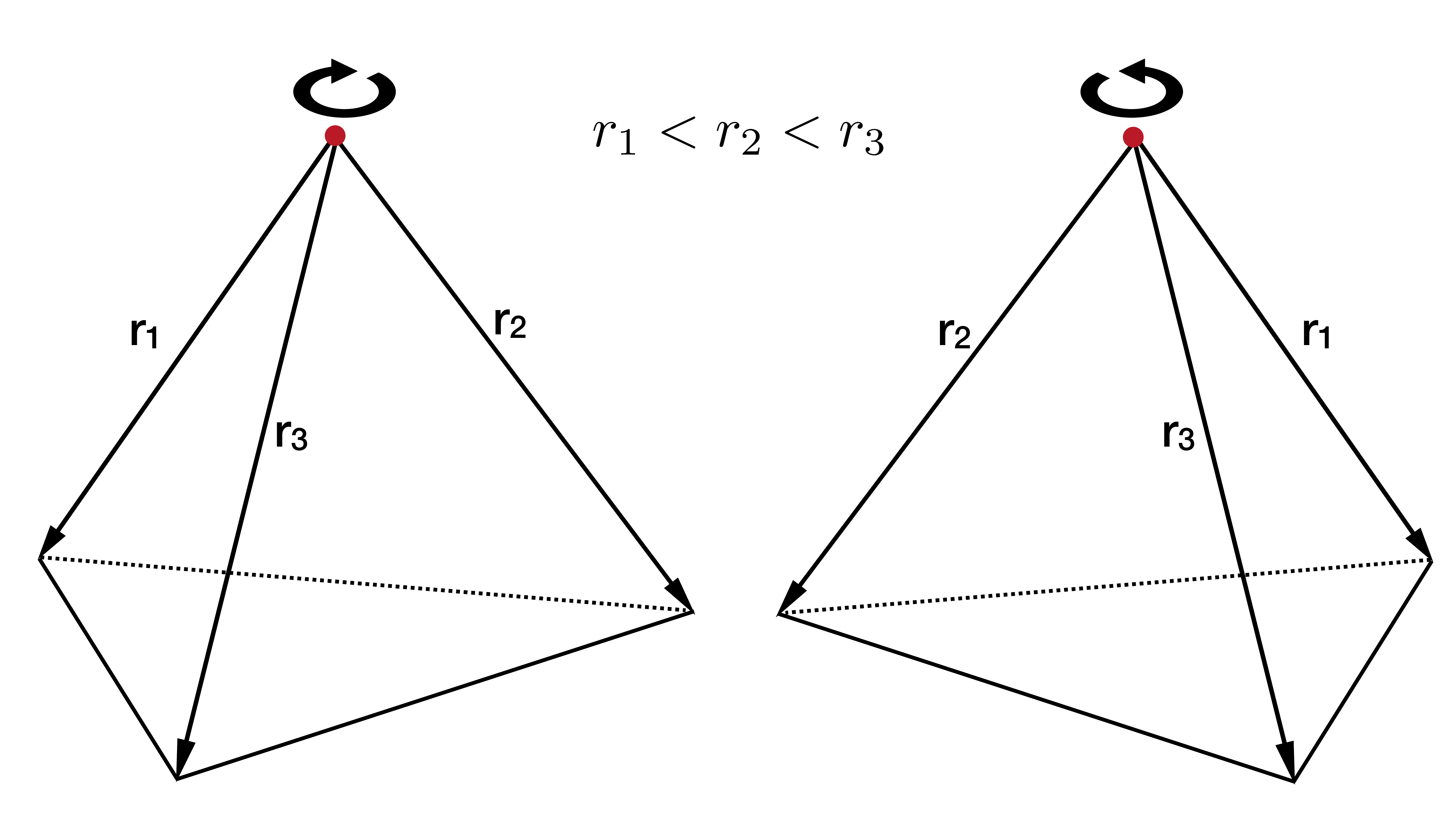}
    \caption{Parity transformation applied to a tetrahedron formed by a quartet of galaxies. Each vertex represents a galaxy. Choosing one galaxy (red dot) as our primary, the quartet is defined by the three vectors to the remaining vertices, $\bfr_1$, $\bfr_2$, and $\bfr_3$. For a quartet, the subscripts are fixed by requiring that $r_1 \leq r_2 \leq r_3$. When viewing the tetrahedron from the primary (red) looking down along each vector $\bfr_i$, the direction in which one reads going from smallest to largest side ($r_1$ to $r_3$) defines a handedness, either clockwise or counterclockwise. Here, the tetrahedron on the left, as viewed from the primary, is clockwise. Parity transformation in 3D is a reflection about a plane and then a $180^{\circ}$ rotation about the vector perpendicular to that plane, and  converts the clockwise tetrahedron at left to the counterclockwise one at right. When one averages over rotations, as in this work, only the mirroring matters.}
    \label{fig:tet_cw_ccw}
\end{figure}

{\color{black} Recently, \citet{Cahn2021parity} made the novel proposal of using the galaxy 4-Point Correlation Function (4PCF) to probe parity violation in 3D large-scale structure.} Four galaxies can be taken as the vertices of a tetrahedron, the lowest-order 3D shape that cannot be rotated into its mirror image, rendering the 4PCF sensitive to parity violation. An illustration of a galaxy quartet and how we define parity on it, is shown in Fig.~\ref{fig:tet_cw_ccw}. Following \cite{Cahn2021parity}, we expand the 4PCF in the isotropic (\textit{i.e.} rotation-averaged) basis functions of \citet{Cahn2020iso}.  
In the standard inflationary paradigm~\citep{Albrecht1982Inflation, Linde1982Inflation, Linde1983ChaoticInflation}, we would not expect a parity-odd 4PCF. {\color{black}The initial density fluctuations are a Gaussian Random Field}~\citep{Bardeen1980:GRF,Starobinsky1982:GRF}, which then evolves under gravity and forms galaxies at late times. Gravity, and even the baryonic physics of galaxy formation, is parity-conserving. Hence, the detection of a parity-odd 4PCF of cosmological origin would be evidence that parity violation was present before the known forces dominated the evolution of the matter distribution.  

We present here a measurement of the parity-odd modes of the 4PCF measured using the Baryon Oscillation Spectroscopic Survey~\citep[BOSS;][]{BOSScollaboration2017} of Sloan Digital Sky Survey (SDSS)-III~\citep{Dawson2013SDSSIII,Eisenstein2011SDSSIII}. \citealt{Philcox2021detection} presented the parity-even 4PCF measurement on the same dataset and found an $8.1\sigma$ detection of a non-Gaussian 4PCF (expected in the standard picture of gravitationally evolved structure formation, \textit{e.g.} \citealt{Bernardeau2002PT}). A progenitor of the algorithm and approach here used was employed to measure the 3-Point Correlation Function (3PCF) of BOSS DR12 CMASS (\citealt{SE_BOSS_3PCF, SE_BOSS_BAO,SE_BOSS_RV}, \citealt{Sugiyama2019FFTrsd,Sugiyama20213pcf}), and extended to use Fourier transforms in \cite{SE_3PCF_FT, Portillo} and to the anisotropic 3PCF in \cite{SE_aniso_3pcf, galactos, Garcia}. The 4PCF has been measured before in just a few works (in 2D; \citealt{Fry_78}; averaged over internal angles \citep{sabiu}, as well as in Fourier space and with a degree of compression (integrated trispectrum;  \citealt{gualdi22}), but never separated into parity-odd modes; the history of {\textcolor{black}{N-Point Correlation Functions} (NPCFs)} is reviewed in \cite{peebles_rvw}.

Given that as yet no detection of parity-odd physics in large-scale structure has been made and that a number of proposed theoretical models can produce it, in this work we pursue a model-independent analysis. Hence we lack an {\it a priori} expectation for the shape of the signal. While using such a model could strengthen any detection significance by correlating data in different modes, it would inevitably tie our detection significance to a particular model, which we wish to avoid. 
In contrast to typical analyses \textit{e.g.} of the 2PCF for Baryon Acoustic Oscillations (BAO) or the 3PCF for BAO or galaxy biasing, in this work we cannot identify systematic errors simply by observing departures from an expected template for the cosmological signal. We must thus pay especial attention to systematics. We make extensive use of both mock catalogs and analytics to assess whether any systematics can produce spurious parity-odd 4PCF modes.

{\color{black}{Our fiducial cosmology here matches that adopted by BOSS \citep{BOSScollaboration2017}. In particular, we take a geometrically flat $\Lambda {\rm CDM}$ model with redshift-zero matter density (in units of the critical density) $\Omega_{\rm m} = 0.31$, baryon density (in units of the critical density) $\Omega_{\rm b} = 0.048$,  Hubble constant $h\equiv H_0/(100\;{\rm km/s/Mpc)} = 0.676$, root-mean-square density fluctuations (within $8\, \mpch$ spheres) of $\sigma_{8}=0.8$, scalar spectral tilt $n_{\rm s}=0.96$ and sum of the neutrino masses $\sum_i m_{\nu, i}=0.06\; \mathrm{eV}$.}}

The present work is organized as follows. \S\ref{sec:guidelines} outlines the multiple analyses of varying complexity used to increase confidence in our measurements. \S\ref{sec:basis} reviews the basis used to decompose the 4PCF. We also present a toy model to illustrate the relation between parity and tetrahedra. \S\ref{sec:data} describes the data, simulations, and covariance matrix. \S\ref{sec:analysis} then presents two different paths to obtaining a detection significance and their outcomes. We also present an analysis of cross-correlations between spatially separated regions. \S\ref{sec:systematics} outlines our systematics tests on mocks as well as analytic work. \S\ref{sec:conclusion} concludes. A number of Appendices present details of the work.

\section{Method for 4PCF Measurement}
\label{sec:basis}
The 4PCF estimator, indicated by a hat, is 
\begin{eqnarray}
 \hat{\zeta}(\bfr_1,\bfr_2,\bfr_3)&\equiv&\av{\delta(\bfs)\delta(\bfs+\bfr_1)\delta(\bfs+\bfr_2)\delta(\bfs+\bfr_3)}\nonumber\\
 &=& \int \frac{d\bfs}{V}\, \delta(\bfs)\delta(\bfs+\bfr_1)\delta(\bfs+\bfr_2)\delta(\bfs+\bfr_3),
 \label{eqn:4pcf_estimator}
\end{eqnarray}
where angle brackets denotes an ensemble average of the density fluctuations field $\delta(\bfs)\equiv \rho(\bfs)/\bar{\rho}-1$, with $\rho(\bfs)$ the density field and $\bar{\rho}$ the average density.\footnote{The 4PCF after rotation-averaging has six degrees of freedom, so we will only require certain combinations of the arguments on the lefthand side of the estimator.} Invoking ergodicity, the ensemble average may be replaced by an integral of spatial position $\bfs$ over the volume $V$. This integration results in a function that depends only on the relative separation vectors $\bfr_1$, $\bfr_2$, and $\bfr_3$. We also average over joint rotations of these vectors. In practice, the density fluctuation field is computed from discrete galaxy data, appropriately weighted. 

Since the 3D distribution of galaxies is assumed to be isotropic on cosmological scales (ignoring redshift-space distortions), the isotropic basis \citep{Cahn2020iso} is an efficient means of systematically extracting cosmological information. The isotropic basis functions required to measure an NPCF are given by products of $(N-1)$ spherical harmonics $Y_{\ell m}(\bfrhat)$ combined according to angular momentum addition. In particular, for the 4PCF ($N=4$) we require the three-argument basis functions, which are 
\begin{eqnarray}
\calP_{\ell_1\ell_2\ell_3}(\bfrhat_1,\bfrhat_2,\bfrhat_3) 
&=& \sum_{m_1m_2m_3} \calC^{\ell_1\ell_2\ell_3}_{m_1m_2m_3} Y_{\ell_1m_1}(\bfrhat_1)Y_{\ell_2m_2}(\bfrhat_2)Y_{\ell_3m_3}(\bfrhat_3).
\label{eqn:isofunc_4pcf}
\end{eqnarray}
The factorizability of these functions is important to the speed-up of the 4PCF algorithm \citep{Philcox2021encore}; in practice, it enables us to compute the 4PCF as a sum over the spherical harmonic coefficients $a_{\ell_i m_i}$ of the density field about a given primary galaxy at $\bfs$. 

Each unit vector $\bfrhat_i$ is associated with one total angular momentum $\ell_i$, with $z$-component $m_i$.\footnote{\color{black} Given the rotational symmetry of the system, the choice of $z$-axis is arbitrary.} 
{\color{black} The key point is simply that spherical harmonics are two-index tensors, and conventionally the total angular momentum and its $z$-component are chosen to represent them. This point is further discussed in \cite{Cahn2020iso} around their equation 2.} The weight is
\begin{eqnarray}
\calC^{\ell_1\ell_2\ell_3}_{m_1m_2m_3} \equiv (-1)^{\ell_1+\ell_2+\ell_3}\six{\ell_1}{\ell_2}{\ell_3}{m_1}{m_2}{m_3}.
\label{eqn:clebsch_gorden_N3}
\end{eqnarray}
The 3-$j$ symbol enforces the triangular inequality $|\ell_1-\ell_2|\leq \ell_3 \leq \ell_1+\ell_2$, because the total angular momentum must be zero for an isotropic function.\footnote{We recall that a 3-$j$ symbol with zeros in the bottom row demands that $\ell_1 + \ell_2 + \ell_3$ be even, but with non-zero $m_i$ there is no such requirement, allowing odd sums of the $\ell_i$ and hence parity-odd basis functions.} Under the parity operator, denoted $\parity$, a spherical harmonic $Y_{\ell m}$ transforms as $(-1)^{\ell}$, so the three-argument isotropic functions transform as
\begin{eqnarray}
\parity\left[\calP_{\ell_1\ell_2\ell_3}(\bfrhat_1,\bfrhat_2,\bfrhat_3)\right] &\equiv& \calP_{\ell_1\ell_2\ell_3}(-\bfrhat_1,-\bfrhat_2,-\bfrhat_3)\nonumber\\
&=& (-1)^{\ell_1+\ell_2+\ell_3}\calP_{\ell_1\ell_2\ell_3}(\bfrhat_1,\bfrhat_2,\bfrhat_3)=\calP^*_{\ell_1\ell_2\ell_3}(\bfrhat_1,\bfrhat_2,\bfrhat_3).
\end{eqnarray}
Thus the basis functions are real if $\ell_1 + \ell_2 + \ell_3$ is even and imaginary if the sum is odd.

The isotropic functions also satisfy an orthonormality relation, which follows  from that of the spherical harmonics. We have
\begin{eqnarray}
\int d\bfrhat_1 d\bfrhat_2 d\bfrhat_3\, \calP_{\ell_1\ell_2\ell_3}(\bfrhat_1,\bfrhat_2,\bfrhat_3)\; \calP^*_{\ell'_1\ell'_2\ell'_3}(\bfrhat_1,\bfrhat_2,\bfrhat_3) = \delta^{\rm K}_{\ell_1\ell'_1}\delta^{\rm K}_{\ell_2\ell'_2}\delta^{\rm K}_{\ell_3\ell'_3}.
\end{eqnarray}
The Kronecker delta $\delta^{\rm K}_{\ell_i \ell_i'}$ is unity when its subscripts are equal and zero otherwise.

The 4PCF estimator defined in Eq.~\eqref{eqn:4pcf_estimator} can be expanded into the basis of isotropic functions (see Eq.~\ref{eqn:isofunc_4pcf}), where the  expansion coefficients depend only on the $r_i$ and are given by orthogonality as
\begin{eqnarray}
\hat{\zeta}_{\ell_1\ell_2\ell_3}(r_1,r_2,r_3) = \int \frac{d\bfs}{V} \, \delta(\bfs) \int d\bfrhat_1 d\bfrhat_2 d\bfrhat_3\,\delta(\bfs+\bfr_1)\delta(\bfs+\bfr_2)\delta(\bfs+\bfr_3) \,\calP^*_{\ell_1\ell_2\ell_3}(\bfrhat_1,\bfrhat_2,\bfrhat_3).
\label{eqn:six}
\end{eqnarray}
To avoid an over-complete basis, the radial arguments $r_i$ are ordered as $r_1\leq r_2\leq r_3$, as further discussed in \citet{Cahn2020iso}. 

To construct a density fluctuation field from the discrete galaxy counts, and also to account for the survey geometry, we use a generalized Landy-Szalay estimator (\citealt{Landy1993estimator, SS_98}, see also \citealt{Kerscher_00}) as first outlined for the angular momentum basis in \cite{SE_3pt} and further developed in \citet{Philcox2021encore, Philcox2021detection}. It is
\begin{eqnarray}
    \hat{\zeta}(\bfr_1,\bfr_2,\bfr_3) =\frac{\mathcal{N}(\bfr_1,\bfr_2,\bfr_3)}{\mathcal{R}(\bfr_1,\bfr_2,\bfr_3)},
    \label{eqn:est}
\end{eqnarray}
where $\mathcal{N} \equiv (D-R)^4$ and $\mathcal{R} \equiv R^4$, and these powers are shorthand for expanding by the binomial theorem and letting each $D$ and $R$ be evaluated at a different spatial position. $D$ means a particle drawn from the ``data'' and $R$ means a particle drawn from the ``random'' catalog (a spatially uniform catalog cut by the survey geometry). As outlined in \cite{SE_3pt}, we may estimate the numerator and denominator separately (\textit{i.e.} compute each separately averaging over the whole survey). Doing so gives optimally weighted estimates of each in the shot-noise limit, as discussed in  \citealt{SE_3pt} \S4, equations 24-26 and surrounding text. Multiplying Eq. (\ref{eqn:est}) through by $\mathcal{R}$, expanding each side of the resulting relation in the isotropic basis, reducing a product of two isotropic basis functions to a sum over single ones, and finally taking an inverse to solve the linear system so obtained \citep{SE_3pt, Philcox2021encore}, we find the edge-corrected 4PCF estimator as
\begin{eqnarray}
\hat{\zeta}_{\ell_1\ell_2\ell_3}(r_1,r_2,r_3) = \sum_{\ell'_1\ell'_2\ell'_3} \left[{\rm {\bf M}}^{-1}\right]_{\ell_1\ell_2\ell_3, \ell'_1\ell'_2\ell'_3}(r_1,r_2,r_3)\, \frac{\calN_{\ell'_1\ell'_2\ell'_3}(r_1,r_2,r_3)}{\calR_{000}(r_1,r_2,r_3)}.
\label{eqn:4pcf_estimator_edge_corrected}
\end{eqnarray}
We note that there is no mixing in the radial variables; survey geometry does not change lengths. Our notation indicates a given element of ${\rm {\bf M}}^{-1}$. This latter is the inverse of the coupling matrix ${\rm {\bf M}}$ describing how survey geometry breaks the orthogonality of our basis functions, much as Fourier modes are orthogonal only on an infinite domain.  $\calN$ denotes a measurement from the ``data-minus-random" catalog, and $\calR_{000}$ is the $\ell_1=\ell_2=\ell_3=0$ expansion coefficient of $\mathcal{R} \equiv R^4$, \textit{i.e.} the randoms' 4PCF. $\calN$ and $\mathcal{R}$ are evaluated by replacing $\delta$ in Eq. (\ref{eqn:six}) with their definitions, given below Eq. (\ref{eqn:est}). 

The coupling matrix has elements
\begin{eqnarray}
{\rm \bf{M}}_{\ell_1\ell_2\ell_3, \ell'_1\ell'_2\ell'_3}(r_1,r_2,r_3) = (4\uppi)^{-3/2} (-1)^{\ell'_1+\ell'_2+\ell'_3} \sum_{L_1L_2L_3}\;\frac{\calR_{L_1L_2L_3}(r_1,r_2,r_3)}{\calR_{000}(r_1,r_2,r_3)} \prod_{i=1}^3 D^{\rm P}_{\ell_iL_i\ell'_i}\; \calC^{\ell_i L_i \ell'_i }_{000}\nine{\ell_1}{L_1}{\ell'_1}{\ell_2}{L_2}{\ell'_2}{\ell_3}{L_3}{\ell'_3},
\label{eqn:4pcf_coupling_matrix}
\end{eqnarray}
with the coefficient
\begin{eqnarray}
\calD^{\rm P}_{\ell_1\ell_2\ell_3} = \sqrt{(2\ell_1+1)(2\ell_2+1)(2\ell_3+1)}
\label{eqn:DP}
\end{eqnarray}
which depends on the product of the \textit{primary} (hence the superscript ``P'') angular momenta.\footnote{Were one to measure an NPCF for $N\geq 5$, one would require isotropic basis functions of four arguments or more, and these basis functions require specification of intermediate angular momenta fixing how the primary momenta are coupled (further detailed in \citealt{Cahn2020iso}). The distinction between primary and intermediate angular momenta is not needed in the present work, but for consistency, we retain the superscript ``P.''} The matrix in curly brackets in Eq. (\ref{eqn:DP}) is a Wigner 9-$j$ symbol. The factor $\mathcal{C}_{000}^{\ell_i L_i \ell_i'}$ (defined in Eq. \ref{eqn:clebsch_gorden_N3} preceding it guarantees that $\ell_i, \ell'_i$, and $L_i$ can be combined to make a zero total angular momentum state. It also requires that $\ell_i+\ell'_i+L_i$ is even.

Regarding the edge correction, we note that formally ${\bf M}$ is infinite, but we have found in practice truncating it at one angular momentum beyond that used for the physical analysis is suitable (\citealt{SE_3pt}, \citealt{Philcox2021encore}). In this work we use $\ell_{\rm max} = 4$ for our analysis but work to $\ell = 5$ on all $\ell_i$ for the edge correction.\footnote{This truncation does not induce spurious parity-odd modes; if it did, we would see them when we edge-correct our mock catalogs.} Further details regarding the suitability of, when performing the edge correction, truncating at an $\ell$ one above that used for the analysis, are in \cite{SE_3pt}. Ultimately this suitability stems from the rough tri-diagonality of the edge-correction matrix (see their \S4.2. and our Fig. \ref{fig:edge_correction}).

\subsection{Illustration With Toy Tetrahedra}
\begin{figure}
     \centering
     \begin{subfigure}[b]{0.45\textwidth}
         \centering
         \includegraphics[width=1\textwidth]{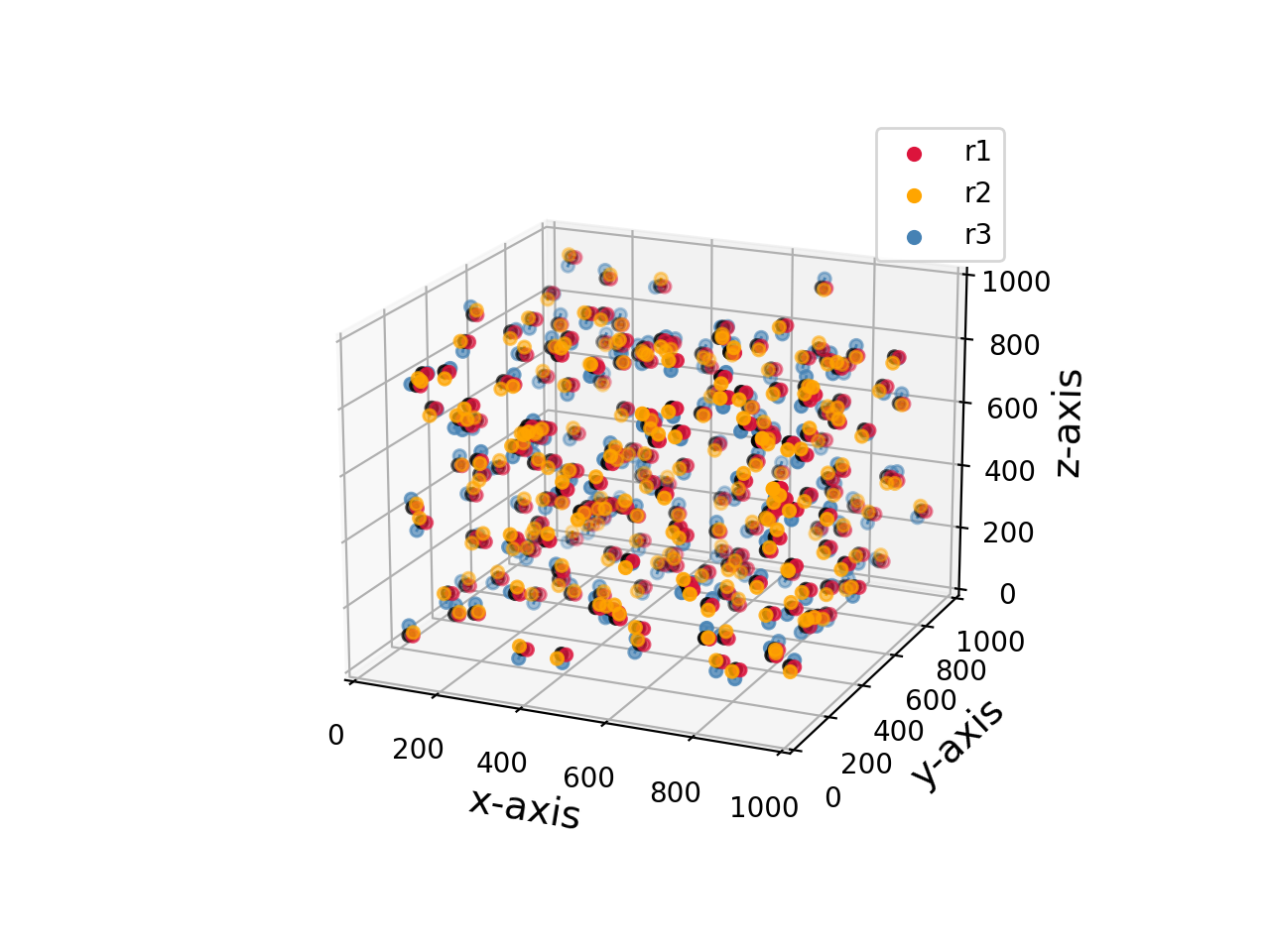}
         \label{fig:Tet_sim_10x20x30}
     \end{subfigure}
     \begin{subfigure}[b]{0.45\textwidth}
         \centering
         \includegraphics[width=1.1\textwidth]{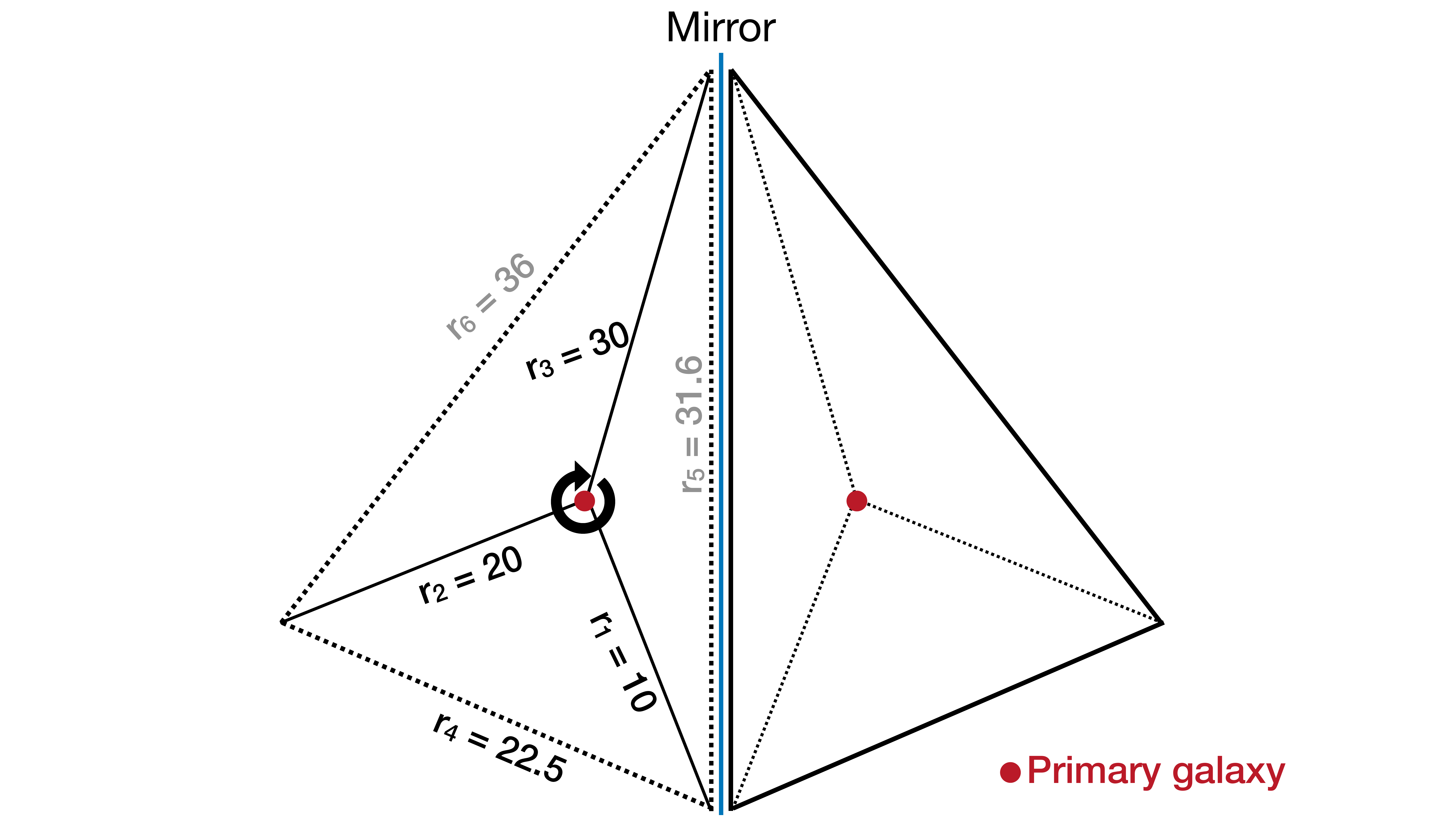}
         \label{fig:tet_sketch}
     \end{subfigure}
     \caption{{\it Left:} A box with side length $L_{\rm box}=1000 \mpch$ filled with tetrahedra (upper left, small panel). Each of them has a {\it unique} primary (black) from which the three sides with respective lengths $r_1\!\sim\!10 \mpch$ (red), $r_2\!\sim\!20 \mpch$ (yellow), and $r_3$$\sim$$30 \mpch$ (blue) extend. The larger panel on the left is a zoom-in on the full box to display the tetrahedra more clearly. {\it Right:} A sketch of a ``clockwise" tetrahedron and  its ``counterclockwise" mirror image. The primary is in red. Our convention on clockwise and counterclockwise is detailed in Fig. \ref{fig:tet_cw_ccw}. On the left, the red point is closer to us than all the others. Thus the tetrahedron on the left is clockwise, as looking down from the primary, we go clockwise as we move from the smallest side to the largest. On the right, the primary (in red) is behind the other galaxies, so looking down from it towards them will reverse the handedness. Thus the rightmost tetrahedron is counterclockwise as viewed from the primary.}
     \label{fig:tet-sim}
\end{figure}

To understand the parity-odd measurement more intuitively, we study cubic boxes of side length $L_{\rm box}=1000\; \mpch$ with tetrahedra tuned to produce particular parity signals. To fill the boxes as fully as possible, yet at the same time have tetrahedra with a minimum side length of order $10\;\mpch$, which is similar to the situation in our BOSS dataset, we choose the three sides extending from the primary to be roughly $r_1\!\sim\!10\; \mpch$, $r_2\!\sim\!20\; \mpch$, and $r_3\!\sim\!30\; \mpch$. We require that the minimum separation between primaries be twice the longest side of the tetrahedron ({\it i.e. $60\;\mpch$}) in order to minimize any overlap between tetrahedra. Finally, we have $N_{\rm tets} \sim\!1,500$ tetrahedra within each cubic box. Fig.~\ref{fig:tet-sim} shows an example box and also an example tetrahedron and its partner under parity transformation. 

In 3D, parity transformation is equivalent to a mirror reflection across a 2D plane (which flips the sign of the coordinate axis perpendicular to that plane) plus a $180^{\circ}$ rotation around this latter axis. For simplicity, in Fig.~\ref{fig:tet-sim} we depict the parity transformation simply as a mirror reflection, since our basis, being isotropic, always averages over 3D rotations and is thus insensitive to a $180^{\circ}$ rotation. Put another way, only the mirroring fundamentally alters the shape of a tetrahedron; the $180^{\circ}$ rotation only changes its orientation in absolute space.\footnote{In general parity transformation and mirror reflection are distinct operations.} In particular, one can imagine the mirroring as taking one side and ``pulling it through'' the tetrahedron from being on one side of the plane formed by the other two sides, to being on the other side of this plane.

In practice, we must allow all four vertices of each tetrahedron a chance to be the primary \citep{Philcox2021encore}. However, for this toy tetrahedron illustration we restricted to bins in side length (radial bins) such that $9\;\mpch< r _i<30\;\mpch$ for all $r_i$ so that nearly always only one of the four vertices will satisfy the radial bins required for each tetrahedron. This means that the contribution to the signal will stem only from the isotropic function evaluated on unit vectors extending from a \textit{single} primary; this renders it easier to understand the measured signal in this toy model. 

We set $\bfrhat_1$ to be along $\bfxhat$, $\bf{\hat{r}}_2$ to be along $\bfyhat$, and $\bf{\hat{r}}_3$ to be along $\bfzhat$. The tetrahedron is therefore clockwise at the only primary allowed in this toy model; our convention is presented in Fig. \ref{fig:tet_cw_ccw}. A parity transformation simply means that we interchange the sides, so that $\bf{\hat{r}}_2$ aligns with the $x$-axis and $\bf{\hat{r}}_1$ aligns with the $y$-axis, while $\bf{\hat{r}}_3$ is unchanged.  Again, characterization as ``clockwise'' or ``counterclockwise'' depends on at which vertex one sits, as discussed in Fig. \ref{fig:tet_cw_ccw}, but here is unambiguous. By restricting the radial bins we use, we force only one galaxy to be the primary; the other choices of primary will not lead to sides extending from them that fall within our chosen bins. Thus, we can ensure that the tetrahedra for this test are perfectly counterclockwise as viewed from the single primary allowed.

We produce toy boxes in three configurations. The first is filled with only counterclockwise tetrahedra (enforced by the radial bin restriction); the second has only clockwise tetrahedra, and the third has an equal mix. To render each toy box somewhat more realistic, we randomly rotate each vertex by an angle $\theta\in[-180^{\circ}, 180^{\circ}]$ around each of the three Cartesian coordinates. We also add random numbers $\Delta r_1\in[0,1]$, $\Delta r_2\in[-2,2]$, and $\Delta r_3\in[-1,0]$ (in $\mpch$) to, respectively, $r_1$, $r_2$, and $r_3$. We choose these ranges such that we always have $r_1 < r_2 < r_3$. Hence the parity of a given tetrahedron will not be flipped by these additions.

\begin{figure}
    \centering
    \begin{subfigure}[b]{0.48\textwidth}
    \includegraphics[width=.9\textwidth]{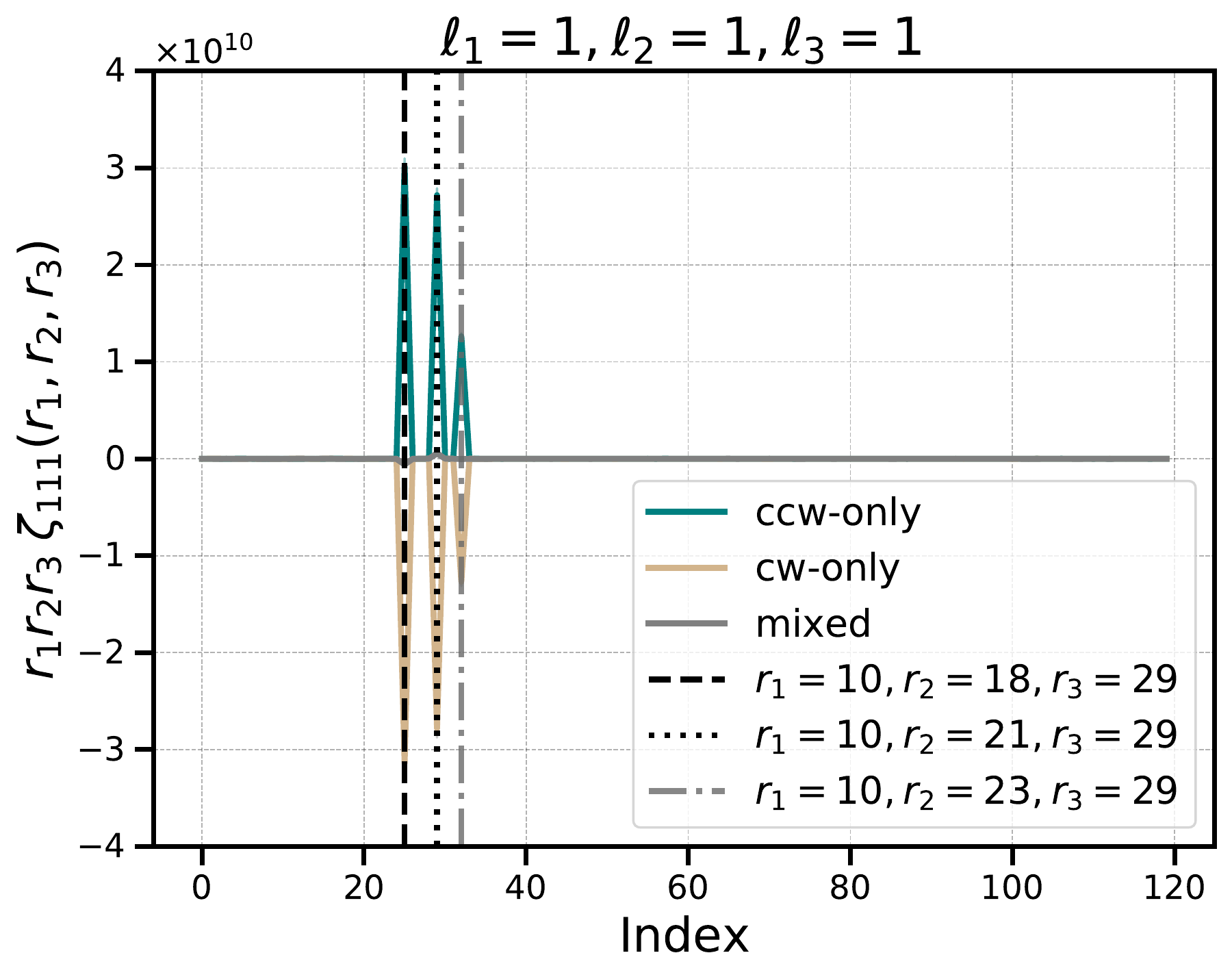}
    \end{subfigure}
    \begin{subfigure}[b]{0.45\textwidth}
    \includegraphics[width=0.95\textwidth]{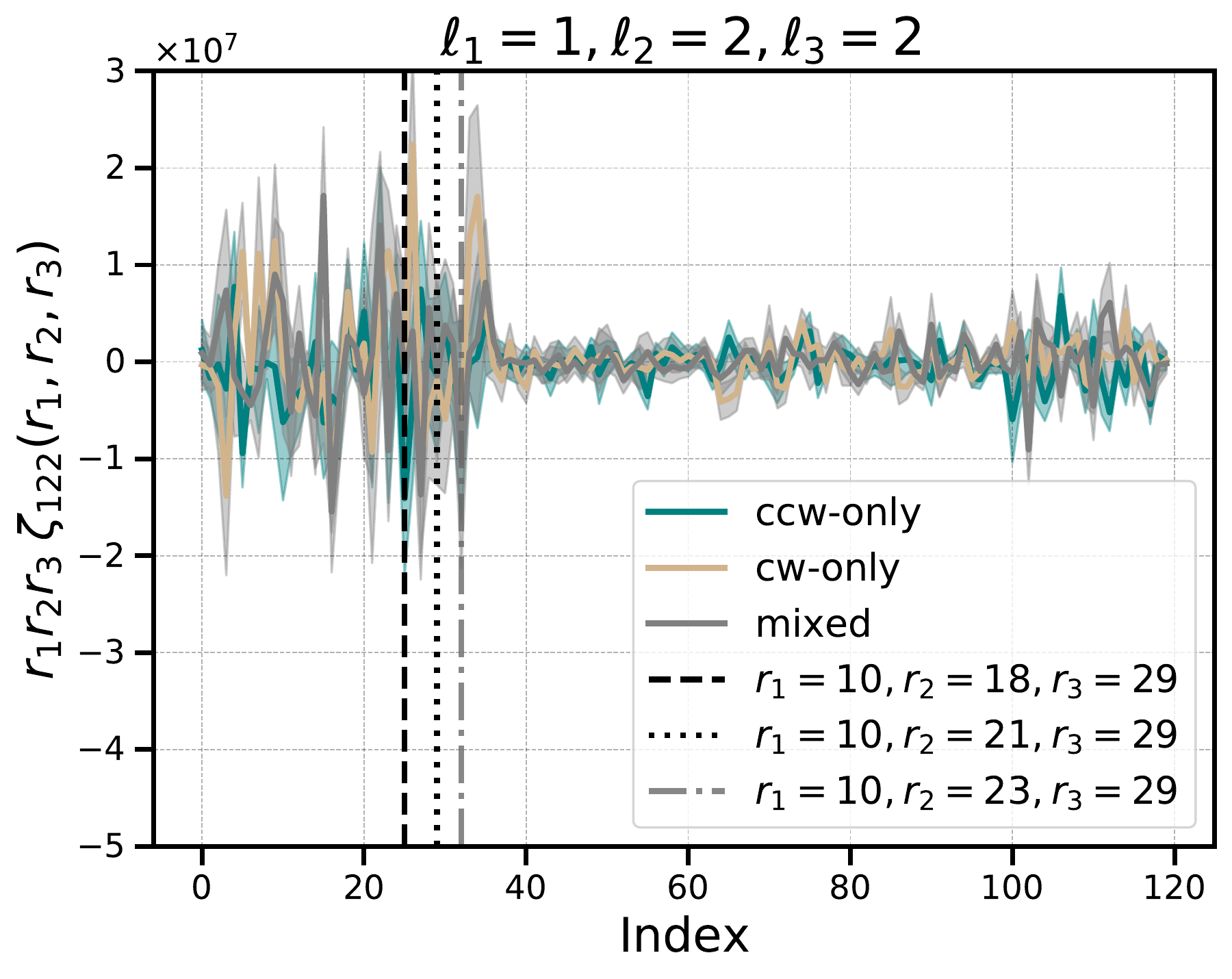}
    \end{subfigure}
    \caption{Here we display 4PCF measurements from the illustrative toy boxes. The left and right panels each show a different channel, as indicated in their titles. In both panels, the  ``clockwise-only'' box results are in blue, the ``counterclockwise-only'' box results are in brown, and the ``mixed'' box results are in grey. {\it Left}: Projection onto $\calP_{111}$.  As discussed in the main text, we expect a negative projection for the clockwise box and a positive projection for the counterclockwise box. The mixed will have on average  zero projection. These expectations are borne out and indeed analytic calculation yields agreement with the measured signals. {\it Right}: Projection onto $\calP_{122}(\bfrhat_1,\bfrhat_2,\bfrhat_3) \propto i\bfrhat_1\cdot(\bfrhat_2\times\bfrhat_3) (\bfrhat_2\cdot\bfrhat_3)$. By construction the scalar triple product $\bfrhat_1 \cdot (\bfrhat_2 \times \bfrhat_3)$ is unity for all three of our illustrative boxes. The amplitudes on the righthand side are not strictly zero because there are still a small number of tetrahedra formed by connecting secondaries around one primary with secondaries around another. Additionally, the amplitude fluctuations divide into two envelopes at bin index $36$. This bin index corresponds to raising the index for $r_1$ (the smallest side and the slowest-varying one).}
    \label{fig:4PCF_measurement_odd_tet_10x20x30_111}
\end{figure}

Fig.~\ref{fig:4PCF_measurement_odd_tet_10x20x30_111} shows the 4PCF of these illustrative boxes. We may approximate each tetrahedron as a sphere about the primary of radius roughly $20 \mpch$, with volume $V_{\rm tet}$, to estimate the expected 4PCF amplitude. We define $n$ as the local number density due to a given tetrahedron, $n = 1/V_{\rm tet}$, and $\bar{n}$ as the average number density in the box, $\bar{n} = N_{\rm tets} /L_{\rm box}^{3}$. We find $\av{\delta^4}=\left[n/\bar{n}-1 \right]^4 \approx 2\times 10^5$. The lowest-lying parity-odd isotropic basis function is $\calP_{111} = -3 i / [\sqrt{2}(4\uppi)^{3/2} ]\; \bfrhat_1\cdot(\bfrhat_2\times\bfrhat_3)$ \citep{Cahn2020iso}. As expected, the counterclockwise-only box has a positive projection onto this function, while the clockwise-only box has a negative projection. The mixed box is consistent with zero projection onto $\calP_{111}$ on average. We can analytically predict the ratios among the 4PCF coefficients for different channels $\ell_1, \ell_2, \ell_3$. We compare the mean ratios of the measured 4PCF coefficients for several combinations to these predictions and find good agreement.\footnote{As an example, the analytically predicted ratio for angular momenta $\{1,1,1\}$ and $\{1,3,3\}$ is $\zeta_{111}/\zeta_{133}=-1.07$, and we measure from the data $\bar{\zeta}_{111}/\bar{\zeta}_{133}=-1.07$, with $\bar{\zeta}$ denoting the bin-averaging.} This also serves as an additional test of our code (the code is further discussed in \S\ref{subsec:data_mocks_code}).

\subsection{Internal Cancellation}
\label{subsec:int_canc}
For a given tetrahedron, in practice (but not in our illustrative boxes above), each of the four vertices gets a chance to serve as the primary about which the isotropic basis function expansion is computed. Some of these vertices will be ``clockwise'', and some ``counterclockwise''. Hence if co-added into the same channel and triple-bin there will be ``internal cancellation'' and consequent reduction of any parity-odd signal. However, if the radial bins are made fine enough then each vertex, in virtue of the presumably unique lengths of the sides extending from it, will be accumulated to a different triple-bin.\footnote{Save for isosceles or equilateral tetrahedra, which we exclude from our analysis in any case.} Thus finer binning can reduce the internal cancellation and increase the signal. This is much the same as in a configuration-space BAO search, where fine enough bins must be chosen that the BAO feature in the 2PCF is not averaged out by all being added into a single bin.

\section{Guiding Principles for the Analysis}
\label{sec:guidelines}

{\color{black}{(This section used to be directly after the introduction; now it has been moved here.)}}

To isolate the potentially parity-violating component of the 4PCF, we expand the correlation function in two distinct sets of isotropic basis functions, one that is parity-even and one that is parity-odd.  These are constructed from products of three spherical harmonics with angular-momentum indices $\ell_1,\ell_2,\ell_3$. Isotropy requires that the $\ell_i$ satisfy the triangular inequality.  If the sum of the $\ell_i$ is even, the product is parity-even and if the sum is odd, the product is parity-odd.  In this analysis, only the parity-odd elements are used.  Each basis element is a function of three radial distances, the length of the sides from a chosen vertex among the four defining a tetrahedron.  In practice, the radial distances are binned.

Ideally, one would like to capture as much information as possible, using many narrow radial bins, and working to some high value of the $\ell_i$. This would increase the difficulty of evaluating all the independent amplitudes, but in fact, the technique of \cite{SE_3pt, Philcox2021encore} makes this manageable, as we discuss in \S\ref{sec:basis}. The real challenge is in determining the covariance matrix. This is especially critical when looking for parity violation, where establishing a statistically significant non-zero signal is the crux and hence understanding the inevitable statistical fluctuations is essential.

A fairly modest choice of separating the radial variable into ten bins and including $\ell_i$ only up to $\ell_{\rm max} = 4$ results in $23 \times  120 = 2,760$ independent amplitudes; we also consider eighteen radial bins, which produces $816\times 120 = 18,768$ 4PCF amplitudes. Determining the covariance matrix is thus a formidable challenge. In order to invert a sampling covariance matrix, as required for calculating $\chi^2$, we need at least many mocks as the dimensionality of the data vector, which is in excess of the 2,000 available to us for BOSS. 

We have chosen three ways to obtain the covariance matrix. First, the NPCF covariance matrix can be calculated analytically under the assumption of a Gaussian Random Field (GRF) as shown in ~\citet{Hou2021analytic}. The GRF does not on average have any parity-violation at the \textit{signal} level, but it can still have non-zero fluctuations in parity-odd modes. Therefore the GRF is still the leading-order contribution to the \textit{covariance} of the parity-odd modes. This is simply the statement that a signal may be zero but its root-mean-square may not. We fit the analytic template covariance matrix by varying the number density and volume with respect to the covariance matrix derived from the mocks.

With this analytic template in hand, we may (i) directly compute the $\chi^2$ of the data using the adjusted analytic covariance matrix. An alternative (ii) is to compress the data vector to reduce its dimensionality. The eigenvectors of the analytical covariance matrix with the smallest eigenvalues represent the linear combinations of basis functions that have the smallest statistical uncertainties. We then expand the measured 4PCF using just the $N_{\rm eig}$ best expansion functions.  We may also determine the $N_{\rm eig}\times N_{\rm eig}$ covariance matrix directly from the mocks. Since $N_{\rm eig}$  is much less than the number of mocks this covariance matrix is invertible. Finally, we may (iii) use the empirical covariance matrix from the mocks directly, with no involvement of the analytic template at all, by considering  many fewer channels than in (i) and (ii) (we lower $\ell_{\rm max}$ to be $\ell_{\rm max} = 2$). A substantial reduction of the number of channels is required to enable inverting the empirical covariance matrix employed in this approach, and so we lose statistical power. We thus treat (iii) as a test rather than as giving us the main result of our analysis. The reliability of all three approaches above can be assessed using the mocks themselves by verifying that their $\chi^2$ (or $T^2$) values match the expected distribution.

\section{Dataset and Covariance}
\label{sec:data}
\begin{figure}
    \centering
    \includegraphics[width=0.4\textwidth]{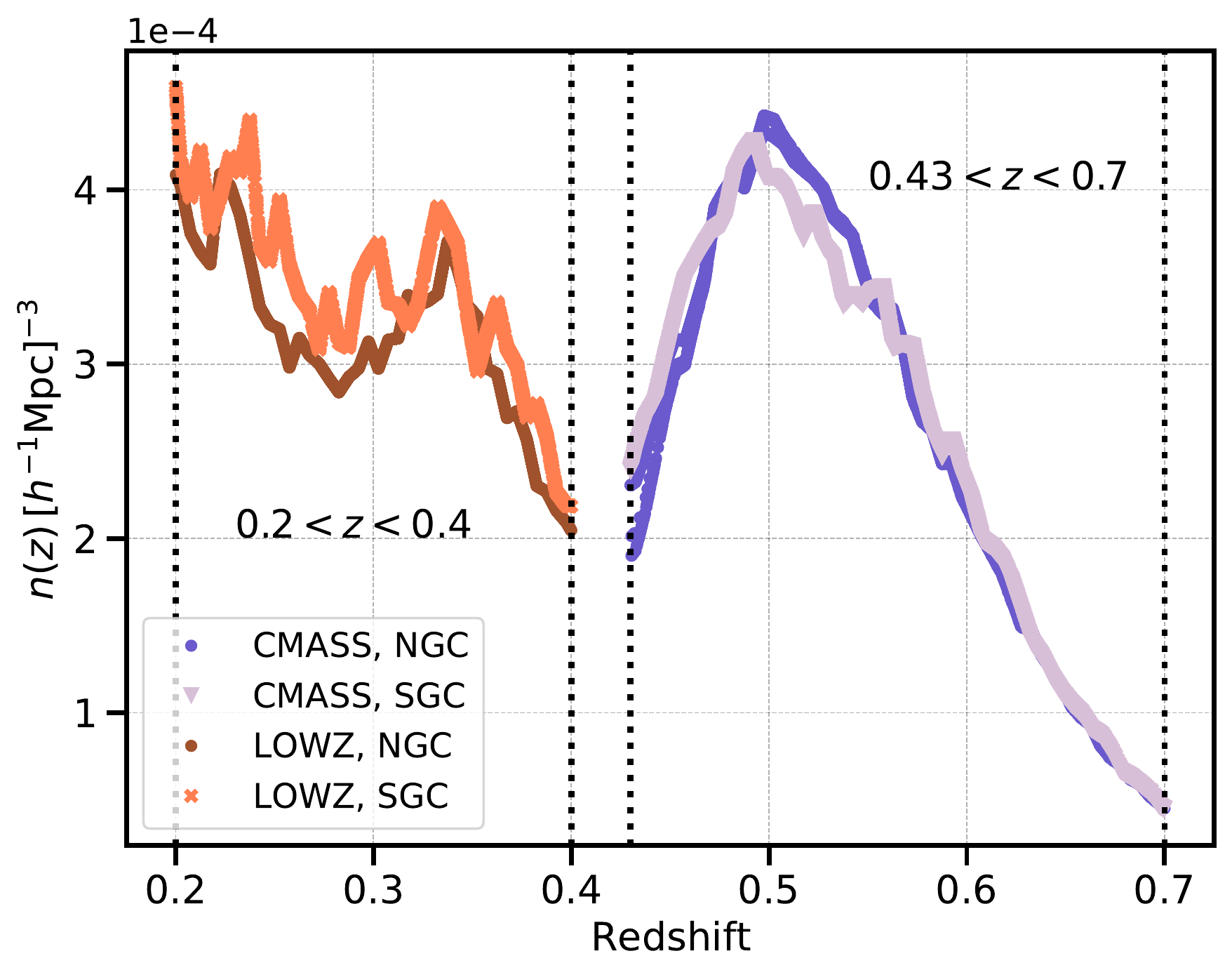}
    \caption{The number density $n$ as a function of redshift $z$ for the two BOSS samples used in this work. The LOWZ ($0.2<z<0.4$) North Galactic Cap (NGC) is brown and South Galactic Cap (SGC) is orange. For CMASS ($0.43<z<0.7$), the NGC is in purple and the SGC is in lavender. We intentionally do not allow redshift overlap between the two samples, and the redshift gap $\Delta z=0.03$ corresponds to a comoving radial separation of $73\, \mpch$. This separation means that the samples are fairly independent; the 2PCF $\xi$ between a point in LOWZ and CMASS would be of order 1\% at this scale. The covariance between two ``worst-case'' tetrahedra, one in each sample (where each is very close to the respective edge), is of order $\xi^4$. Few tetrahedra are near enough on either edge to be significantly correlated with those in the other slice. The plot shows that LOWZ has both a more uniform selection function and a somewhat higher average number density than CMASS. It also lacks the strongly decaying tail with an increasing redshift that CMASS displays. These points are important when assessing the possible impact of systematics on each sample and when addressing  any differences between the detection significances in the two samples.}
    \label{fig:n_of_z_CMASSxLOWZ}
\end{figure}

We use the final galaxy catalog of the Baryon Oscillation Spectroscopic Survey (BOSS), from the twelfth data release (DR12) of the Sloan Digital Sky Survey-III (SDSS-III). The catalog is split into the North  Galactic Cap (NGC) and the South Galactic Cap (SGC). The catalog contains two samples, CMASS and LOWZ, which were selected via the SDSS multicolor photometry and cover a redshift range of $0.15 < z < 0.7$. CMASS and LOWZ use similar target selection algorithms~\citep{Eisenstein2001BOSStarget,Cannon20062dfSDSS}. The target selection algorithm provides samples that are mainly composed of Luminous Red Galaxies (LRGs). For CMASS the selection algorithm is further tuned to select massive objects uniformly in redshift~\citep{Reid2016boss}, which results in an approximately mass-limited sample down to a stellar mass $M\!\sim\!10^{11.3}\, h^{-1} {\rm M}_{\odot}$~\citep{Maraston2013stellar,Leauthaud2016stellarmass,Saito2016stellarmass,Bundy2017stellarmass}. The majority of CMASS is LRGs ($\sim\!74$ percent), while the rest is late-type spirals~\citep{Masters2011BOSSmorphology}. LOWZ consists primarily of LRGs~\citep{Parejko2013LOWZ}. Despite the difference in target selection, the LRGs in the two samples have similar stellar mass distribution~\citep{Maraston2013stellar}.
We apply a redshift cut of $0.43<z<0.7$ to CMASS, which results in a redshift tail from LOWZ at redshift $0.43<z<0.5$. This tail ($\sim\!15\%$ of the entire ``CMASS'' sample) slightly raises the purity of the CMASS sample by adding more LRGs. To ensure that the LOWZ sample as used here is independent of the CMASS one, we apply a redshift cut of $0.2<z<0.4$ to the former. This cut produces a separation of about $70\mpch$ between the lower edge of CMASS and the upper edge of LOWZ. Fig.~\ref{fig:n_of_z_CMASSxLOWZ} shows the number density for each sample as a function of redshift. 
Finally, we note that the early LOWZ target selection was not uniform due to the use of different iterations of the galaxy-star separation algorithm~\citep{Reid2016boss}. Therefore we do not include those early chunks in this analysis.

\subsection{4PCF from Data and Mocks}
\label{subsec:data_mocks_code}
For our main analyses, we considered tetrahedra with side lengths $r_1,\;r_2$ and $r_3$ from the primary ranging from $20\, h^{-1} {\rm Mpc}$ to $160\, h^{-1} {\rm Mpc}$ inclusive, split into ten linearly-spaced bins, and also from $20\, h^{-1} {\rm Mpc}$ to $164\, h^{-1} {\rm Mpc}$ inclusive, split into eighteen linearly-spaced bins. We also explored a coarser binning (six bins), presented as a test in Appendix \ref{sec:coarse}. As a result, the sides of the tetrahedron that do not include the primary can range from zero to $320\, h^{-1} {\rm Mpc}$. We expand the parity-odd 4PCF in the 23 angular channels with $\ell_i \leq 4$ given in Cartesian form in Appendix \ref{app:fns}. For the edge corrections, we include all functions with $\ell_i\leq 5$, as further discussed in \S\ref{sec:basis}. We also compute the even-parity modes (for which we do not reproduce the basis functions here), as they are needed within the edge-correction, despite that our actual analysis focuses solely on the parity-odd ones. 

To each galaxy, we apply a total weight $w$ given by
\begin{eqnarray}
w = w_{\rm fkp} w_{\rm sys} (w_{\rm noz}+w_{\rm cp}-1),
\end{eqnarray}
with the systematic weight $w_{\rm sys}$ (a combined weight for stellar density and seeing), the redshift failure weight $w_{\rm noz}$, and the fiber collision weight $w_{\rm cp}$ (subscript ``cp'' for ``close pairs''). The FKP weight~\citep{Feldman1994FKP} is $w_{\rm fkp} = \left[1+ n(z)P_0\right]^{-1}$, with $P_0 = 10^4\, [h^{-1}{\rm Mpc}]^3$; $n(z)$ is the weighted number density at the given redshift. We use the public random catalog provided by BOSS\footnote{\url{https://data.sdss.org/sas/dr12/boss/lss/}}, which is fifty times the size of the data catalog. We split it into 32 chunks such that each chunk is about 1.5 to 2 times the size of the data; the reason is discussed in \cite{SE_3pt} and \cite{Philcox2021detection} though the detailed splitting does not affect the 4PCF algorithm speed notably. Each chunk is first randomly shuffled and then normalized so that its weighted sum matches the sum of the completeness weights $w_{\rm c}$ of the data.

We use 2,000 public MultiDark \patchy lightcone mocks~\citep[\patchy hereafter;][]{Kitaura2016patchy,Rodriguez-Torres2017patchy}. These mocks use second-order Lagrangian Perturbation Theory (2LPT) plus a spherical collapse prescription calibrated on full N-body simulations, and the mocks are additionally  calibrated to match the 2-point and some 3-point statistics of the observed BOSS CMASS and LOWZ samples. The \patchy mocks include realistic survey geometry as well as many observational effects, further detailed in \citet{Rodriguez-Torres2017patchy}.

The 4PCF measurement for the BOSS data catalog and the 32 $(D-R)$ catalogs alone takes $0.33$ GPU-hours for the measurement where we use ten radial bins, and $0.65$ GPU-hours for the measurement where we use eighteen radial bins; both timings are using 69 NVIDIA A100 GPUs (further discussed below) and include edge-correction, which is done on the CPU linked to the GPU and takes negligible extra time. 

All calculations were done with the GPU-accelerated NPCF code \textsc{CADENZA} ~\citep{Slepian2021gpu}, built on the CPU-based NPCF code \textsc{ENCORE}~\citep{Philcox2021encore}. \textsc{CADENZA} uses the representation of the basis functions in spherical harmonics, for reasons outlined in \S\ref{sec:basis}. Tests of \textsc{ENCORE} are described in \citep{Philcox2021encore}, and \textsc{CADENZA} was verified to obtain exactly the same outputs on a fiducial data set used for testing. As noted already, we used 69 NVIDIA A100 GPUs simultaneously on the HiPerGator cluster at the University of Florida. Overall, all of the computations performed for this work would have taken roughly 7.5 million CPU hours, or about 20 months of continuously running on a cluster of 500 CPU cores. In practice, since \textsc{CADENZA} is about 140$\times$ faster than \textsc{ENCORE} for the 4PCF, we required about 54,000 GPU-hours, or about 1 month if one ran continuously on 69 GPUs.

\begin{figure}
    \centering
    \begin{subfigure}[b]{1\textwidth}
    \includegraphics[width=1.0\textwidth]{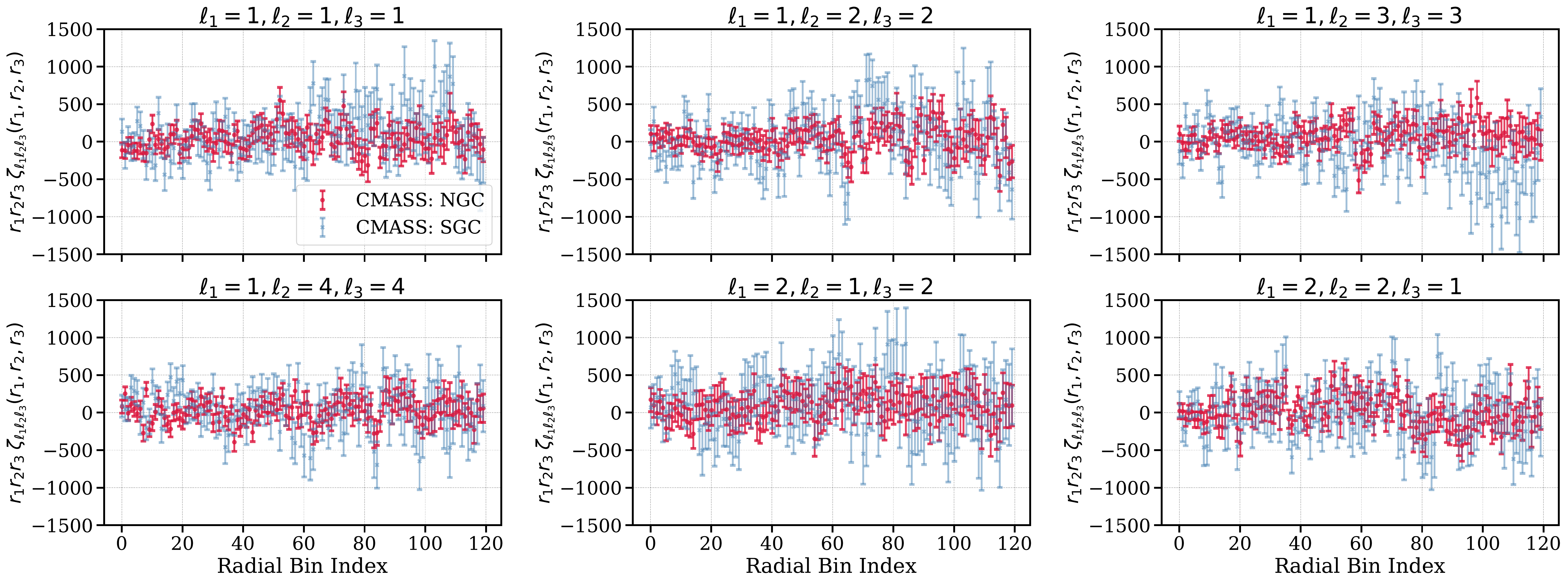}
    \end{subfigure}\vfill
    \begin{subfigure}[b]{1\textwidth}
    \includegraphics[width=1.0\textwidth]{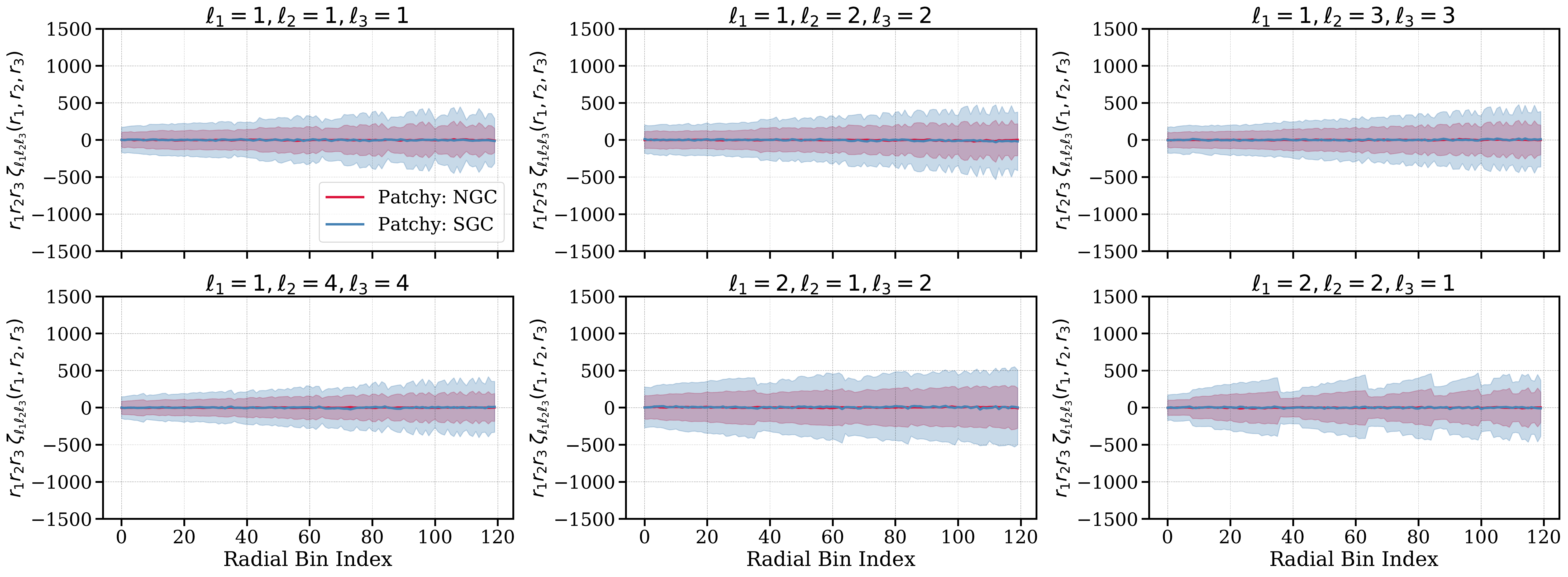}
    \end{subfigure}
    \caption{{\it Upper}: The parity-odd 4PCF for the BOSS CMASS data with {\bf ten} radial bins, with NGC in red and SGC in blue, showing only the six lowest-lying channels. The error bars are the root-mean-square of the \patchy mocks (over our set of 2,000). Here for legibility we focus on the ten-bin results; the eighteen-bin results are in Appendix \ref{app:data_all_angular_ell}, as are the remaining channels for the ten-bin analysis. We have mapped the three radial bins to a single index, with $r_3$ varying the fastest and $r_1$ the slowest. This is done in all similar plots that follow. {\it Lower}: The mean of $2,000$ \patchy mocks (solid curves) for both NGC (red) and SGC (blue). The shaded region is the rms expected for a single mock; this set of panels is intended to display the region around zero that the 4PCF of a dataset with no true parity-odd signal might inhabit.}
    \label{fig:six_channels_lowz}
\end{figure}

\begin{figure}
    \centering
    \begin{subfigure}[b]{1\textwidth}
    \includegraphics[width=1\textwidth]{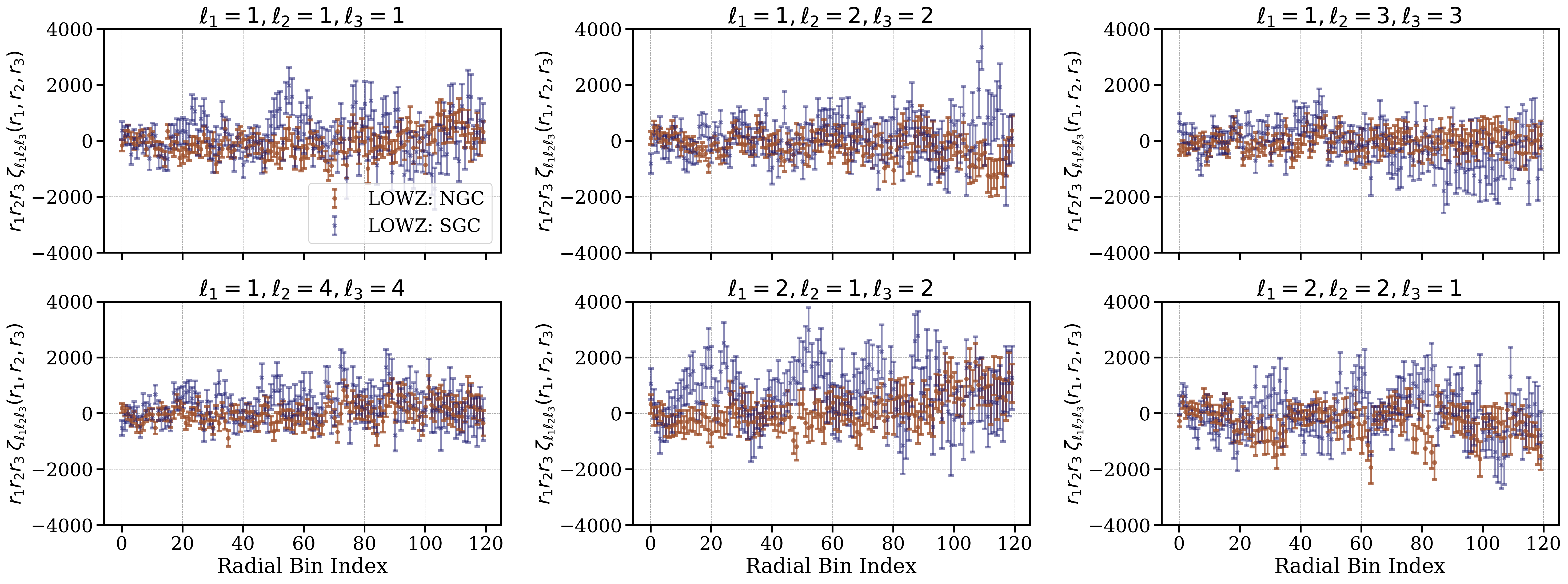}    
    \end{subfigure}\vfill
    \begin{subfigure}[b]{1\textwidth}
    \includegraphics[width=1\textwidth]{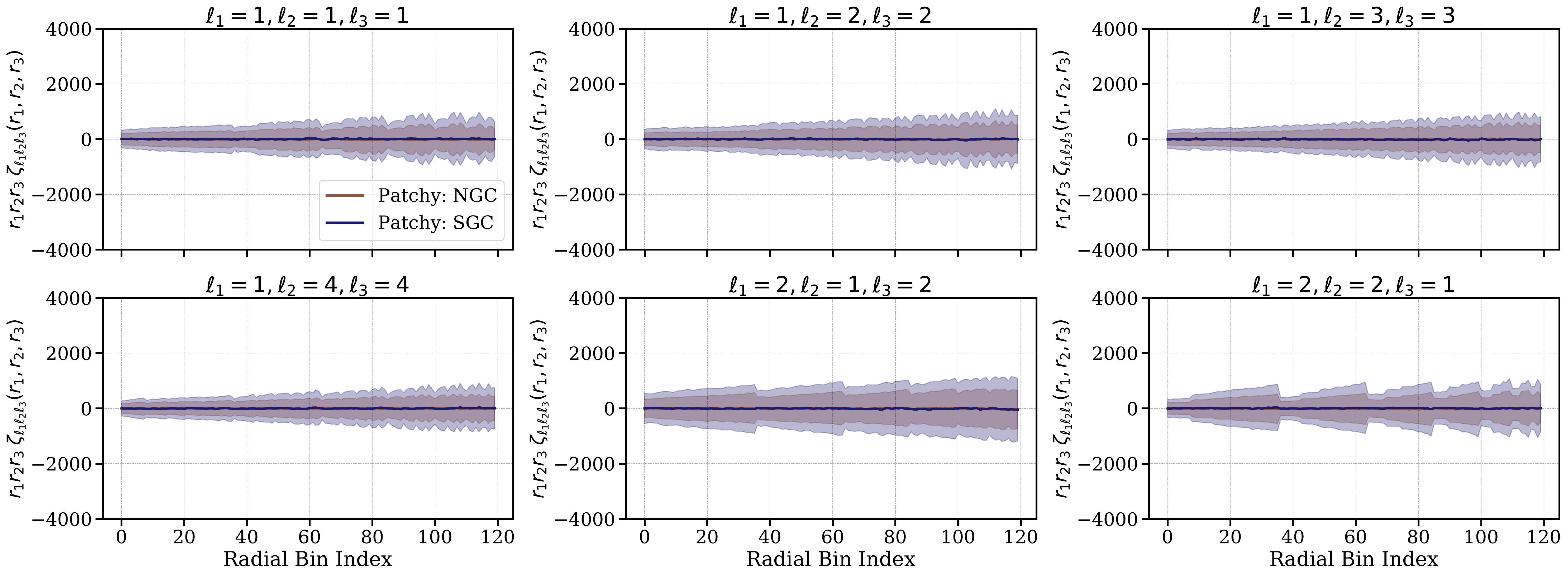}    
    \end{subfigure}
    \caption{\textit{First and second rows}: measurement of the parity-odd 4PCF using {\bf ten} radial bins for the BOSS LOWZ data in the NGC (brown) and SGC (blue). We show the same six channels as in Fig. \ref{fig:six_channels_lowz}, and the error bars are again the rms over the mocks. Again for legibility, we focus on the ten-bin results; the eighteen-bin results are in Appendix \ref{app:data_all_angular_ell}, as are the remaining channels for the ten-bin analysis. \textit{Third and fourth rows}: The mean of $2,000$ \patchy mocks (solid curves) for both NGC (brown) and SGC (blue). The shaded region is the rms expected for a single mock.}
    \label{fig:4PCF_measurement_odd_boss_patchy_lowz}
\end{figure}

\subsection{Covariance Matrix}
\label{subsec:covariance_matrix}
\subsubsection{Analytic Covariance} 
The covariance matrix for the 4PCF may be computed analytically if we assume that the density is a Gaussian Random Field \citep{Hou2021analytic}. It is expressed in terms of $f$-functions 
\begin{eqnarray}
f_{\ell_1\ell_2\ell_3}(r_1,r_2,s)\equiv\int\frac{k^2 dk}{2\uppi^2}\;P(k)j_{\ell_1}(kr_1)j_{\ell_2}(kr_2)j_{\ell_3}(ks),
\label{eqn:f3l-tensor}
\end{eqnarray}
with $j_{\ell}(kr)$ the spherical Bessel function of order $\ell$ and $P(k)$ the galaxy power spectrum. The power spectrum is taken to be isotropic; the effect of RSD (after averaging over the line of sight) is simply to boost the amplitude, which is embedded in the input power spectrum and also partially absorbed in our later fitting for the best amplitude. Fast approaches do exist to evaluate these $f$-integrals (\textit{e.g.} \citealt{rot_meth}) but here we use direct computation.

The analytic covariance matrix couples each vertex of an (unprimed) tetrahedron to a vertex in another (primed) tetrahedron. Since in our approach, the primary vertex is distinguished from the other three, in the covariance expression there are two pieces: in one the two primary vertices, unprimed and primed, are paired with each other; in the other, they are not. The final result will be a  sum given by products involving Wigner-$nj$ symbols and the $f$-functions of Eq. (\ref{eqn:f3l-tensor}). Below we simply reproduce the final result derived in~\citet{Hou2021analytic}, for brevity here we define $ \Lambda \equiv \left\{{\ell_1, \ell_2, \ell_3} \right\}$ and $\Lambda' \equiv \left\{ \ell_1', \ell_2', \ell_3'\right\}$. 
\begin{eqnarray}
\mathrm{Cov}_{\Lambda,\Lambda'}(r_1,r_2,r_3; r'_1,r'_2,r'_3)= \mathrm{Cov}_{\Lambda,\Lambda'}^{\rm I}(r_1,r_2,r_3; r'_1,r'_2,r'_3) + \mathrm{Cov}_{\Lambda,\Lambda'}^{\rm II}(r_1,r_2,r_3; r'_1,r'_2,r'_3).
\label{eqn:fun_covar}
\end{eqnarray}

\begeqar\label{eqn:final-cov-caseI}
    \mathrm{Cov}_{\Lambda,\Lambda'}^{\rm I}(r_1,r_2,r_3; r'_1,r'_2,r'_3) &=& (4\uppi)^4 \sum_G (-1)^{\Sigma(\Lambda)(1-\calE_G)/2}\sum_{L_1L_2L_3} \calD^{\rm P}_{L_1L_2L_3}\,\mathcal{C}^{L_1L_2L_3}_{000} \begin{Bmatrix} \ell_{G1} & \ell_{G2} & \ell_{G3}\\ \ell_1' & \ell_2' & \ell_3'\\ L_1 & L_2 & L_3\end{Bmatrix}\non
    &&\,\times\, \int \frac{s^2ds}{V} \prod_{i=1}^3 \left[(-1)^{(-\ell_{Gi}-\ell_i'+L_i)/2}\,\calD^{\rm P}_{\ell_i\ell'_iL_i}\mathcal{C}^{\ell_{Gi}\ell'_i L_i}_{000} 
    \xi(s)f_{\ell_{Gi}\ell_i'L_i}(r_{Gi},r_i',s) \right], 
\endeqar

\begeqar\label{eqn:final-cov-caseII}
    \mathrm{Cov}_{\Lambda,\Lambda'}^{\rm II}(r_1,r_2,r_3; r'_1,r'_2,r'_3) &=& (4\uppi)^4\sum_{G,H}(-1)^{\Sigma(\Lambda')(1-\calE_H)/2}
    \sum_{L_1L_2L_3}\calD^{\rm P}_{L_1L_2L_3}\calC^{L_1L_2L_3}_{000}\, \begin{Bmatrix} \ell_{G1} & \ell_{G2} & \ell_{G3}\\ \ell_{H1}' & \ell_{H2}' & \ell_{H3}'\\ L_1 & L_2 & L_3\end{Bmatrix}\non
    &&\,\times\,\int\frac{s^2ds}{V}\prod_{i=1}^3\left[
    (-1)^{(-\ell_{Gi}-\ell'_{Hi}+L_i)/2}\,
    \calD^{\rm P}_{\ell_{Gi}\ell'_{Hi}L_i} \calC^{\ell_{Gi}\ell'_{Hi}L_i}_{000} \right]\\
    &&\,\times\, f_{\ell_{G1}0\ell_{G1}}(r_{G1},0,s)f_{0\ell_{H1}'\ell_{H1}'}(0,r_{H1}',s)f_{\ell_{G2}\ell_{H2}'L_2}(r_{G2},r_{H2}',s)f_{\ell_{G3}\ell_{H3}'L_3}(r_{G3},r_{H3}',s).\nonumber
\endeqar
In Case I, the sum over $G$ includes all six permutations, while in Case II the sum over $G$ includes only cyclic permutations. The sum over $H$ includes all six permutations. $\calE_{G}$ is the Levi-Civita symbol, giving a negative sign if the permutation $G$ is odd. $\Sigma(\Lambda') \equiv \ell_1' + \ell_2' + \ell_3'$. The coefficients $\calC^{\ell_1\ell_2\ell_3}_{m_1m_2m_3}$ and $\calD^{\rm P}_{\ell_1\ell_2\ell_3}$ are given in Eq.~\eqref{eqn:clebsch_gorden_N3} and Eq.~\eqref{eqn:DP}, respectively. 
For the parity-even modes of the 4PCF, the phases in Eqs. (\ref{eqn:final-cov-caseI}) and (\ref{eqn:final-cov-caseII}) are identically unity because the sum of the three angular momenta is always even. However, these phases do matter for the parity-odd 4PCF. 

We also note that, apart from the leading Gaussian contribution, the covariance should also contain non-Gaussian contributions as well as the coupling of the redshift-space power spectrum multiple moments to different angular channels of the covariance. Any combination of multipoles that after multiplying out is rotation-invariant (\textit{i.e.} has total angular momentum zero) can contribute. An expression for this RSD-added analytic covariance is in \citet{Hou2021analytic} Appendix E., where it is also shown that these redshift-space induced coupling can also increase the off-diagonals of the covariance. \citet{Hou2021analytic} showed that the rms of the off-diagonals after comparison of the analytic covariance  to the mock-based covariance is exactly the same in real-space and redshift-space, both for lognormal mocks and for \textsc{quijote} simulations ({\it cf}. figure 6, 7 and figure 10, 11). Therefore, off-diagonal enhancement by redshift-space-induced coupling is in principle not significant enough to greatly affect our analysis. Furthermore, some of the redshift-space-induced effects can be absorbed in the overall normalization, as much of their effect after isotropic averaging is to rescale the overall power spectrum. Nevertheless, we caution that missing these other terms could in principle reduce the accuracy of our analytic covariance relative to that from the mocks.

Fitting the volume and number density to match the mocks' covariance can be done without inverting this latter, important to avoid as that inverse will be biased unless the number of degrees of freedom is much smaller than the number of mocks. For the fitting, we maximize a likelihood based on the Kullback-Leibler (KL) divergence~\citep{Kullback1951}. One can show that with the likelihood Eq. (55) of \cite{Hou2021analytic}, the optimal volume at a given number density is uniquely determined, so the optimization is only a 1D problem, making it efficient. There is no guarantee that the best-fitting number density and volume for the covariance of the even-parity 4PCF (\textit{e.g.} as found in \citealt{Philcox2021detection}) and for the covariance of the parity-odd 4PCF should be the same. We summarize the values obtained for each sample and each radial binning in Table~\ref{tab:summary}. \textcolor{black}{For comparison, we also give the effective volume estimated from BOSS using equation (50) of~\citet{Reid2016boss}; the survey area used in that equation is taken from Table 2 of~\citet{Reid2016boss}}.
\textcolor{black}{As discussed above, the template covariance employs several approximations, including the assumption that the density field is purely Gaussian Random, and the omission of higher power spectrum multipole moments (w/rt line of sight). Such simplifying assumptions may cause an underestimate of the true covariance. To address this issue, an overall pre-factor is introduced and is interpreted as the effective volume $V_{\rm eff}$, where this fitted effective volume could be lower than the one reported in~\citet{Reid2016boss}. However, we want to point out that this volume-like prefactor is only a highly approximated estimation. To fully account for the effective volume requires including geometry effects in the analytic covariance template.}

\begin{figure}
    \centering
    \includegraphics[width=0.8\textwidth]{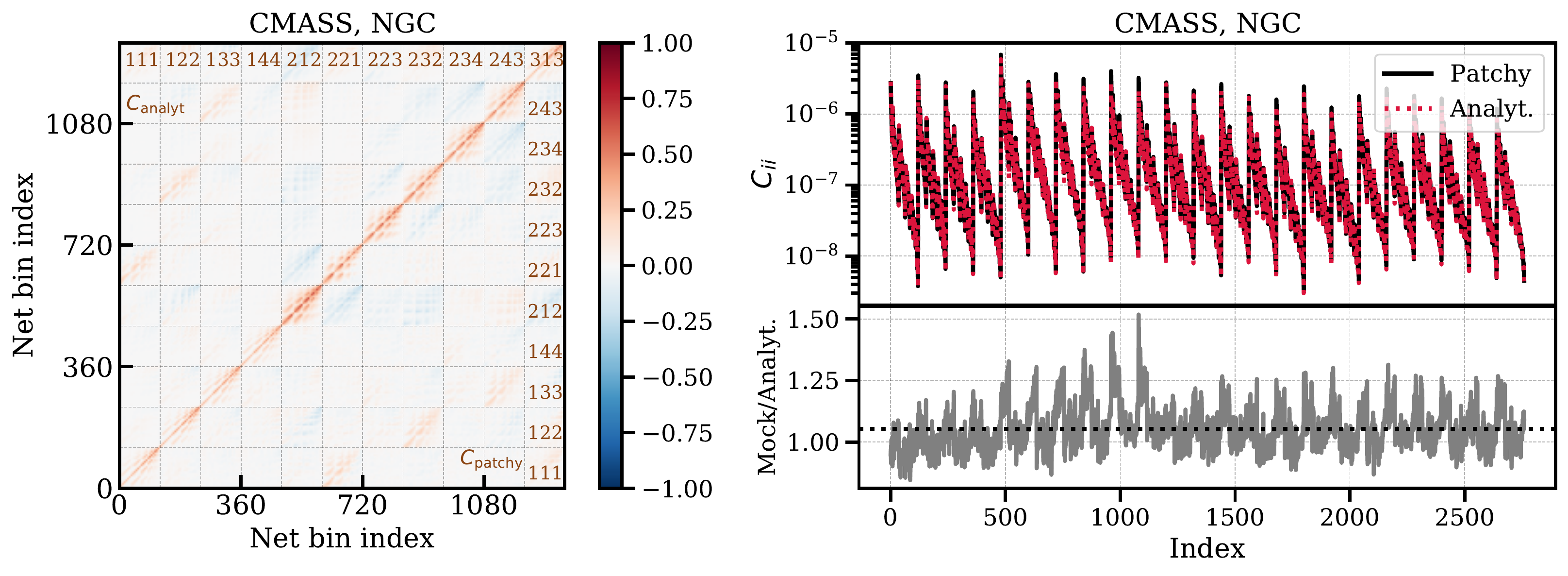}
    \caption{Comparisons of the analytic parity-odd 4PCF covariance with that from the \patchy mocks, all for CMASS NGC and ten radial bins. For greater legibility, we show only 11 of the 23 channels used in our analysis, and we have mapped each unprimed combination of angular momenta and radial bins to a 1D index, and the same for each primed; the full set of variables and indices involved is given in Eq. (\ref{eqn:fun_covar}), and the volume and number density here used are given in Table \ref{tab:summary}. {\it Left}: analytic correlation matrix (upper triangle) and correlation matrix from the \patchy mocks (lower triangle). The sub-blocks represent the covariance between a pair of channels; each is $120 \times 120$ as for ten radial bins, there are 120 radial bin combinations in each channel. These sub-blocks show how the covariance changes as the side lengths of each tetrahedron vary at fixed channel pairs. The similarity between the upper and lower triangles shows that the analytic covariance captures the cross-covariance between different radial bins as well as between different channels.
    {\it Right}: diagonals of the \patchy covariance matrix (solid black) and the analytic covariance (dashed red) for CMASS NGC. The average ratio is $1.05$.}
    \label{fig:corr_matrix_comparison_analyt_patchy_CMASS_ngc_odd_nells11}
\end{figure}

\begin{figure}
    \centering
    \includegraphics[width=0.8\textwidth]{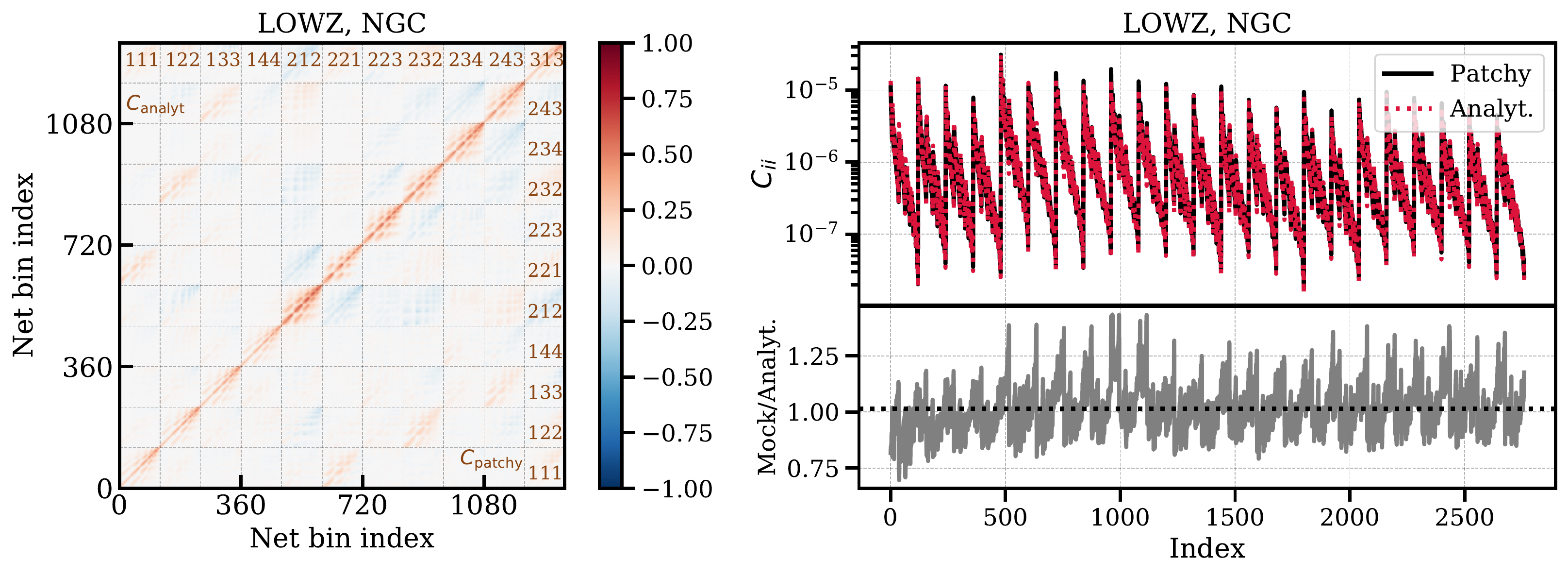}
    \caption{Here we show the same comparisons as made in Fig.~\ref{fig:corr_matrix_comparison_analyt_patchy_CMASS_ngc_odd_nells11} but for the LOWZ NGC. Again, in the left panel, we see that the analytic and mock correlation matrices have very similar patterns. The average ratio between the mock and analytic covariance diagonals is $1.01$ as indicated by the dotted black horizontal line in the lower right panel.}
    \label{fig:corr_matrix_comparison_analyt_patchy_LOWZ_ngc_odd_nells11}
\end{figure}

In Fig.~\ref{fig:corr_matrix_comparison_analyt_patchy_CMASS_ngc_odd_nells11} we compare for CMASS NGC the parity-odd 4PCF analytic covariance matrix to that from \patchy for ten radial bins. We have mapped the channel and triple-bin to a 1D index, so the covariance may be plotted as a 2D matrix (one 1D index for the unprimed channels and triple-bins, another for the primed). In the left panel, we show a direct comparison of the correlation matrix, which is the covariance matrix normalized by its diagonal. The similarity between the analytic correlation matrix (upper triangle) and the mock correlation matrix (lower triangle) shows that the analytic approach can well capture the covariance's features. In the right panel, we compare the diagonal elements of the covariance matrices and also show the ratio between the mock and analytic covariance diagonals. Again this uses our mapping of the angular momenta and radial bins into a 1D index. Despite the similarity between the two diagonals in their behaviour with increasing index, there is non-negligible variation. On average, the mean of the ratio between the mock and analytic covariance diagonal elements is roughly $1.05$ for CMASS NGC. Fig.~\ref{fig:corr_matrix_comparison_analyt_patchy_LOWZ_ngc_odd_nells11} shows the same comparisons for the LOWZ NGC. Again, we see that the analytic covariance can well describe that of the mocks. The mean of the ratio is $1.01$ for the LOWZ NGC.

To further quantify the similarity between the analytic and mock covariance, we define a {\bf half-inverse test} matrix as
\begin{eqnarray}
    \Phi \equiv \mathsf{C}^{-1/2}_{\rm analytic}\mathsf{C}_{\rm patchy}\mathsf{C}^{-1/2}_{\rm analytic} - \mathbb{1},
    \label{eqn:half_inv}
\end{eqnarray}
where $\mathsf{C}^{-1/2}_{\rm analytic}$ is the square root of the inverse of the analytic covariance matrix, and $\mathsf{C}_{\rm patchy}$ is the covariance from the mocks. We use half-inverses so that the test is symmetric. Were the two covariance matrices identical, $\Phi$ would be the zero matrix (we note that the identity matrix is subtracted). One can show that the optimal volume at a given number density (as discussed above) will imply that $\Phi$ is traceless. 

\begin{figure}
     \centering
     \begin{subfigure}[b]{0.45\textwidth}
         \centering
         \includegraphics[width=1\textwidth]{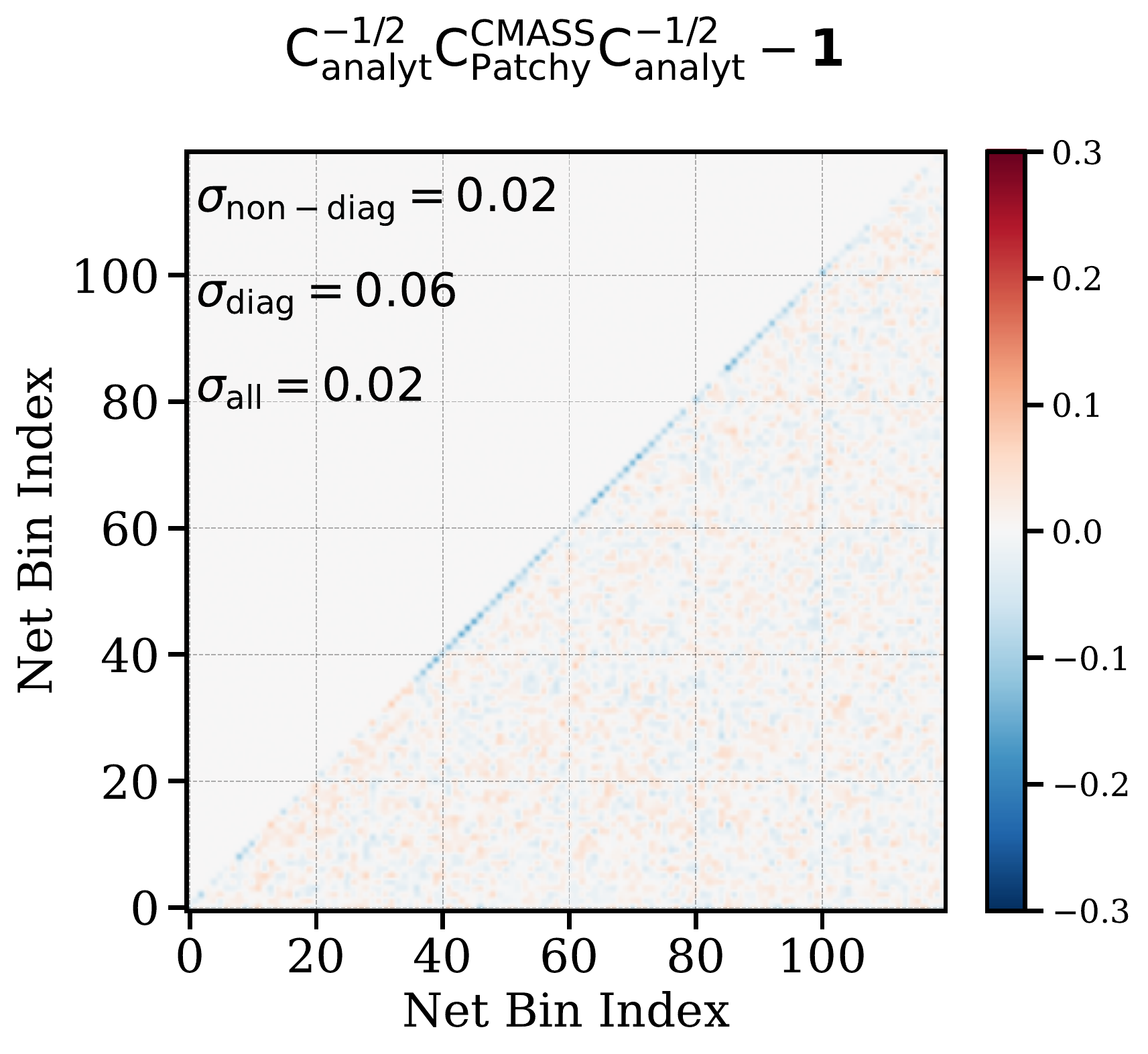}
         \label{fig:half_inv_CMASS_ngc_odd_nells23}
     \end{subfigure}
     \begin{subfigure}[b]{0.45\textwidth}
         \centering
         \includegraphics[width=1\textwidth]{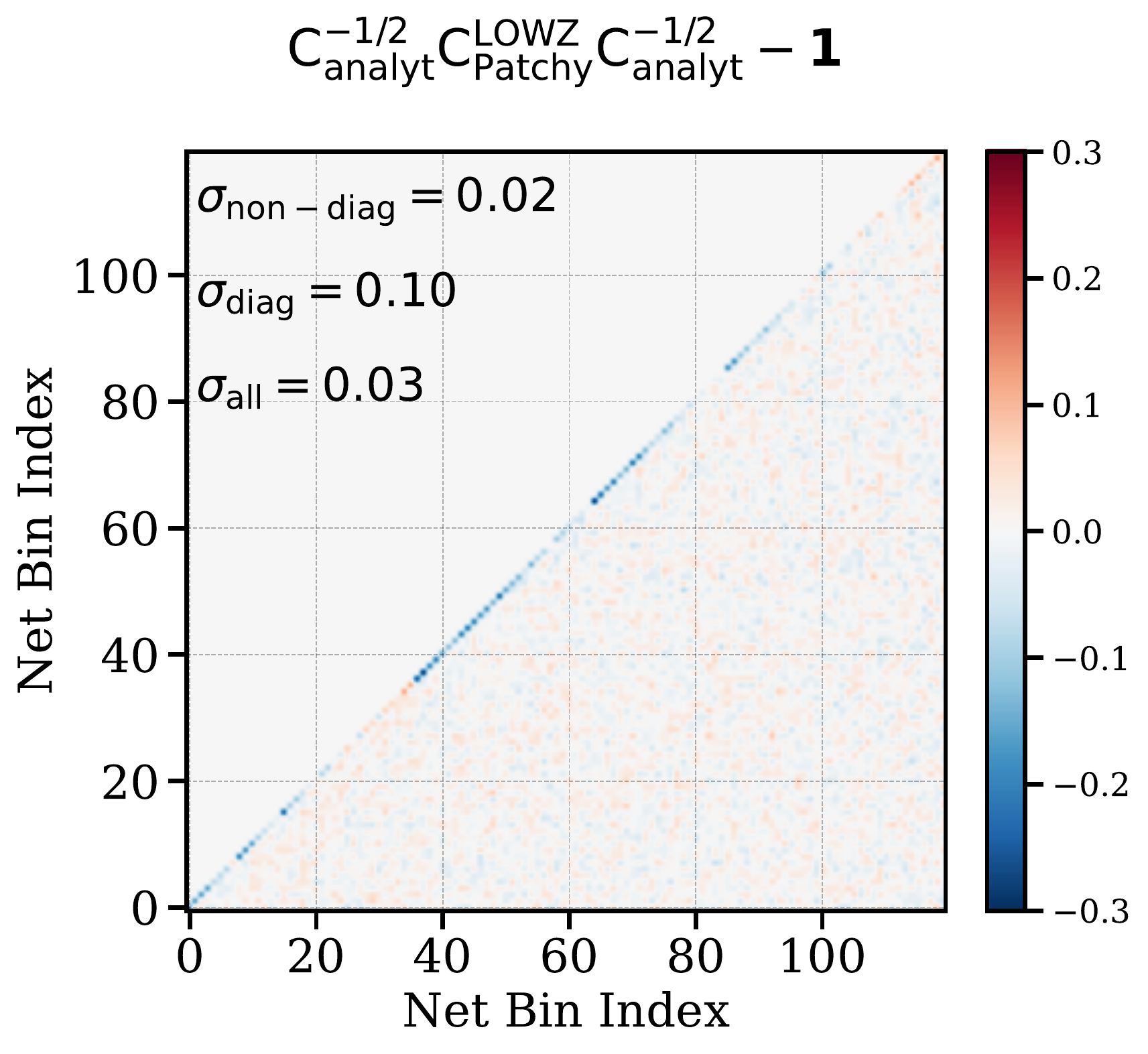}
         \label{fig:half_inv_LOWZ_ngc_odd_nells23}
     \end{subfigure}
     \caption{Half-inverse test $\Phi$ (Eq. \ref{eqn:half_inv}) of the analytic covariance matrix at $\{\Lambda, \Lambda'\} = \{111,111\}$ for CMASS NGC (\textit{left}) and LOWZ NGC (\textit{right}). We show only half of the matrix in the lower triangle and indicate the standard deviation of the matrix elements in the upper triangle; the test is symmetric so there is no information lost by deleting the latter. Here we compare the analytic covariance matrix to that from $2,000$ \patchy mocks; the overall standard deviation is roughly $ 1/\sqrt{N_{\rm mocks}} \approx 0.02$. It is expected that the standard deviation of the diagonal is double, as further discussed in the main text; we see 
a slight deviation from this relation in both panels.} 
     \label{fig:half_inv_BOSS_ngc_odd_nells23}
\end{figure}

Fig.~\ref{fig:half_inv_BOSS_ngc_odd_nells23} shows the lower half of the matrix $\Phi$ for the half-inverse test of CMASS NGC (left) and LOWZ NGC (right), with the standard deviation shown on the upper triangle. The standard deviation for all elements is $\sigma_{\rm all}=0.06$ for both CMASS NGC and LOWZ NGC, which is expected as it scales as $1/\sqrt{N_{\rm mocks}}$ (we used $N_{\rm mocks} = 2,000$ for this test). Each element of the covariance matrix follows a \citet{Wishart1928Distribution} distribution, and the variance of the diagonal elements for this distribution is two times that of the off-diagonal elements. We do find that the standard deviation $\sigma_{\rm diag}$ is approximately two times that of the off-diagonal, $\sigma_{\rm non-diag}$. However, we also find residuals in the diagonal elements in $\Phi$ for both the left and right panels, and larger residuals for LOWZ NGC than for CMASS NGC. These indicate that fitting the volume and number density to best match the mock covariance is imperfect. As mentioned in \S\ref{sec:guidelines}, one of our analysis approaches (``compressed'') will only require a smooth estimate of the covariance and turns out to be insensitive to its amplitude. Furthermore, even when employing the analytical covariance matrix directly, we may compare the $\chi^2$ from data only to the distribution of $\chi^2$ values for the mocks, both computed using the same covariance matrix. This comparison should minimize any bias in the detection significance computed using the analytic covariance.

\subsubsection{Mock Covariance Matrix}
\label{subsub:covar_sys}
In this work, we assume that the \patchy mocks match well the covariance structure of the observed data. This is a reasonable assumption as the \patchy mocks are calibrated to match the 2-point and some 3-point statistics measured from BOSS. Here we outline in detail aspects of \patchy that could cause a mismatch between the mock covariance and the data's true covariance. In the end, we do not find any strong reason to believe that such a mismatch is present, but that is after performing a number of tests motivated by the below points.
\begin{enumerate}
    \item \patchy uses approximate methods to simulate nonlinear structure formation, and underestimates small-scale power ($s\leqq 20 \;\mpch$), especially for galaxies with high stellar mass (\citealt{Kitaura2016patchy}; see their Fig. 4). Lacking power at smaller scales could have two impacts on the covariance. First, it reduces the ``squeezed'' tetrahedra (where the side lengths from the primary are close to each other in size and hence galaxies can become very close to each other as well as the angles at the primary vertex become small). This could produce an underestimate of the covariance on small scales. Second, this lack of small-scale power also may reduce the derived covariance at all scales, as the covariance comes from averaging the product of unprimed and primed tetrahedra over all possible separations between them, including small separations wherein points on the primed one may be close to those on the unprimed one. Indeed, this is the origin of $s$ in the integral Eq.~\eqref{eqn:f3l-tensor}. 
    
    To assess the impact of the above, we do an analysis where we force each side to be different in length from the others by at least one full bin width; this at least addresses the first point in (i). We still find a significant detection of parity-odd 4PCF in all samples. The second point is more challenging to deal with, although we did not have a direct test against it; its impact can be inferred from the parity-even \textit{connected} 4PCF measurement (which tracks only the contribution due to non-linear evolution). The level of agreement between \patchy and BOSS CMASS suggests that \patchy actually does well-capture the non-linearity on the scales relevant to our analysis.
    \item \patchy may not reproduce all systematics that may be present in data. We computed the 2PCF and found that the LOWZ SGC has an amplitude that is higher on average than the mean of the \patchy mocks by $1\sigma$. However, it is quite consistent with the spread we see in the measured 2PCF for a set of 1,000 mocks. In addition, as discussed in \citet{Ross2016BOSS} and \citet{Kitaura2016patchy} there are deviations between \patchy and the BOSS 2PCF at scales greater than roughly 100 $\mpch$. These 2PCF bins are highly correlated though and the actual deviation may be less severe than the visual impression. Again we found that the spread of the 2PCF of 1,000 mocks covered the BOSS data well. Nonetheless, to make our analysis robust, we also performed our parity-odd analysis with a cut on the maximum side lengths from the primary so that no side of the tetrahedron could exceed 160 $\mpch$. We still found evidence for a parity-odd 4PCF consistent with what one would expect given the significant reduction in the constraining power in this analysis due to the much smaller number of triple-bins it permits. 
    
    We also briefly consider that the number density of the survey is not equal to the true number density of the Universe; this would be a failure of the total integral constraint.  Such a failure would produce a correction to the observed power spectrum, such that a term $P_{\rm ic} \geq 0$ is subtracted (see equation (29) of \citealt{Beutler2014BOSS}). The \patchy mocks would not have this correction, and hence have a larger power spectrum than the observed data. Thus, the covariance estimated from them would be {\it larger} than it should be. Since this error, if present, would cause an {\it over-estimation} of the covariance (and thus a spuriously decreased detection significance), and also since given BOSS's large volume it is expected to be a small effect, we do not explore it further.
\end{enumerate}

\subsubsection{Consistency of Parity-Even Modes}
\label{subsub:parity_even}
\paragraph*{Impact of possible systematics}\mbox{}\\
As will be discussed further in \S\ref{sec:systematics}, distortions in the radial or the angular directions that are not captured in the mocks can mean that a mock-based covariance underestimates the true covariance. Here we study the sensitivity of the parity-even sector to such contamination. For a distortion of the radial selection function, we apply the contamination directly to the mocks and find their mean detection significance of a connected-even parity 4PCF (when computed with an undistorted covariance, as would be the case if the data had an unaccounted-for change)  is shifted by $\sim\! 0.6\sigma$ for ten radial bins and $\sim\! 2.4\sigma$ for eighteen radial bins. We also explore angular contamination as described in \S\ref{subsubsec:zfail}; this increases the even-parity detection significance difference between data and mean of the mocks to $0.9\sigma$ for the ten bin and $1.8\sigma$ for the eighteen bin. 

\paragraph*{Self-calibration of covariance}\mbox{}\\
Our detection significance relies on correctly estimating the covariance. Since the \patchy mocks have no intrinsic parity-violating mechanism, they cannot be used to check our pipeline regarding signal, but only to estimate the covariance. Yet it is possible that the mocks do not contain all the systematics present in the real data, and this lack could result in mis-estimation of the covariance. 

However, any unknown data-only systematic might well impact the parity-even modes as well as the parity-odd modes. Hence, we can see if the mocks and data are consistent in the parity-even sector at the signal level, where both do have a signal due to non-linear evolution. If we see that the even-parity signal from the data is consistent with that from the mocks, this suggests that there are unlikely to be unaccounted-for systematics. It also would argue that we have reasonably estimated the even-sector covariance, as an underestimate of it would show up as a severe tension between the signal in the mocks and in the data. Since our procedures for obtaining the even and odd-sector covariances are exactly the same, such a finding would in turn build confidence that our odd-sector covariance is correct.

Furthermore, as a highly conservative approach, we can ask what factor the covariance in the even-sector would need to be rescaled by to force the data to agree within \textit{e.g.} $3\sigma$ or $1\sigma$ with the mocks. We can then apply this rescaling to our odd-parity covariance and ask how it impacts our detection significance in the odd sector. We emphasize that this is a highly conservative check, not a correction procedure; it is perfectly possible that the even-parity data signal is \textit{e.g.} $3\sigma$ different from that in the mocks just by chance, without implying that there is any error that should be corrected in either sector. Thus, while we summarize the results of this idea in Table~\ref{tab:rescaling-detection}, one should not take these rescalings too literally.

For CMASS with ten radial bins, we found excellent agreement between the data and mock distribution both using the analytic covariance and the compressed method (also see~\cite{Philcox2021detection}, where a slightly different sample selection was applied). Therefore, we do not present any rescaling factor for the detection significance for the parity-odd modes. 

For CMASS with eighteen radial bins, we found a discrepancy in the even-parity sector between the data and the mock distribution of $4.9\sigma$ when using the analytic covariance. Rescaling the even covariance such that the data and mocks agree at the 3$\sigma$ level, we then propagate the same factor to the parity-odd measurement. We also repeat this procedure to enforce the agreement at the 1$\sigma$ level in the even sector. We summarize these results in Table~\ref{tab:rescaled_detection}. 

Regarding the discrepancy found in the eighteen-bin test, we ascribe it to the possible breakdown of the GRF assumption of the analytic covariance as one goes to smaller scales, especially for the even sector,  as discussed in \S\ref{subsec:covariance_matrix}. In particular, the parity-even covariance requires including the leading higher-order covariance and connected estimator contribution. These contributions are naturally absent when estimating the parity-odd covariance. Further, when using the compressed method with $N_{\rm eig}=800$, the discrepancy is reduced to $1.3\sigma$. 

\begin{table}
\centering
\caption{Summary of the datasets and binning schemes used. We give the effective redshift $z_{\rm eff}$ and the fitted effective volume $V_{\rm eff, fit}$ and number density $\bar{n}$ derived by fitting our analytic covariance to that from the mocks. The fitted effective volume and number density are different for each number of radial bins because we re-fit the covariance for each radial binning scheme. In the last row, we also list the effective volume using equation (50) from~\citet{Reid2016boss} for comparison.}
\label{tab:summary}
\begin{tabular}{cccccc} 
\toprule
\multirow{2}{*}{} & \multirow{2}{*}{} & \multicolumn{2}{c}{CMASS} & \multicolumn{2}{c}{LOWZ} \\ 
\cline{3-6}
 &  & NGC & SGC & NGC & SGC \\ 
\hline
Bins &  & \multicolumn{2}{c}{$z_{\rm eff}=0.57$} & \multicolumn{2}{c}{$z_{\rm eff}=0.32$} \\ 
\hline
\multirow{2}{*}{$18$} & \begin{tabular}[c]{@{}c@{}}$V_{\rm eff, fit}$\\ $[h^{-1}{\rm Gpc}]^3$\end{tabular} & $2.50$ & $0.79$ & $0.31$ & $0.12$ \\ 
\cline{2-6}
 & \begin{tabular}[c]{@{}c@{}}$\bar{n} \times 10^{4} $\\ $[h^3 {\rm Mpc}^{-3}] $\end{tabular} & $1.4$ & $1.4$ & $2.0$ & $2.0$ \\ 
\hline
\multirow{2}{*}{$10$} & \begin{tabular}[c]{@{}c@{}}$V_{\rm eff, fit}$\\ $[h^{-1}{\rm Gpc}]^{3}$\end{tabular} & $1.71$ & $0.54$ & $0.23$ & $0.09$ \\ 
\cline{2-6}
 & \begin{tabular}[c]{@{}c@{}}$\bar{n} \times 10^{4} $\\ $[h^3 {\rm Mpc}^{-3}] $\end{tabular} & $1.8$ & $1.8$ & $2.6$ & $2.6$ \\ 
\hline
\multirow{2}{*}{$6$} & \begin{tabular}[c]{@{}c@{}}$V_{\rm eff, fit}$\\ $[h^{-1}{\rm Gpc}]^3$\end{tabular} & $1.36$ & $0.44$ & - & - \\ 
\cline{2-6}
 & \begin{tabular}[c]{@{}c@{}}$\bar{n} \times 10^{4} $\\ $[h^3 {\rm Mpc}^{-3}] $\end{tabular} & $2.2$ & $2.2$ & - & - \\
\hline
\multicolumn{1}{l}{} & \begin{tabular}[c]{@{}c@{}}$V_{\rm eff}$\\$[h^{-1}{\rm Gpc}]^3$\end{tabular} & \multicolumn{1}{l}{$2.50$} & \multicolumn{1}{l}{$0.95$} & \multicolumn{1}{l}{$0.66$} & \multicolumn{1}{l}{$0.29$} \\ \bottomrule
\end{tabular}
\end{table}

\begin{table}
\centering
\setlength{\extrarowheight}{0pt}
\addtolength{\extrarowheight}{\aboverulesep}
\addtolength{\extrarowheight}{\belowrulesep}
\setlength{\aboverulesep}{0pt}
\setlength{\belowrulesep}{0pt}
\caption{The detection significances using the analytic covariance for the split sky (into NGC and SGC) and joint sky (NGC + SGC). The latter treats a given channel and radial bin combination in N and S as two separate degrees of freedom, \textit{i.e.} we assume vanishing covariance between N and S.  All of the quoted detection significances are from comparing the $\chi^2$ of the data to a Gaussian fit to the histogram of the mocks' $\chi^2$; see lower rightmost panels of Figs. \ref{fig:chi2s_cmass_ns}-\ref{fig:chi2s_lowz_ns_18bins_LMAX4} (ten and eighteen bin results) and of Fig. \ref{fig:chi2s_cmass_ns_6bin} (six bin results, CMASS only).}
\label{tab:sigs}
\begin{tabular}{cccccc} \toprule
 &  & \multicolumn{2}{c}{CMASS} & \multicolumn{2}{c}{LOWZ} \\ \midrule
Bins & \begin{tabular}[c]{@{}c@{}}$\sigma_{\rm eff}$, Detection \\ Significance\end{tabular} & NGC & SGC & NGC & SGC \\ \midrule
\rowcolor[rgb]{0.898,0.898,0.898} {\cellcolor[rgb]{0.898,0.898,0.898}} & split & $4.7$ & $5.4$ & $2.5$ & $2.0$ \\
\rowcolor[rgb]{0.898,0.898,0.898} \multirow{-2}{*}{{\cellcolor[rgb]{0.898,0.898,0.898}}$18$} & joint & \multicolumn{2}{c}{$7.1$} & \multicolumn{2}{c}{$3.1$} \\ \midrule
\multirow{2}{*}{$10$} & split & $3.3$ & $2.6$ & $1.8$ & $1.9$ \\
 & joint & \multicolumn{2}{c}{$4.0$} & \multicolumn{2}{c}{$2.7$} \\ \midrule
\rowcolor[rgb]{0.898,0.898,0.898} {\cellcolor[rgb]{0.898,0.898,0.898}} & split & $-0.7$ & $1.2$ & - & - \\
\rowcolor[rgb]{0.898,0.898,0.898} \multirow{-2}{*}{{\cellcolor[rgb]{0.898,0.898,0.898}}$6$} & ~joint & \multicolumn{2}{c}{$0.4$} & \multicolumn{2}{c}{-} \\ \bottomrule
\end{tabular}
\end{table}

\begin{table}
\centering
\caption{Correlation coefficients between NGC and SGC for CMASS and LOWZ samples; see \S\ref{subsec:cross} and Figs. \ref{fig:pearson_corr_cmass_NGCxSGC_odd_stacked23ell}-\ref{fig:dist_pearson_corr_NGCxSGC_odd_stacked23ell_nbin18_LMAX4_drCut0_rmax200}. We do not see significant cross-correlation between NGC and SGC in either binning scheme or sample; the possible reason is further discussed in the main text, as is te reason we cannot perform a CMASS $\times$ LOWZ analysis.}
\label{tab:correlation}
\begin{tabular}{cccccc} \toprule
 &  & \multicolumn{2}{c}{CMASS} & \multicolumn{2}{c}{LOWZ} \\ \midrule
Bins & Correlation & NGC & SGC & NGC & SGC \\ \midrule
\multirow{2}{*}{$18$} & $r_{\rm p}$ & \multicolumn{2}{c}{0.002} & \multicolumn{2}{c}{-0.017} \\
 & $p$-value & \multicolumn{2}{c}{0.752} & \multicolumn{2}{c}{0.017} \\\midrule
\multirow{2}{*}{$10$} & $r_{\rm p}$ & \multicolumn{2}{c}{-0.014} & \multicolumn{2}{c}{0.000} \\
 & $p$-value & \multicolumn{2}{c}{0.460} & \multicolumn{2}{c}{0.985} \\ \bottomrule
\end{tabular}
\label{tab:corrln}
\end{table}

\begin{table}
\centering
\caption{Here we summarize the rescaled detection significance in the parity-odd modes by forcing agreement in the parity-even sector at the levels indicated in the leftmost column. These results are for CMASS and eighteen radial bins; see \S\ref{subsub:parity_even}. We explore this rescaling both fr our analytic covariance analysis and our compressed analysis with 800 eigenvalues; in the last line, no rescaling is required to attain $3\sigma$ consistency in the even sector. Generally, it is notable that an odd detection more or less remains in all cases even after rescaling.}
\label{tab:rescaled_detection}
\begin{tabular}{@{}cccc@{}}
\toprule
\multicolumn{1}{l}{\begin{tabular}[c]{@{}l@{}}Consistency in \\ Even-Parity Sector\end{tabular}} & \begin{tabular}[c]{@{}c@{}}CMASS\\ 18 bins\end{tabular} & \begin{tabular}[c]{@{}c@{}}Rescaling \\ Factor\end{tabular} & \begin{tabular}[c]{@{}c@{}}Rescaled\\ Odd Detection Significance\end{tabular} \\ \midrule
\multirow{2}{*}{\begin{tabular}[c]{@{}c@{}}$1$ standard\\ deviation\end{tabular}} & \begin{tabular}[c]{@{}c@{}}Analytic\\ Covariance\end{tabular} & 0.88 & $2.0 \sigma$ \\
 & $N_{\rm eig}=800$ & 0.98 & $4.0\sigma$\vspace{1mm} \\
\cline{2-4}
\multirow{2}{*}{\begin{tabular}[c]{@{}c@{}}$3$ standard\\ deviation\end{tabular}} & \begin{tabular}[c]{@{}c@{}}Analytic\\ Covariance\end{tabular} & 0.94 & $4.6\sigma$ \\
 & $N_{\rm eig}=800$ & - & $4.4\sigma$ \\ \bottomrule
\end{tabular}
\label{tab:rescaling-detection}
\end{table}

\section{Results on CMASS and LOWZ Data}
\label{sec:analysis}

Here, we first compute the significance of our observed parity-odd 4PCF using the methods briefly outlined in \S\ref{sec:guidelines}, and then explore the cross-correlation of NGC and SGC in each sample.

\subsection{Statistical Significance}
We quantify the statistical significance of the parity-violating amplitudes using three approaches: a direct analysis harnessing the analytic covariance calibrated against \patchy (\S\ref{subsub:direct}), a compressed analysis using the analytic covariance just to select a reduced basis with few enough degrees of freedom that the mocks can then provide the covariance (\S\ref{subsub:comp}), and finally, by lowering $\ell_{\rm max}$ enough so that we may use the direct approach but with covariance from the mocks (\S\ref{subsub:red}).

\subsubsection{Direct Analysis}
\label{subsub:direct}
With the analytic covariance, adjusted in density and volume to best fit the covariance given by the mocks, we evaluate 
\begin{eqnarray}
\chi^2 = \left(\hat{\zeta} - \zeta_{\rm model}\right)^{\rm T} \mathsf{C}^{-1} \left(\hat{\zeta} - \zeta_{\rm model}\right).
\end{eqnarray}
where $\hat{\zeta}$ and $\zeta_{\rm model}$ are respectively the various parity-violating amplitudes, $\hat{\zeta}_{\ell_1\ell_2\ell_3}(r_1,r_2,r_3)$ and $\zeta_{\rm model}$ is a model for them. To investigate the null hypothesis that there is no parity violation, we set all elements of $\zeta_{\rm model}$ to zero. $\mathsf{C}^{-1}$ is the inverse covariance matrix. 
If the underlying data have Gaussian behavior with $N_{\rm d}$ degrees of freedom, the resulting distribution is the $\chi^2$ distribution
\begin{eqnarray}
f(\chi^2) = \frac{1}{2^{N_{\rm d}/2}\;\Gamma(N_{\rm d}/2)}\; \left({\chi^2}\right)^{N_{\rm d}/2} e^{-\chi^2/2}.
\label{eqn:chisq_dist}
\end{eqnarray}

The adequacy of the covariance obtained by fitting the analytic covariance to the data sets can be assessed by examining the distribution of $\chi^2$ values from the mocks when run through our analysis, and seeing if it matches the predicted $\chi^2$ distribution.\footnote{We expect that radially-binning the 4PCF amplitudes results in a multivariate Gaussian distribution on each triple-bin. Thus a $\chi^2$ distribution is appropriate here.} 

\subsubsection{Compressed Analysis}
\label{subsub:comp}
The alternative means of obtaining an invertible covariance matrix by reducing the dimensionality of the problem, the ``compressed'' analysis, is described in \S\ref{sec:guidelines}. This scheme was introduced by \citet{Scoccimarro1999bispectrum} to quantify detection significance, and the same method was  applied to the even-parity connected 4PCF analysis of~\citet{Philcox2021detection}. We diagonalize the analytic covariance, writing $\mathsf{C} = \mathsf{O} \mathsf{D} \mathsf{O}^{-1}$ where $\mathsf{D}$ is diagonal and $\mathsf{O}$ is an orthogonal matrix (we note that $\mathsf{O}^{-1} = \mathsf{O}^{\rm T}$). 

The columns of $\mathsf{O}$ then provide an eigenbasis for the analytic covariance. We would like to select the subset of these eigenvectors that will be most useful in detecting a signal. If one has a model for the expected signal, one can rank the eigenvectors by their signal-to-noise, and choose $N_{\rm eig}$ of them with the best S/N. In the absence of an anticipated signal, we choose the $N_{\rm eig}$ eigenvectors associated with the \textit{smallest} eigenvalues. These eigenvectors have the lowest noise. The analysis is then restricted to the space spanned by these eigenvectors.  With $N_{\rm eig}\ll N_{\rm mocks}$, the covariance matrix for this restricted analysis can be determined entirely from the mocks, while retaining the domain with the greatest information.

While this procedure provides an invertible covariance matrix, 
the inverse of the mock covariance matrix is unbiased only when $ N_{\rm eig} \ll N_{\rm mocks}$, where here $N_{\rm eig}$ is the number of degrees of freedom, \textit{i.e.} plays the role of $N_{\rm d}$ in \S\ref{subsub:direct}. 
Consequently, the $\chi^2$-distribution we used in \S\ref{subsub:direct} is modified.  To distinguish this distribution from the $\chi^2$ distribution, we refer to the variable $T^2$, defined analogously to $\chi^2$, as
\begin{eqnarray}
T^2 = \left(\hat{\zeta} - \zeta_{\rm model}\right)^{\rm T} \mathsf{C}^{-1} \left(\hat{\zeta} - \zeta_{\rm model}\right).
\end{eqnarray}
whose distribution is~\citep{Sellentin2016tdist} 
\begin{eqnarray}
f(T^2) = \frac{\Gamma\left(N_{\rm mocks}/2\right)}{\Gamma(N_{\rm eig}/2)\;\Gamma\left([N_{\rm mocks} - N_{\rm eig}]/2\right)}\;\frac{(N_{\rm mocks} -1)^{-N_{\rm eig}/2}\;(T^2)^{N_{\rm eig}/2-1}}{(T^2/(N_{\rm mocks}-1)+1)^{N_{\rm mocks}/2}}.
\label{eqn:T2-pdf}
\end{eqnarray}
This reduces to the $\chi^2$ distribution for $N_{\rm mocks}\rightarrow\infty$.

Since the data from CMASS and LOWZ in the NGC and SGC  are  statistically independent given the large physical separations of both NGC and SGC and of CMASS and LOWZ, we can simply add their values of $\chi^2$ or $T^2$, while increasing the number of degrees of freedom correspondingly. 

\subsubsection{Detection Significances from Direct \& Compressed Methods}
\label{subsub:ds_detxn}

\begin{figure}
    \centering
    \includegraphics[width=1\textwidth]{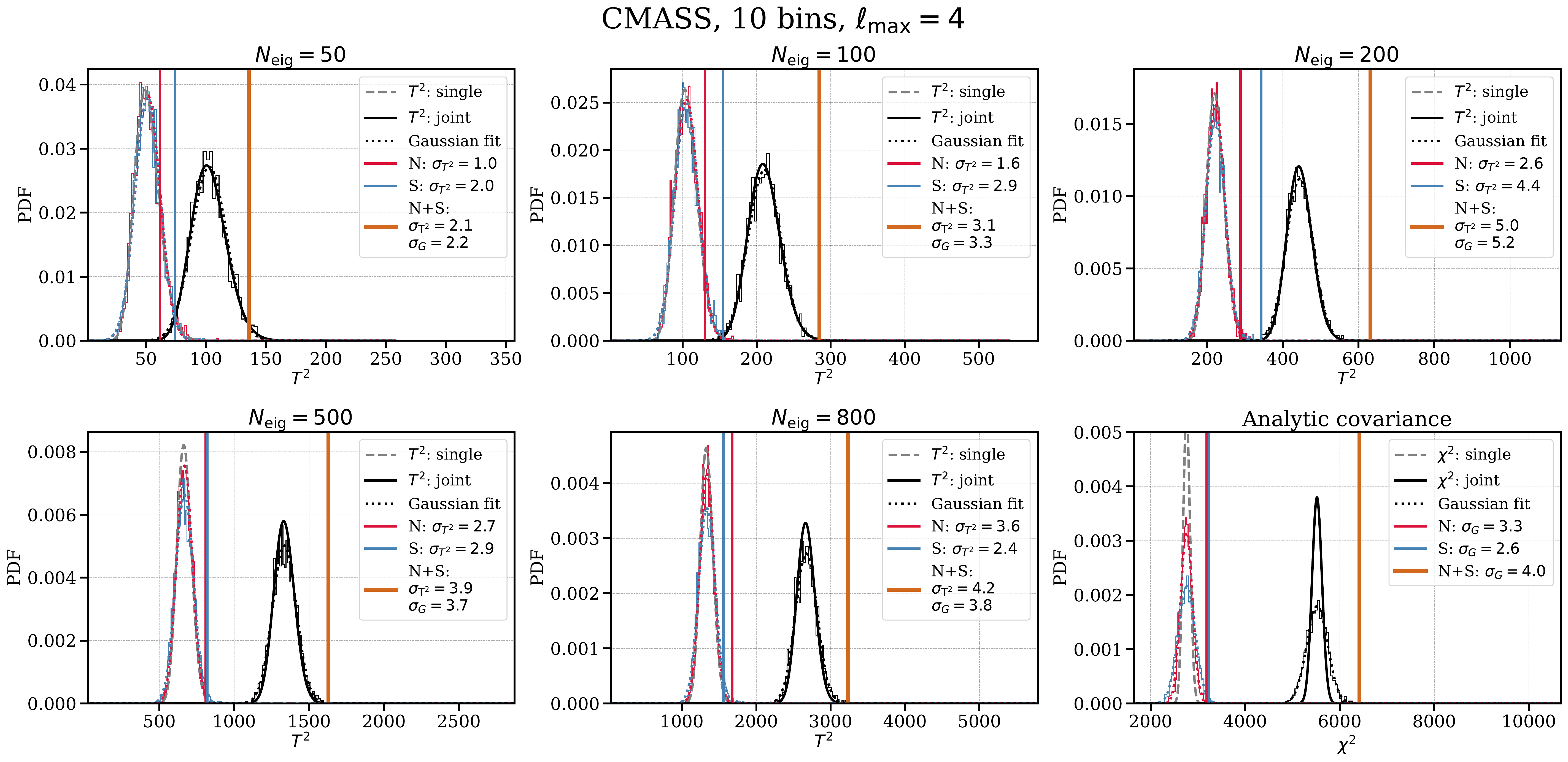}
    \caption{$T^2$ and $\chi^2$ distributions for CMASS data using {\bf ten} radial bins. The {first five} plots are obtained using the data compression scheme of \S\ref{subsub:comp} with different numbers of eigenvalues $N_{\rm eig}$ as indicated in the plot titles. If there is no parity-breaking, we expect a $T^2$ distribution (Eq. \ref{eqn:T2-pdf}) with $N_{\rm eig}$ degrees of freedom (solid black). The vertical lines in the first five plots are the $T^2$ values from CMASS for the NGC (red), SGC (blue), and joint NGC and SGC (black). For comparison, we show histograms of the $T^2$ values from the \patchy mocks for each case. We interpret the increasing deviations of the mocks' histograms from the $T^2$ distribution as $N_{\rm eig}$ rises as evidence that the covariance becomes less well-determined for larger $N_{\rm eig}$. This is expected as for $N_{\rm eig} = 800$, there are only about three mocks per degree of freedom (we use 2,000 mocks).  In the lower rightmost plot, we use the direct approach of \S\ref{subsub:direct}, which employs the analytic covariance. Here, the expected distribution if there is no parity-breaking is a $\chi^2$ distribution (Eq. \ref{eqn:chisq_dist}; solid black). We show a histogram of the mocks' $\chi^2$ values for comparison (orange) as well as a Gaussian fit to it (dot-dashed). This histogram is wider than the predicted $\chi^2$ distribution (which the mocks should match as they have no parity-breaking). We interpret this as an indication that the analytic covariance may not be unbiased. To be conservative, we compute a detection significance $\sigma_G$ by comparing the data $\chi^2$ values to the width of the Gaussian fit to the mocks. In fact, we quote $\sigma_G$ in all panels above, but it generally agrees with that computed from the $T^2$ distribution, $\sigma_{T^2}$. We do not quote a detection significance computed with respect to the $\chi^2$ distribution (as would naively be appropriate in the lower rightmost panel) because the distribution of mocks is clearly mismatched with a $\chi^2$ distribution.}
    \label{fig:chi2s_cmass_ns}
\end{figure}

\begin{figure}
    \centering
    \includegraphics[width=1.\textwidth]{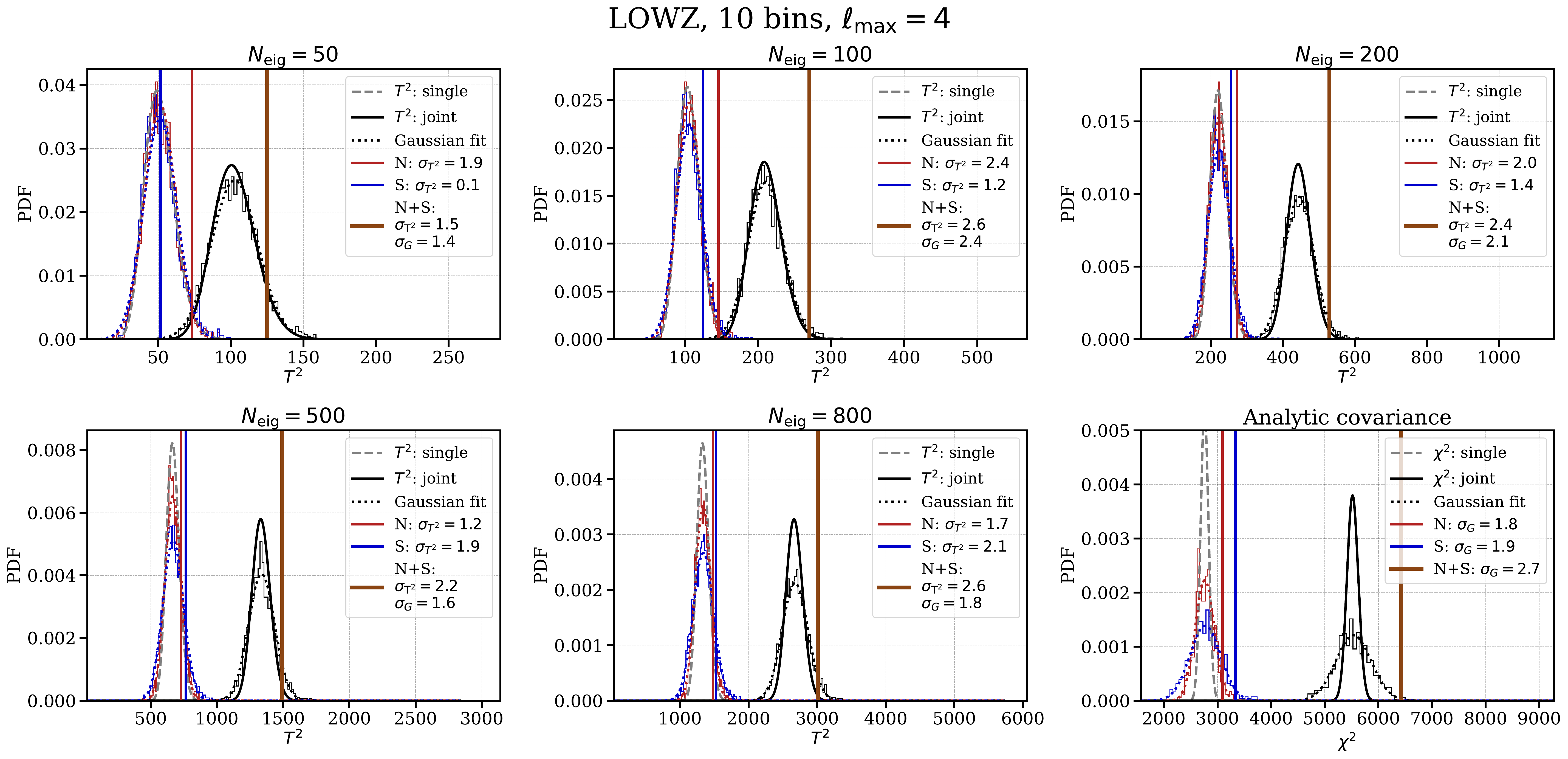}
    \caption{Similar to Fig.~\ref{fig:chi2s_cmass_ns} but for LOWZ; again using {\bf ten} radial bins. As for CMASS, in the compressed approach, the detection significance first increases with an increasing number of eigenvalues and then drops. In general, the detection significance is comparable to CMASS for both compressed and direct analyses and for split and joint sky. Again in the direct method, we see a noticeable deviation of the mocks' histogram from a $\chi^2$ distribution; this again suggests that the analytic covariance may not be unbiased. We also see noticeable deviation at $N_{\rm eig} = 800$ (and even at $N_{\rm eig} = 500$), to a much greater extent than for the same $N_{\rm eig}$ values in CMASS (\textit{cf.} Fig. \ref{fig:chi2s_cmass_ns}).}
    \label{fig:chi2s_lowz_ns}
\end{figure}

Fig.~\ref{fig:chi2s_cmass_ns} displays the $T^2$ or $\chi^2$  distributions for CMASS (both NGC and SGC) for the two approaches described above (direct and compressed). The detection significance changes as we vary the number of eigenvalues, but this is expected: as $N_{\rm eig}$ rises, more information is added (unless the S/N is pathological), but at the same time the potential bias of the inverse covariance is increasing. This latter can be assessed by looking for when the distribution of the mocks' $T^2$ values begins to deviate from the $T^2$ distribution expected (Eq. \ref{eqn:T2-pdf}). Fig.~\ref{fig:chi2s_lowz_ns} shows the same information but for LOWZ, which  generally shows comparable detection significances compared to CMASS both at each number of eigenvalues in the compressed analysis, and in the direct analysis. 

\subsubsection{Using Finer Binning}
\label{subsub:18bins}
\begin{figure}
    \centering
    \includegraphics[width=1\textwidth]{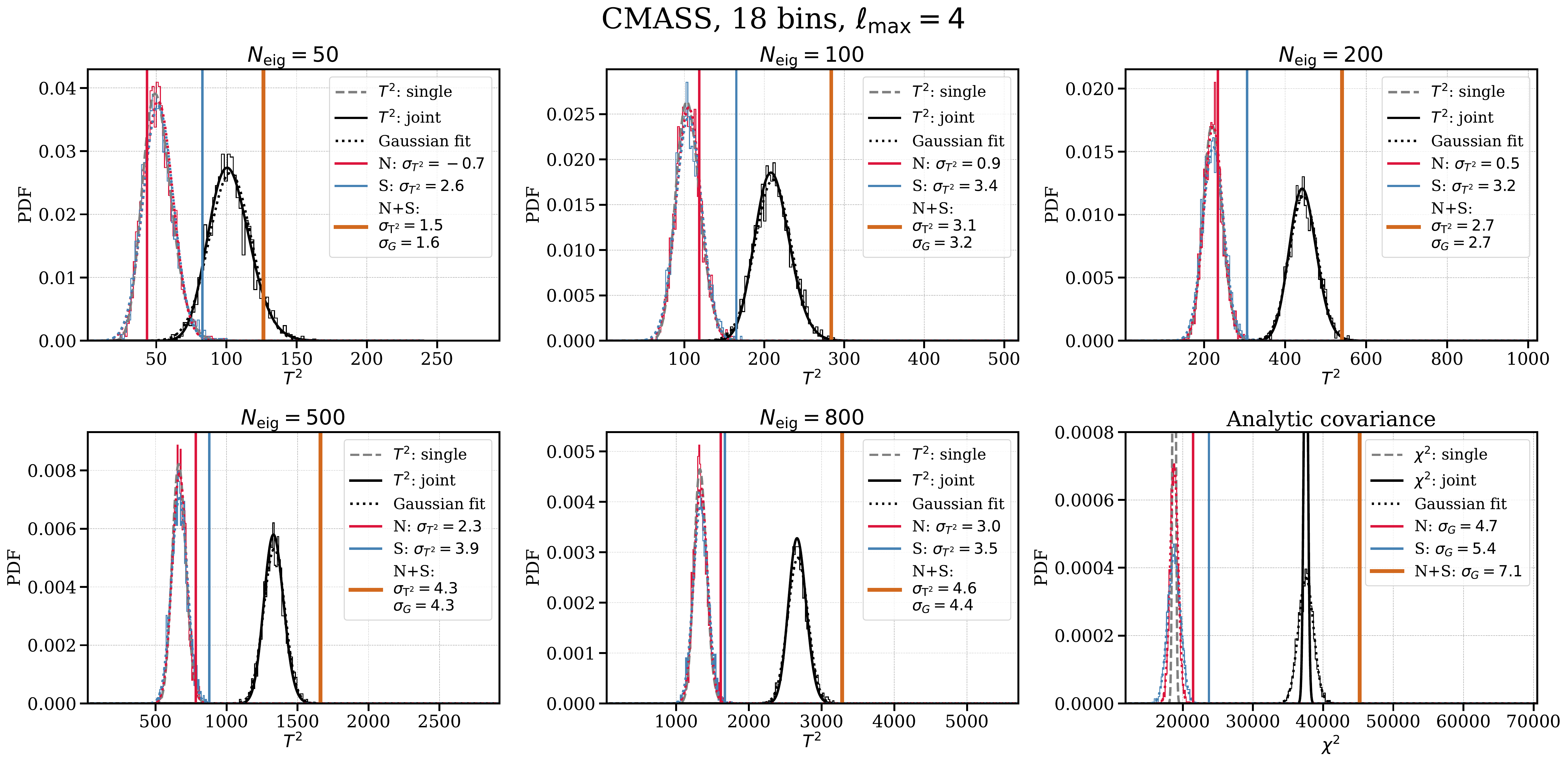}
    \caption{Similar to Fig.~\ref{fig:chi2s_cmass_ns} but for CMASS with {\bf eighteen} bins. Using finer binning we find even higher detection significance, both in our compressed and direct analyses ({\it cf.} Fig. \ref{fig:chi2s_cmass_ns}). Unlike the ten-bin CMASS analysis, which peaked at $N_{\rm eig} = 200$, here
    the detection significance monotonically increases with $N_{\rm eig}$. This behavior is likely due to how the covariance matrix structure and signal-to-noise in each channel and bin combination balance with the penalty paid for adding more degrees of freedom if they do not contain much additional S/N. Given that this balance can change depending on the radial binning, we would not have expected that the behavior of the ten-bin and eighteen-bin compressed analyses with $N_{\rm eig}$ would be uniform.}
    \label{fig:chi2s_cmass_ns_18bins_LMAX4}
\end{figure}

\begin{figure}
    \centering
    \includegraphics[width=1\textwidth]{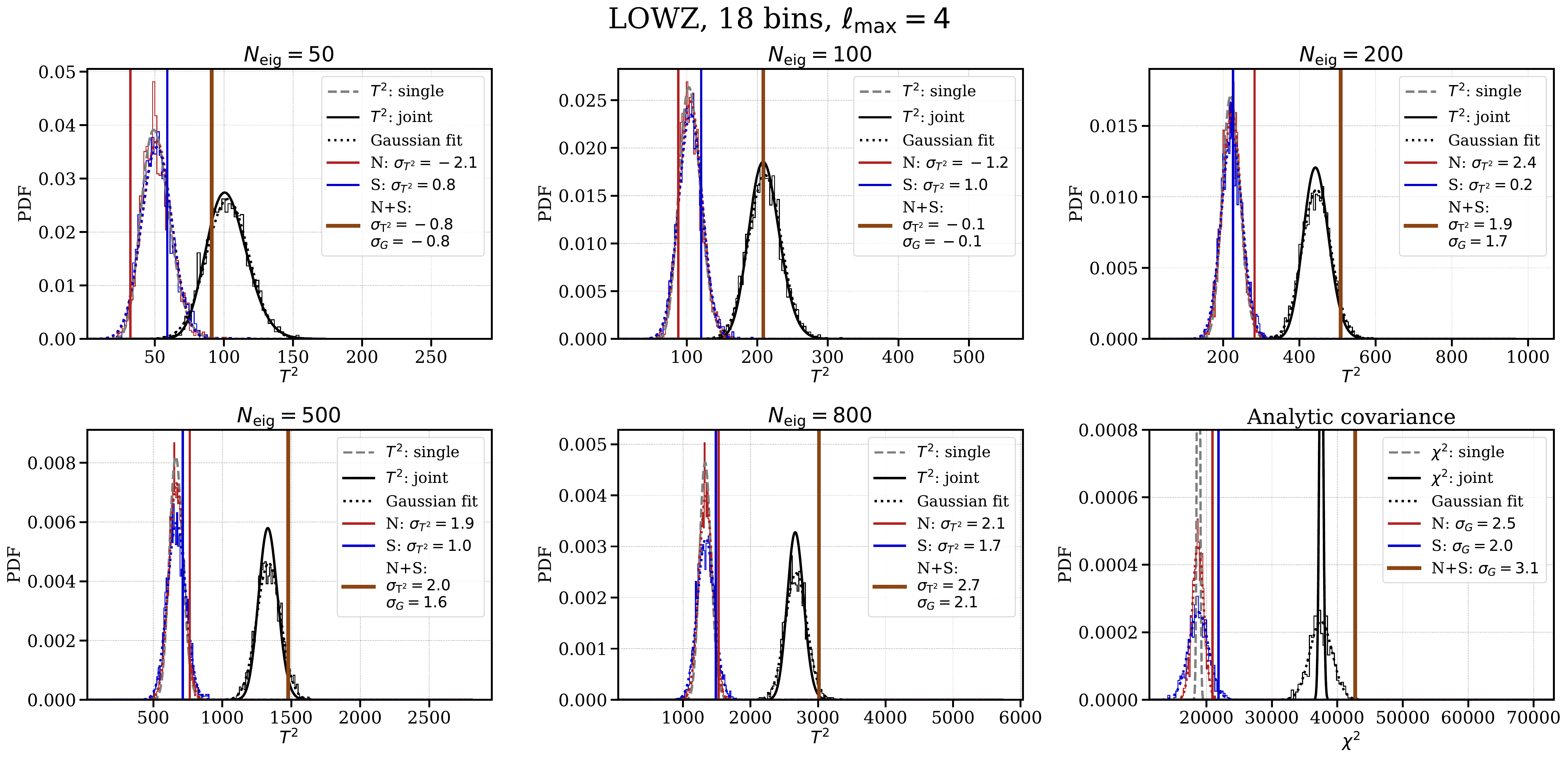}
    \caption{Similar to Fig.~\ref{fig:chi2s_lowz_ns} but for LOWZ with {\bf eighteen} bins. As is the case for CMASS, the detection significance increases when using the finer binning ({\it cf.} Fig. \ref{fig:chi2s_lowz_ns}). However, the increase is more moderate than the increase for CMASS. The detection significance rises monotonically with $N_{\rm eig}$, in agreement with the trend for the ten-bin LOWZ analysis of Fig. \ref{fig:chi2s_lowz_ns}.}
    \label{fig:chi2s_lowz_ns_18bins_LMAX4}
\end{figure}

Since two canceling contributions can occur from a single tetrahedron when they fall into the same radial bins (\S\ref{subsec:int_canc}), increasing the number of bins could increase the power of the analysis, at the cost of further enlarging the covariance matrix. So motivated, we now use eighteen linearly-spaced radial bins from $r_{\rm min}=20\;\mpch$ to $r_{\rm max}=164\;\mpch$, leading to a bin width of $8\;\mpch$, roughly double the radial resolution of our previous analyses. We refit the volumes and number densities for the covariance matrices of CMASS NGC and SGC and LOWZ NGC and SGC as in \S\ref{subsec:covariance_matrix}, and these numbers are in Table \ref{tab:summary}. Other than these new inputs to the covariance matrix, all else is the same in our analysis, and we display the results in Fig. \ref{fig:chi2s_cmass_ns_18bins_LMAX4} (CMASS) and Fig. \ref{fig:chi2s_lowz_ns_18bins_LMAX4} (LOWZ). Generally, for each sample, the detection significance rises at each $N_{\rm eig}$ relative to the ten-bin analysis, and the same is true for the ``direct'' significances (where the analytic covariance was used). This increase is expected given our arguments about ``internal cancellation''. We summarize the significances in Table \ref{tab:sigs}.

\subsubsection{Reduced-$\ell_{\rm max}$ Analysis}
\label{subsub:red}
A pure-mock covariance is ideal for deriving the detection significance, yet the high dimensional data vector leads to a non-invertible sampling matrix due to the limited number of mocks. Reducing the number of channels considered makes it possible to use a covariance matrix taken directly from the mocks, with no need for the analytic one. 

\begin{figure}
    \centering
    \begin{subfigure}[b]{1\textwidth}
    \includegraphics[width=\textwidth]{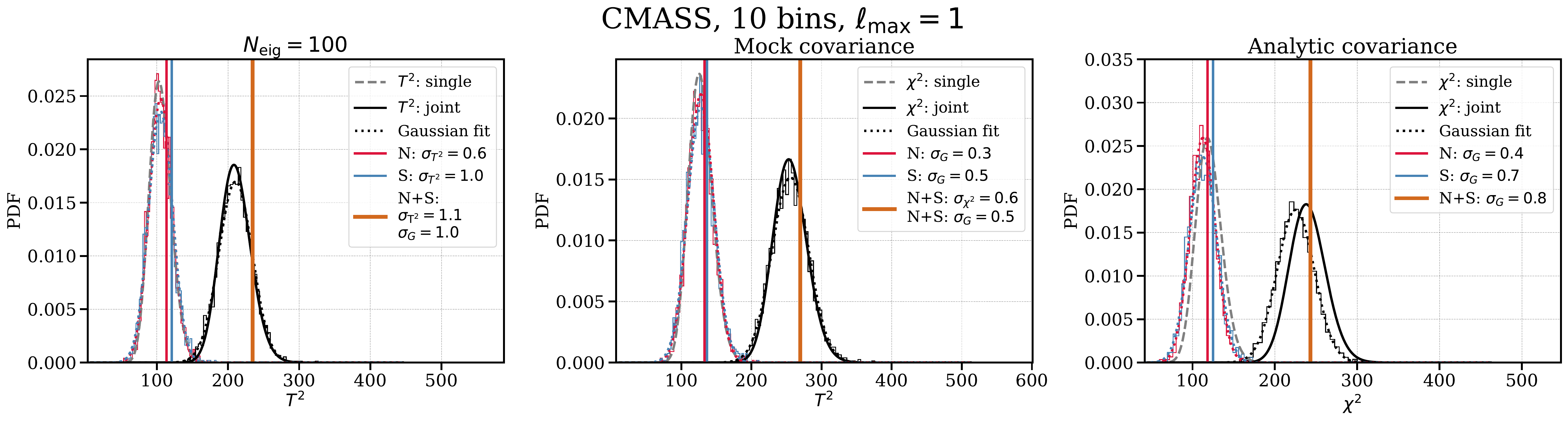}
    \end{subfigure}
    \begin{subfigure}[b]{1\textwidth}
    \includegraphics[width=\textwidth]{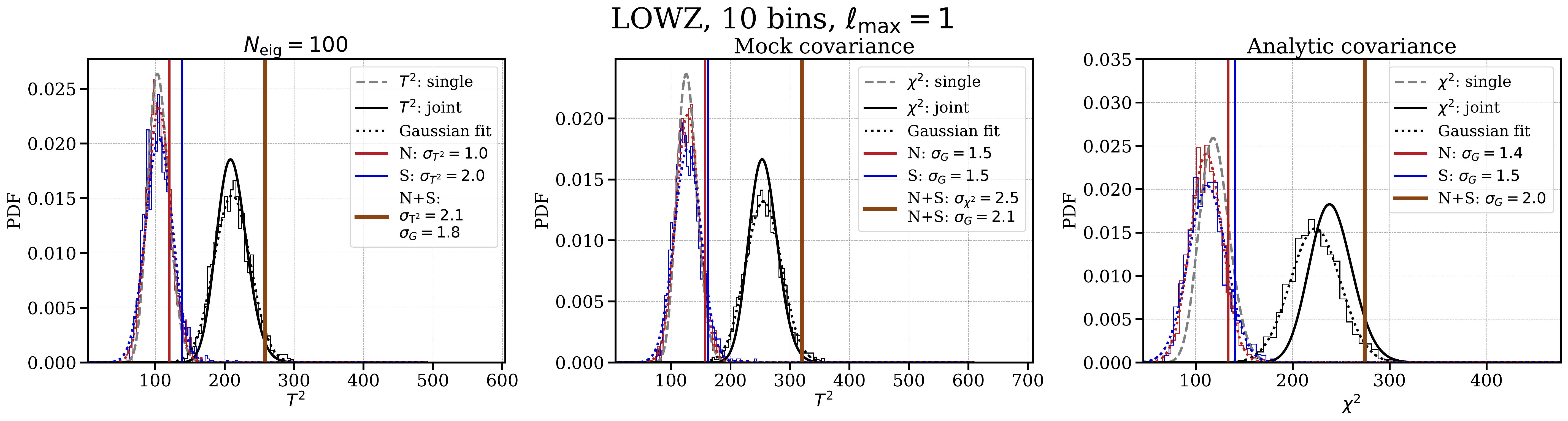}
    \end{subfigure}
    \begin{subfigure}[b]{1\textwidth}
    \includegraphics[width=\textwidth]{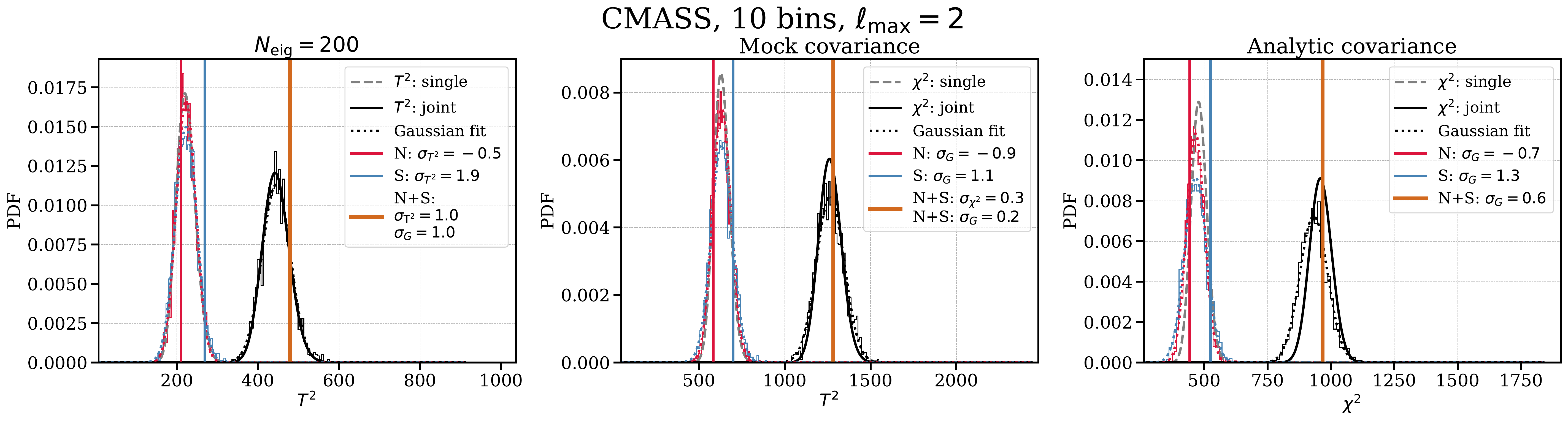}
    \end{subfigure}
    \begin{subfigure}[b]{1\textwidth}
    \includegraphics[width=\textwidth]{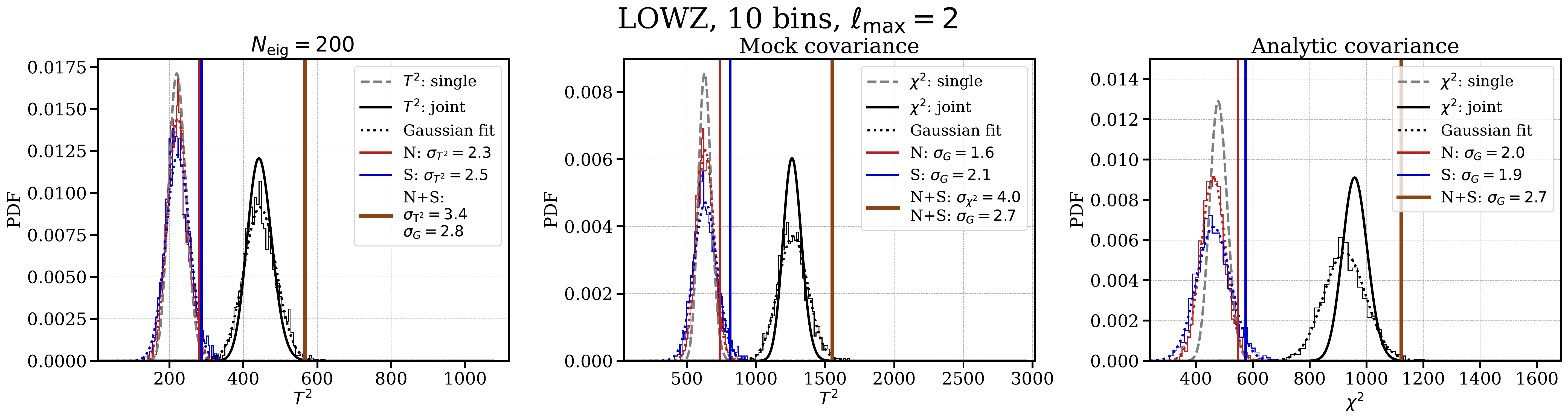}
    \end{subfigure}
    \label{fig:chi2s_boss_ns_LMAX12}
    \caption{{\bf Reduced} $\ell_{\rm max}$ analysis for {\bf ten} radial bins. We use three different methods of obtaining the covariance matrix but always with  $\ell_{\rm max}$ restricted to 1 or 2. The left column uses the compression scheme. The central column uses the covariance matrix obtained directly from the mocks, which is possible for this reduced $\ell_{\rm max}$ analysis. The third column uses the covariance obtained from adjusting the analytical covariance to fit, as well as possible, the purely mock-based covariance. The solid curves show the expected distributions for $T^2$ or $\chi^2$ while the histograms show the distributions obtained from the mocks. The solid lines indicate results from the data. There is overall good agreement in detection significance between the purely mock-based covariance (central column) and the analytic covariance (right column).}
    \label{fig:chi2s_ns_LMAX1_2}
\end{figure}

For $\ell_{\rm max}=1$, there is just one parity-odd channel, and for $\ell_{\rm max}=2$ there are four. Fig.~\ref{fig:chi2s_ns_LMAX1_2} shows the detection significances for CMASS and LOWZ with $\ell_{\rm max}=1$ in the first and second rows and $\ell_{\rm max}=2$ in the third and fourth rows. For $\ell_{\rm max}=1$ the histograms of the mocks' $T^2$ values agree well with the expected $T^2$  distribution, both for the compressed and the pure mock-based covariance approaches. However, the histogram of the mocks' $\chi^2$ values from the direct approach with analytic covariance deviates from the expected $\chi^2$ distribution, which indicates that the analytic covariance is imperfect. Nonetheless,
all three methods show consistent detection significance. For $\ell_{\rm max}=2$, all three methods are still consistent. However, we notice the changes in detection significance of CMASS and LOWZ are different when going from $\ell_{\rm max}=1$ to $\ell_{\rm max}=2$. We suspect this is due to the difference in the two samples, given their redshifts and number density, which we will leave for future exploration. After all, this reduced $\ell$ analysis serves as a robustness check of the pure analytic covariance matrix approach.

\subsection{Cross-Correlation Between NGC and SGC in Each Sample}
\label{subsec:cross}
A cosmological parity-odd correlation is expected to appear isotropically when decomposed using our choice of basis function.  This would manifest itself as a correlation between the signals in the SGC and NGC.   Noise in the observed 4PCF, would however not be correlated between hemispheres or between samples, and would reduce any underlying  cross-correlation.
We compute the Pearson correlation coefficient $r_{\rm p}$ between the signals in the SGC and NGC,  with some modifications which we now describe. The naive Pearson coefficient does not incorporate covariance between different points \textit{within} each data stream being cross-correlated, nor does it account for varying errors from point to point. For instance, two neighboring triple-bins within a given channel in NGC are likely to be much more correlated with each other than two very different triple-bins. Thus, if it happens that one such triple-bin is highly correlated with its analog in SGC, then the neighboring combination in NGC is also likely correlated with the neighboring combination in SGC. This is not unphysical, but it should be taken as less strong evidence for correlation than if two highly independent triple-bins in NGC \textit{each} showed a strong correlation with their analogs in SGC. The Pearson coefficient would not distinguish. This motivates us to rotate our measured data streams from NGC and from SGC to a basis where each data vector (\textit{within} the NGC or SGC) is independent; we thus work in the eigenbasis of the analytic covariance matrix. This procedure will be affected by any issues with the covariance matrix estimation, which is present for both analytic and mock covariance. 

Another issue with naive Pearson correlation is that some channels and triple-bins have larger statistical errors than others. They may therefore more easily be outliers and drive a spurious Pearson correlation. To correct this, once we are in the ``independent'' basis as above, taking the square root of the associated covariance matrix eigenvalue as its error bar, we divide each value by its error bar.

For the above procedure, if we had the true covariance, rotating into the true eigenbasis would preserve any correlations perfectly. However, as discussed in \S\ref{subsec:covariance_matrix} the analytic covariance is likely imperfect. We believe it to be close enough to the true covariance to preserve correlations within a given sample (CMASS or LOWZ) after rotation, but that trying to rotate into the different eigenbases implied by the covariances of the two different samples and then cross-comparing would not be robust. In particular, the different number densities in CMASS and LOWZ (see Table \ref{tab:summary}) will non-trivially impact the eigenvectors for each, likely randomizing any underlying correlations. Given this issue (explored more fully in Appendix \ref{app:covar_compare}), we do not report the results of a CMASS cross LOWZ analysis in this work.

Having outlined how we render our measured 4PF coefficients suitable for a cross-correlation analysis, we now define the statistic we use, the Pearson correlation coefficient $r_{\rm p}$. It is\footnote{The Pearson correlation coefficient is best for quantifying bivariate normally-distributed data. In the scaled, decorrelated basis, the data vectors indeed satisfy this assumption. We also quote Spearman's rank correlation coefficient, which identifies correlations between any two monotonic datasets by searching for correlations in the space of their ranks. It is more general than Pearson but also more difficult to manipulate analytically (as we do with the Pearson correlation coefficient in Appendix \ref{app:pears}). The results of both are consistent.}
\begin{eqnarray}
r_{\rm p} &=& \frac{\sum_{i=1}^{ N_{\rm d}} \left(d_{1,i}-\av{\vd_1}_{\rm all}\right)\left(d_{2,i}-\av{\vd_2}_{\rm all}\right)}{\sqrt{\sum_{i=1}^{ N_{\rm d}} \left(d_{1,i}-\av{\vd_1}_{\rm all}\right)^2 \sum_{i=1}^{ N_{\rm d}} \left(d_{2,i}-\av{\vd_2}_{\rm all}\right)^2 }},
\label{eqn:corr_rs}
\end{eqnarray}
where $d_{i}$ is the $i^{th}$ element of the 4PCF data vector $\vd_i$ with $i=1$ for the NGC and $i=2$ for the SGC.
Both are in the decorrelated, normalized basis for their respective hemisphere. $\av{\ldots}_{\rm all}$ denotes the average over the elements of the vector, and $N_{\rm d}$ is the number of degrees of freedom.

The PDF for the Pearson correlation coefficient if there is no correlation is~\citep{Kenney1947mathstat, Hotelling1953CorrCoeff}
\begin{eqnarray}
f(r_{\rm p}) = \frac{{(1-r_{\rm p}^2)}^{(N_{\rm d}-4)/2}}{{\rm B}\left(1/2, (N_{\rm d}-2)/2\right)},
\label{eqn:pear_f}
\end{eqnarray}
where ${\rm B}$ is the Beta function. 

\begin{figure}
    \centering
    \includegraphics[width=.7\textwidth]{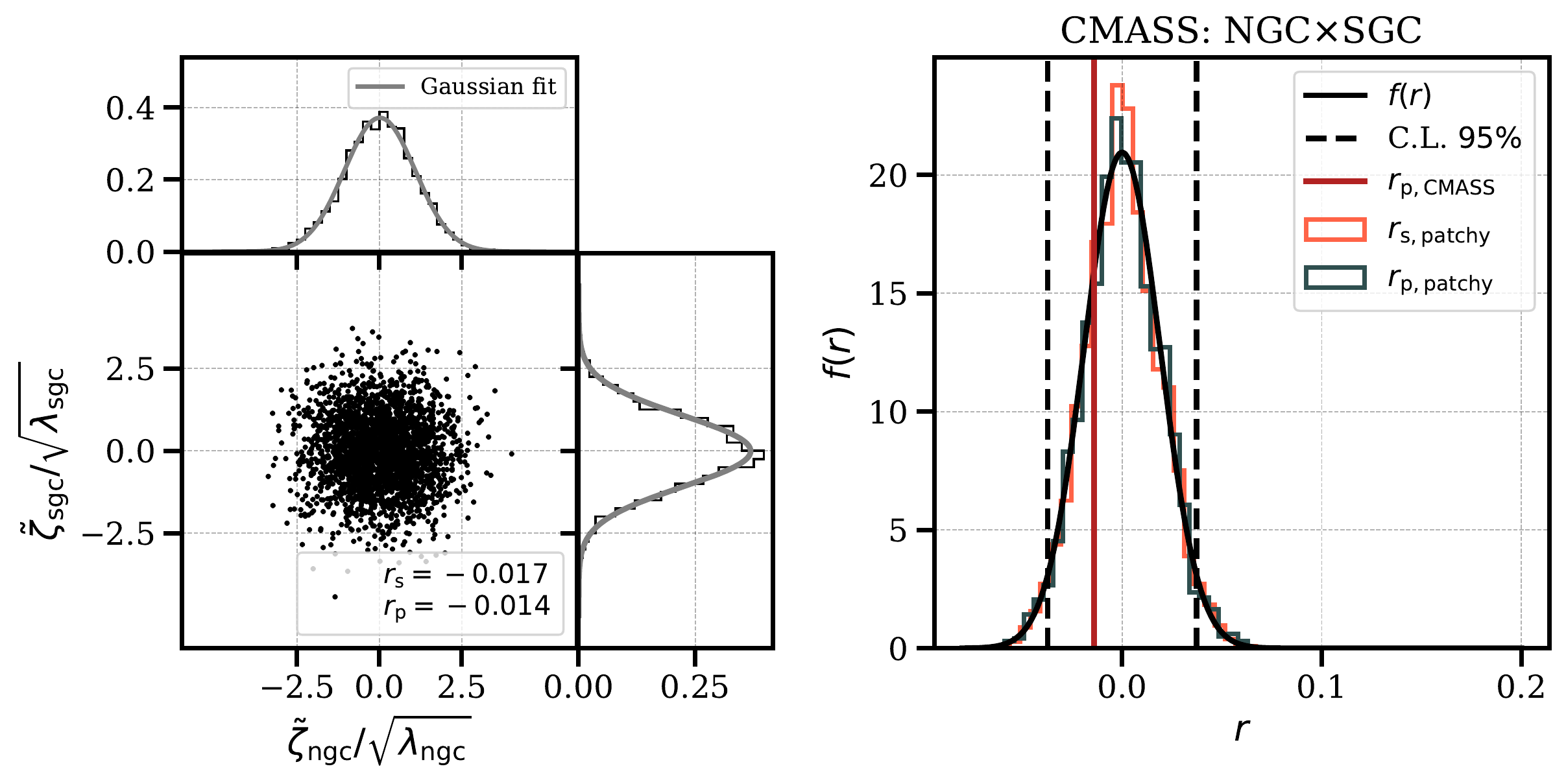}
    \caption{{\it Left}: Scatter plot for the 4PCF with {\bf ten bins} measured from CMASS in the decorrelated basis (described in \S\ref{subsec:cross}) and scaled by the covariance matrix eigenvalues, with both Pearson ($r_{\rm p}$) and Spearman ($r_{\rm s}$) correlation coefficients given in the legend. 
    The extended panes in the left panel display the histogram overplotted with a Gaussian fit. {\it Right}: PDF $f(r)$ (black curve; Eq. \ref{eqn:pear_f}) for the Pearson correlation coefficient and the Pearson and Spearman correlation coefficients obtained from \patchy (orange and blue histograms). The 95$\%$ confidence level (dashed vertical lines) corresponds to an $r_{\rm p} = 0.037$. Our Pearson $r_{\rm p}$ of -0.014 corresponds to a $p$-value of  $0.460$, so there is no statistically significant correlation detected.}
    \label{fig:pearson_corr_cmass_NGCxSGC_odd_stacked23ell}
\end{figure}

Fig.~\ref{fig:pearson_corr_cmass_NGCxSGC_odd_stacked23ell} shows the correlation analysis results for CMASS using ten bins. The left panel is a scatter plot of the (decorrelated and scaled) data for CMASS NGC and SGC. The histograms in the extended panels can be well-fit by a Gaussian, showing this assumption of the Pearson coefficient is satisfied. We find no statistically significant correlation at the 95\% confidence level. We also report Spearman's rank coefficient, $r_{\rm s}$, in the Figure (see footnote 10). The right panel shows the PDF for $r_{\rm p}$ (black curve; see Eq. \ref{eqn:corr_rs}) if the two datasets are both normally-distributed. As noted above, a correlation at some level might be expected if the signal were of cosmological origin. Meanwhile, there are some systematics where such a correlation would \textit{not} be expected, and the same goes for some instrumental artifacts. On the other hand, artifacts at the telescope \textit{could} correlate NGC and SGC.

\begin{figure}
    \centering
    \includegraphics[width=.7\textwidth]{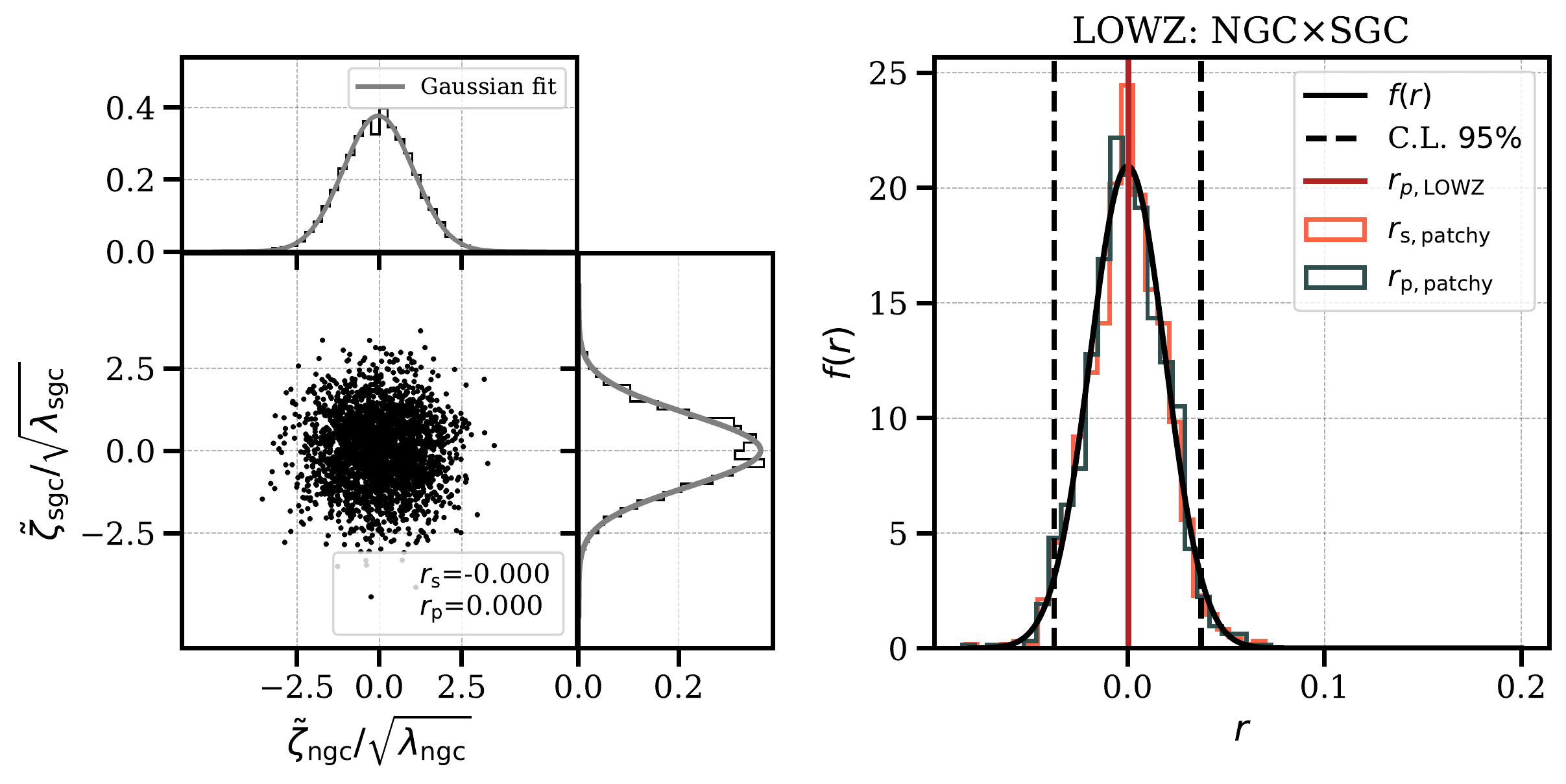}
    \caption{Same as Fig.~\ref{fig:pearson_corr_cmass_NGCxSGC_odd_stacked23ell} but for LOWZ, again with {\bf ten bins}. Again we detect no statistically significant correlation.}
    \label{fig:pearson_corr_lowz_NGCxSGC_odd_stacked23ell}
\end{figure}

\begin{figure}
    \centering
    \begin{subfigure}[b]{0.45\textwidth}
    \includegraphics[width=.9\textwidth]{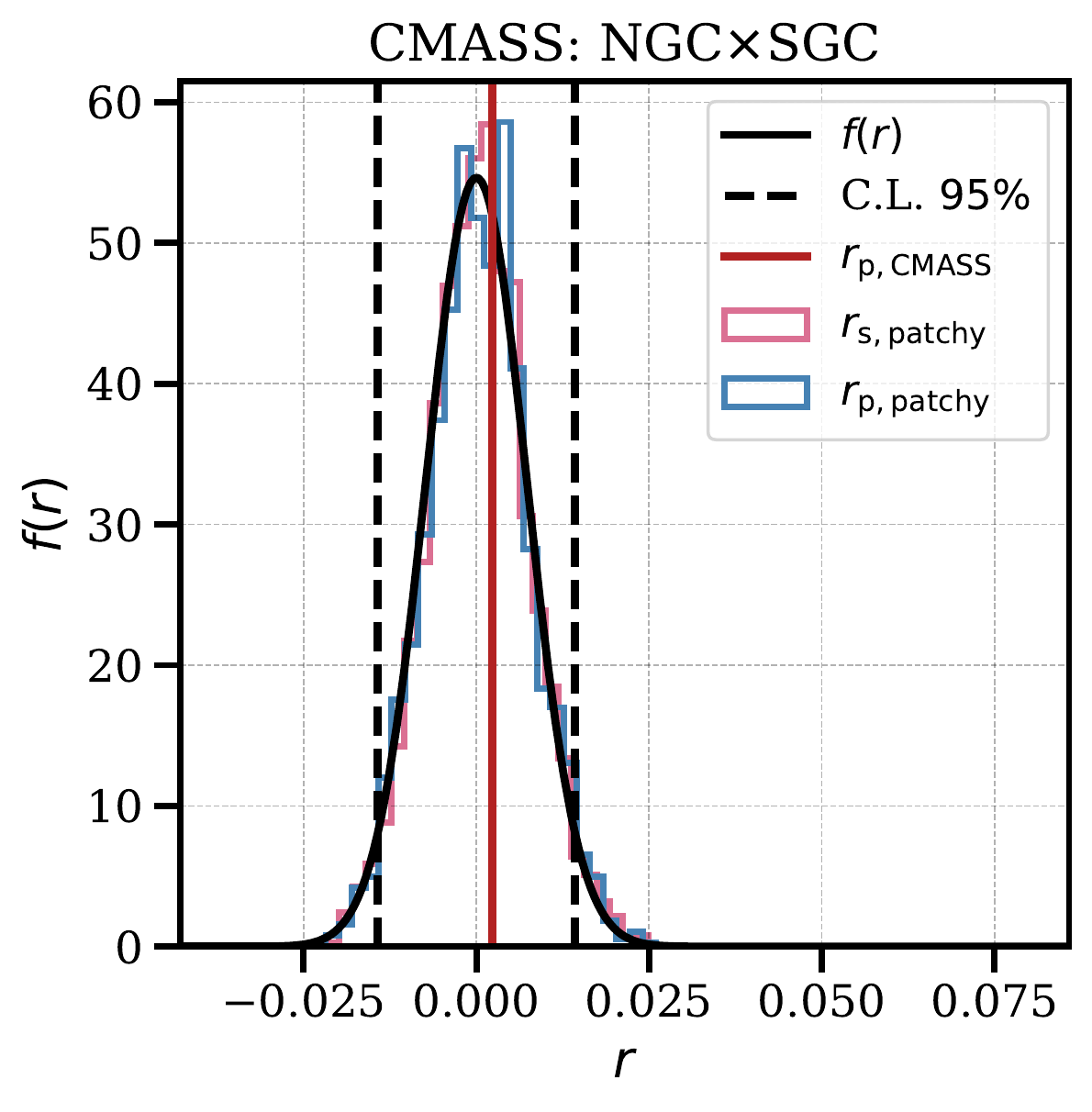}    
    \end{subfigure}
    \begin{subfigure}[b]{0.45\textwidth}
    \includegraphics[width=.9\textwidth]{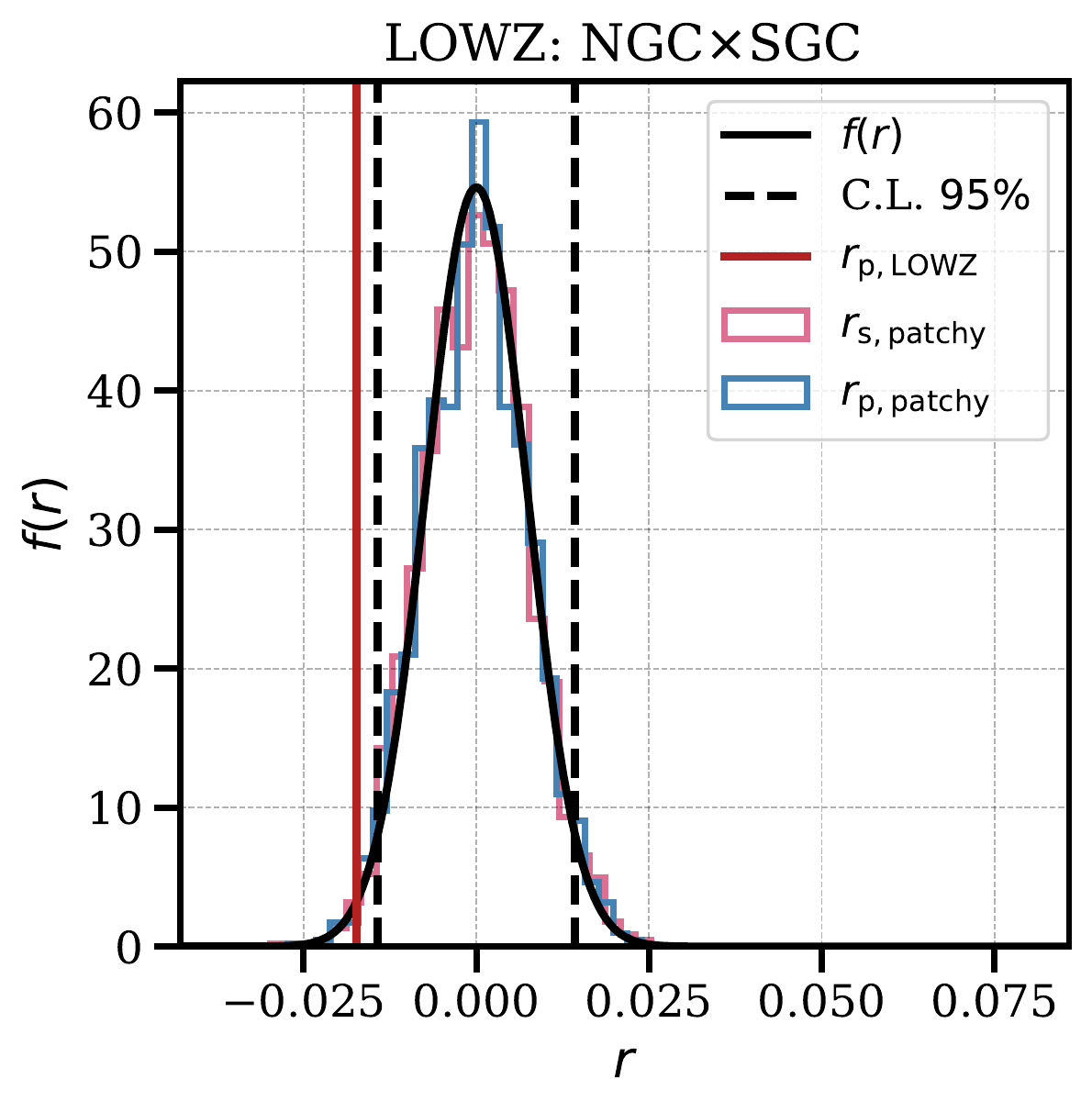}    
    \end{subfigure}
    \caption{PDF of $f(r_{\rm p})$ for {\bf eighteen bins}; the left panel displays results using CMASS, the right using LOWZ. The correlation coefficient for CMASS is consistent with zero, but we see a negative correlation for LOWZ; this is further discussed in the main text.}
    \label{fig:dist_pearson_corr_NGCxSGC_odd_stacked23ell_nbin18_LMAX4_drCut0_rmax200}
\end{figure}

Fig.~\ref{fig:pearson_corr_lowz_NGCxSGC_odd_stacked23ell} is similar to Fig.~\ref{fig:pearson_corr_cmass_NGCxSGC_odd_stacked23ell} but for the LOWZ sample.
Fig.~\ref{fig:dist_pearson_corr_NGCxSGC_odd_stacked23ell_nbin18_LMAX4_drCut0_rmax200} shows the correlation analysis results using eighteen bins, where the data here also follow a Gaussian distribution (although we do not show them explicitly). In the right panel, for LOWZ, the correlation coefficient is a negative number at more than $95\%$ confidence. However, we also find that the correlation coefficients are sensitive to the covariance matrix. Changing the number density used in the analytic covariance matrix by 10\% can flip the LOWZ correlation coefficient to a positive number (but one consistent with zero). Overall, we thus conclude that there is no evidence for a positive correlation between the N and S for LOWZ or for CMASS when using eighteen bins. Our correlation study results for ten and eighteen bins and LOWZ and CMASS are summarized in Table \ref{tab:corrln}.

\subsubsection{Linear Toy Model in the Eigenbases to Explore $\chi^2$ and $r_{\rm p}$}
To gain an intuition for the high detection significance and the low correlation between the NGC and SGC in the LOWZ, we construct a toy model. Henceforth we use the Pearson correlation coefficients and  $\chi^2$ obtained from ten bins results. The purpose is not to constrain real physics, but simply to gain an intuition for whether the values of $\chi^2$ and $r_{\rm p}$ found in our data are plausibly consistent with each other. 

We assume the signal is given by a toy model that is linear in the eigenbases of the analytic covariance matrices parametrized by a slope $m$ and an intercept $b$ as $s = mx + b$, where $x$ is the \textit{index} of each eigenvalue. We produce two sets of fake data. Each is drawn randomly from a Gaussian distribution, the mean of which is the linear model, and the width of which is the square root of the eigenvalue of the analytic covariance matrix for either CMASS or LOWZ as appropriate.

We separately maximize the log-likelihood for the data vectors ${\bf d}_{\rm cmass}=\{r_{\rm p, cmass},\; \chi^2_{\rm cmass}\}$ and ${\bf d}_{\rm lowz}=\{r_{\rm p, lowz},\; \chi^2_{\rm p,\; lowz}\}$, respectively. If there is a true primordial signal, we might expect that the parity-odd 4PCF in LOWZ has an overall amplitude 2$\times$ that in CMASS, folding in the difference in growth rates and linear biases. Therefore we search over different ranges in the slope and the intercept for each: $m_{\rm cmass}\in [1\times 10^{-8},\; 2\times 10^{-7}]$, the intercept $b_{\rm cmass}\in [1\times 10^{-5},\; 5\times 10^{-4}]$ for CMASS and $m_{\rm lowz}\in [2\times 10^{-8}, 4\times 10^{-7}]$, the intercept $b_{\rm lowz}\in [2\times 10^{-5},\; 1\times 10^{-3}]$. In Fig.~\ref{fig:lik_fit_pearson_chi2_cmass_lowz_vec2d} is the log-likelihood plotted for slope as a function of intercept. We find the best fit slope and intercept (the bluest region in the log-likelihood): $m_{\rm cmass}=5.3\times 10^{-8}$ and $b_{\rm cmass}=1.1\times 10^{-4}$ for CMASS; $m_{\rm lowz}=1.5 \times 10^{-7}$ and $b_{\rm lowz}=2.7 \times 10^{-4}$ for LOWZ. In Fig.~\ref{fig:pearson_chi2_cmass_lowz} we show a comparison of the correlation coefficients and $\chi^2$ values obtained from the real data to the distribution of those from the fake data. The fake data distribution in different colors is generated with (i) a zero signal (ii) a signal with the best-fitting parameters and (iii) a signal with both a steeper slope and a higher intercept. This plot shows that it is possible to explain the low correlation and noticeable detection significance with a simple linear model.\footnote{The linear model can be regarded as a first-order Taylor expansion of some true physical model. We imagine some true model in the space of the physical 4PCF amplitudes; then rotate this model into the eigenbasis of the covariance and take a Taylor series for it there.} This model's purpose is {\bf not} to constrain any non-standard inflationary physics. Rather we wish to demonstrate that it is generically possible to have high $\chi^2$ and low correlation coefficients, even with a simple functional form for the model.

\begin{figure}
    \centering
    \includegraphics[width=\textwidth]{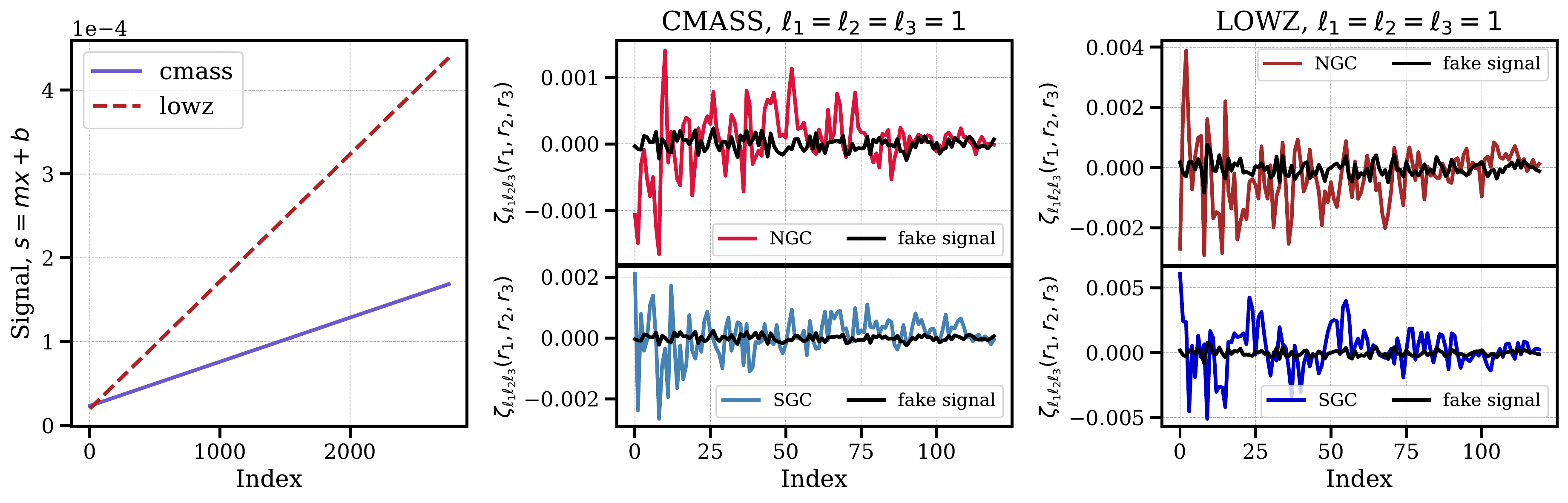}
    \caption{Exploration of a linear model for the underlying signal (vs. the eigenvalue index) for both CMASS and LOWZ (with {\bf ten} radial bins). {\it Left}: Toy models for CMASS with $m=5.3\times 10^{-8}$, $b=1.1\times 10^{-4}$ (dashed red) and LOWZ (solid purple) with $m=1.5 \times 10^{-7}$, $b=2.7\times 10^{-4}$ (these are our best-fit values; see Fig. \ref{fig:lik_fit_pearson_chi2_cmass_lowz_vec2d} for discussion of the fitting and likelihood). {\it Middle}: Toy model (black curve; ``fake signal'') transformed back to the observed basis plotted with the measured CMASS 4PCF for $\ell_1=\ell_2=\ell_3$; upper row is NGC and lower is SGC. {\it Right}: Analog of the middle panel but for LOWZ.}
    \label{fig:fake_signal_cmass_lowz}
\end{figure}

\begin{figure}
    \centering
    \includegraphics[width=.8\textwidth]{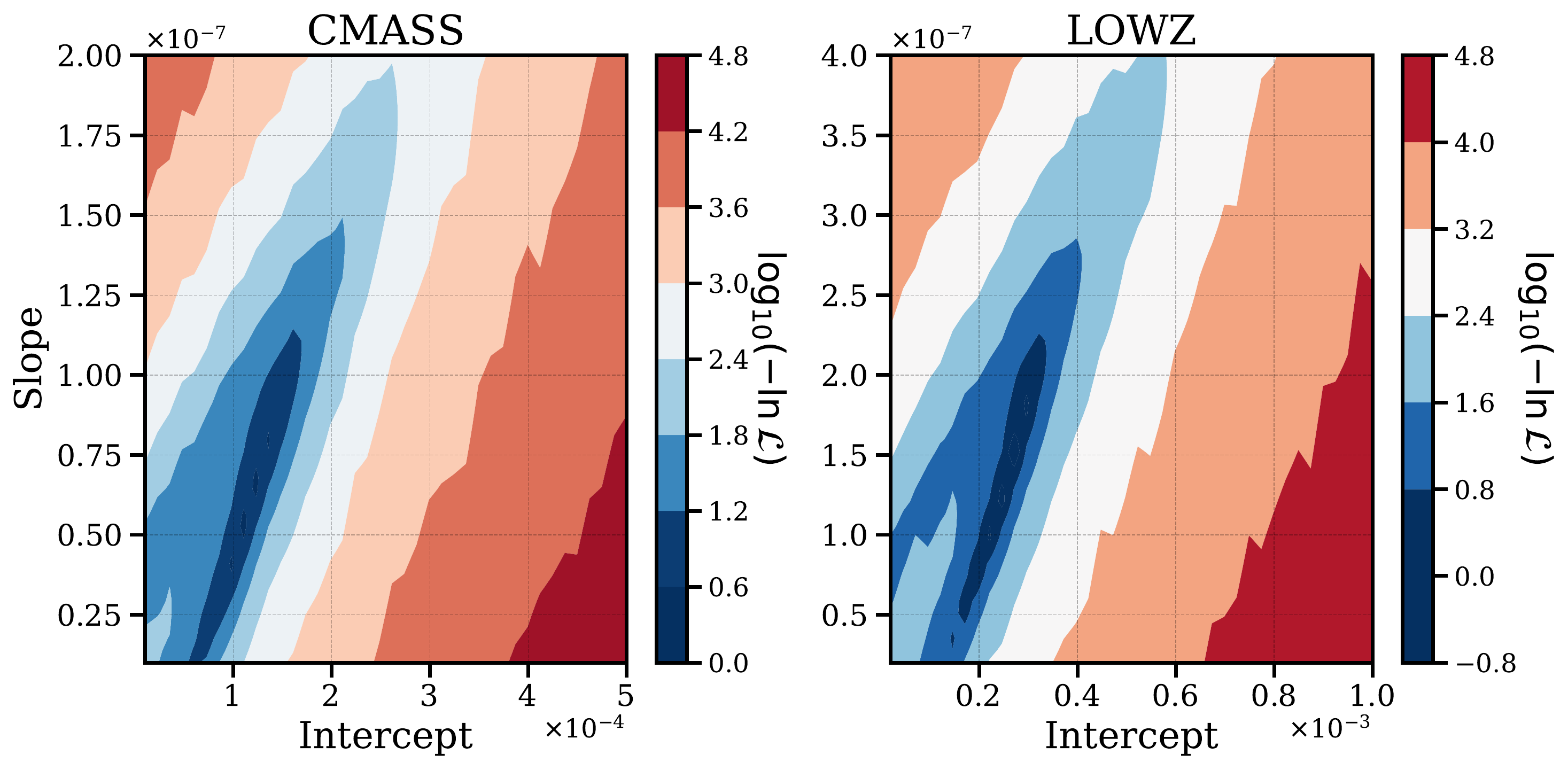}
    \caption{Log-likelihood for our two-parameter linear toy model $s = mx + b$ (``fake signal''). We search for $m$ and $b$ so as to make the observed $\chi^2$ and $r_{\rm p}$ consistent with each other in each sample. The fitting is independent for CMASS and LOWZ. In the eigenbasis of the 4PCF analytic covariance for either CMASS or LOWZ, we produce our linear model and then add Gaussian noise drawn using these eigenvalues to produce $200$ realizations of our ``fake signal'', each with an associated $\chi^2$ and $r_{\rm p}$. This then lets us obtain the covariance matrix of $\chi^2$ and $r_{\rm p}$ in this toy model. We may then compute, for a given $m$, and $b$, their likelihood given the actual observed $\chi^2$ and $r_{\rm p}$ from the data. The blue  region is the highest likelihood. We display these results for both CMASS NGC cross SGC and LOWZ NGC cross SGC. It is notable that the best-fit parameters for LOWZ are roughly two times those for CMASS; this did not have to be the case, as the fitting is independent for each sample, but is in fact the scaling we would naively expect were the parity-odd signal truly of primordial origin. Our interpretation of this linear toy model is further discussed in \S\ref{subsec:cross}.}
    \label{fig:lik_fit_pearson_chi2_cmass_lowz_vec2d}
\end{figure}

\begin{figure}
    \centering
    \includegraphics[width=.7\textwidth]{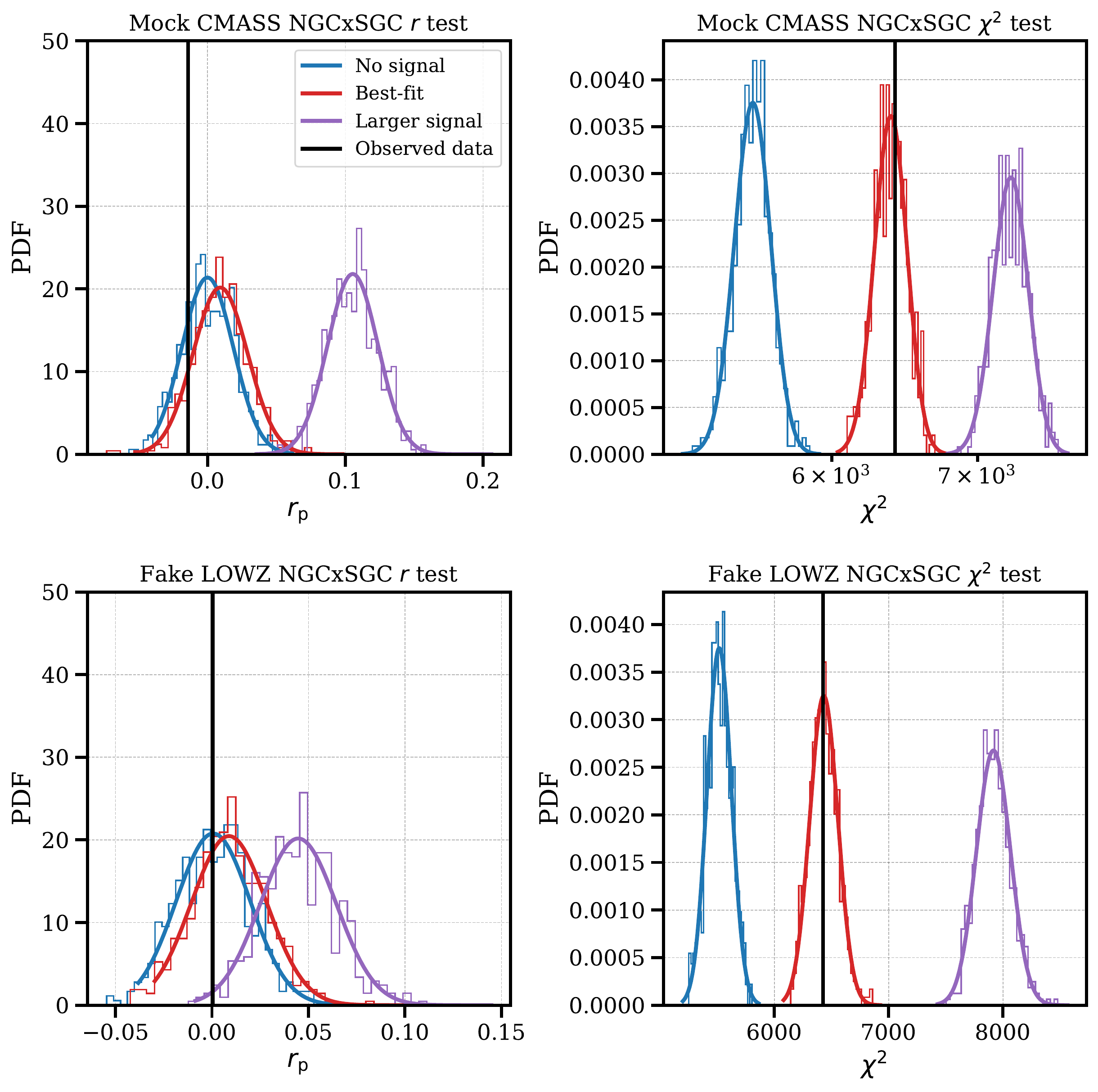}
    \caption{The Pearson correlation coefficients $r_{\rm p}$ (\textit{left column}) and $\chi^2$ values (\textit{right column}) observed in our data (black lines) shown over the distribution of the same from the ``fake signal'' generated by our ``linear-in-eigenbasis'' toy  model. We show results from $m = b = 0$ in blue, from the best-fit $m$ and $b$ in red (Fig. \ref{fig:fake_signal_cmass_lowz} displays these models and gives the best-fit values in the caption, and the likelihood used is shown in Fig. \ref{fig:lik_fit_pearson_chi2_cmass_lowz_vec2d}), and from a larger ``fake'' signal in purple to indicate the possible range of outcomes. The fake data at each index (see Fig. \ref{fig:fake_signal_cmass_lowz}) is drawn from a Gaussian distribution with mean given by the linear model and width given by the square-root of the eigenvalue of the analytic covariance matrices. CMASS results are shown in the upper row and LOWZ in the lower. We see that there exists a ``fake signal'' that does explain the observed $r_{\rm p}$ and $\chi^2$ from the real data. While our ``fake signal'' should not be taken too literally, this does indicate that, generically, our observed $\chi^2$ and $r_{\rm p}$ are not necessarily in tension with each other.}
    \label{fig:pearson_chi2_cmass_lowz}
\end{figure}

\section{Potential Systematics}
\label{sec:systematics}
Parity-breaking in large-scale structure is not expected in the standard cosmological paradigm. Hence it is important to assess whether systematic errors could produce an apparent parity-breaking signal. We must consider systematics both at the signal level and at the covariance level. These latter must be considered because they could cause an underestimation of the covariance, thus leading to an overestimation of the detection significance.

We divide these systematics into three categories: survey-related effects, observer-related effects, and procedure-related effects. We discuss these in respectively \S\ref{subsec:survey_sys}, \S\ref{subsec:observe_sys}, and \S\ref{subsec:procedure}. We have already considered the possible impact of systematics on the covariance matrix in \S\ref{subsub:covar_sys}.

\subsection{Survey-Related Effects}
\label{subsec:survey_sys}
Spectroscopic surveys provide galaxies' 3D positions, but estimating the correlation functions requires the excess probability \textit{over random} of finding an N-tuplet of galaxies in a given configuration. This in turn demands precise knowledge of the selection function.

\subsubsection{A Toy Model for Errors in the Selection Function}
\label{subsub:toy_model}

Despite that great care has been taken in BOSS to model the impact of astrophysical foregrounds (\textit{e.g.} stellar density), or observational conditions on the survey selection function, we still seek an intuition for the impact of an imperfectly-modeled selection function on the parity-odd modes.
In particular, our goal is to see if an even-parity 4PCF is converted into an odd parity one by an improperly corrected selection function. For this goal, it is sufficient to begin with a spatially unclustered density field (which would produce only an even-parity 4PCF).

The selection function encapsulates how our selection criteria (\textit{e.g.} color and magnitude cuts) samples the underlying distribution of galaxies in the universe.
The underlying density field can be inferred from the observed number density $n(\bfs)$ and the selection function $f^{\rm obs}(\bfs)$. Since we have no access to the underlying distribution, we cannot disentangle how our selection criteria shape the sample we end up observing from any underlying variation in the universe (usually specifically along the line of sight).
In this discussion we allow the deviation of the selection function to be general and depend on 3D position $\bfs$.

Our estimate of the true number density of objects is
\begin{eqnarray}
    \hat{n}^{\rm true} (\bfs)= \frac{n(\bfs)}{\hat{f}^{\rm obs}(\bfs)},
\end{eqnarray}
where $n$ is the observed number density and $\hat{f}^{\rm obs}$ our estimate of the selection function. The actual true number density is
\begin{eqnarray}
    n^{\rm true}(\bfs) = \frac{n(\bfs)}{f^{\rm obs}(\bfs)},
\end{eqnarray}
where $f^{\rm obs}$ is the true selection function.

Hence we may relate our estimated true number density to the actual true number density as
\begin{eqnarray}
\hat{n}^{\rm true}(\bfs) = \frac{f^{\rm obs}(\bfs)}{\hat{f}^{\rm obs}(\bfs)} n^{\rm true}(\bfs) \equiv g(\bfs) n^{\rm true} (\bfs).
\end{eqnarray}
Our density fluctuation field is then
\begin{eqnarray}
\delta(\bfs) = \frac{\hat{n}^{\rm true}(\bfs) -\bar{n}}{\bar{n}} = g(\bfs) -1.
\end{eqnarray}
We have assumed that we correctly estimated $\bar{n}$ and also that $n^{\rm true}(\bfs) \equiv \bar{n}$, \textit{i.e.} in this simple toy model, there is no underlying clustering. We require that the integral of $g$ over the survey will be unity; we use this condition to normalize $g$ shortly.

We now make a toy model for $g$ to enable assessment of its possible impact. We take it that it is factorizable as 
\begin{eqnarray}
g(\bfs) = g_x(x) g_y(y) g_z(z)
\end{eqnarray}
and that along each direction it is a power law as
\begin{eqnarray}
&&g_x = I^{-1/3}\left(x_0 + \Delta x\right)^{p_x}, \;\;\; g_y = I^{-1/3}\left(y_0 + \Delta y\right)^{p_y}, \;\;\;
g_z = I^{-1/3}\left(z_0 + \Delta z\right)^{p_z},
\label{eqn:pl_mod}
\end{eqnarray}
where we have written $\Delta x$, $\Delta y$, and $\Delta z$ as displacements around a central point $(x_0,y_0, z_0)$ and have defined the normalization $I$ as
\begin{eqnarray}
I \equiv \frac{L_{x}^{p_x+1}L_{y}^{p_y+1}L_{z}^{p_z+1}}{(p_x+1)(p_y+1)(p_z+1)}. 
\label{eqn:normalization_I}
\end{eqnarray}
This normalization ensures that the integral of $g$ over the survey is unity and the $L_{\rm i}$ are the lengths of the box sides in each direction.

The primary galaxy for our computation of the spherical harmonic coefficients $a_{\ell m}$ (see \S\ref{sec:basis}, as well as \citealt{Philcox2021encore} for a detailed discussion of how these coefficients enter our algorithm) is at $\bfs_0$ and each secondary galaxy is displaced by $\bfr_i$ from the primary. If we now assume that the true density field is uniform, its change due to our modulation will be (for the primary)
\begin{eqnarray}
&&\delta(\bfs_0) = g(\bfs_0)-1.
\end{eqnarray}
For the secondaries (in this case, the three field positions around the primary, which, with it, build up that primary's contribution to a given 4PCF channel) we have, by taking a Taylor series for $g$ to leading order:
\begin{eqnarray}
\delta(\bfs_i) &=& g(\bfs_0) + \nabla g(\bfs)|_{\bfs_0} \cdot \bfr_i - 1,
\end{eqnarray}
with $\bfr_i \equiv \bfs_i - \bfs_0$. We now assess whether this systematic can contribute to the parity-odd 4PCF. When we multiply out the three secondaries, we can produce terms involving one, two, or three factors of $\nabla g$. All of the parity-odd basis functions are proportional to $\bfrhat_{1} \cdot (\bfrhat_{2} \times \bfrhat_{3})$, so only the term involving three factors of $\nabla g$ may possibly contribute.

We now compute the projection of this term onto our basis. Denoting it $\zeta_g$, we have
\begin{eqnarray}
\zeta_g = \int d\bfrhat_1 d\;\bfrhat_2\; d\bfrhat_3\; (\nabla g(\bfs)|_{\bfs_0}\cdot \bfr_1) (\nabla g(\bfs)|_{\bfs_0}\cdot \bfr_2) (\nabla g(\bfs)|_{\bfs_0}\cdot \bfr_3) \;\bfrhat_{1} \cdot (\bfrhat_{2} \times \bfrhat_{3}) \nonumber\\
= r_1 r_2 r_3 \;\nabla g(\bfs)|_{\bfs_0}\cdot (\nabla g(\bfs)|_{\bfs_0}\times \nabla g(\bfs)|_{\bfs_0}) = 0,
\end{eqnarray}
where we used a vector integral identity to obtain the second line.

\begin{figure}
    \centering
    \includegraphics[width=0.4\textwidth]{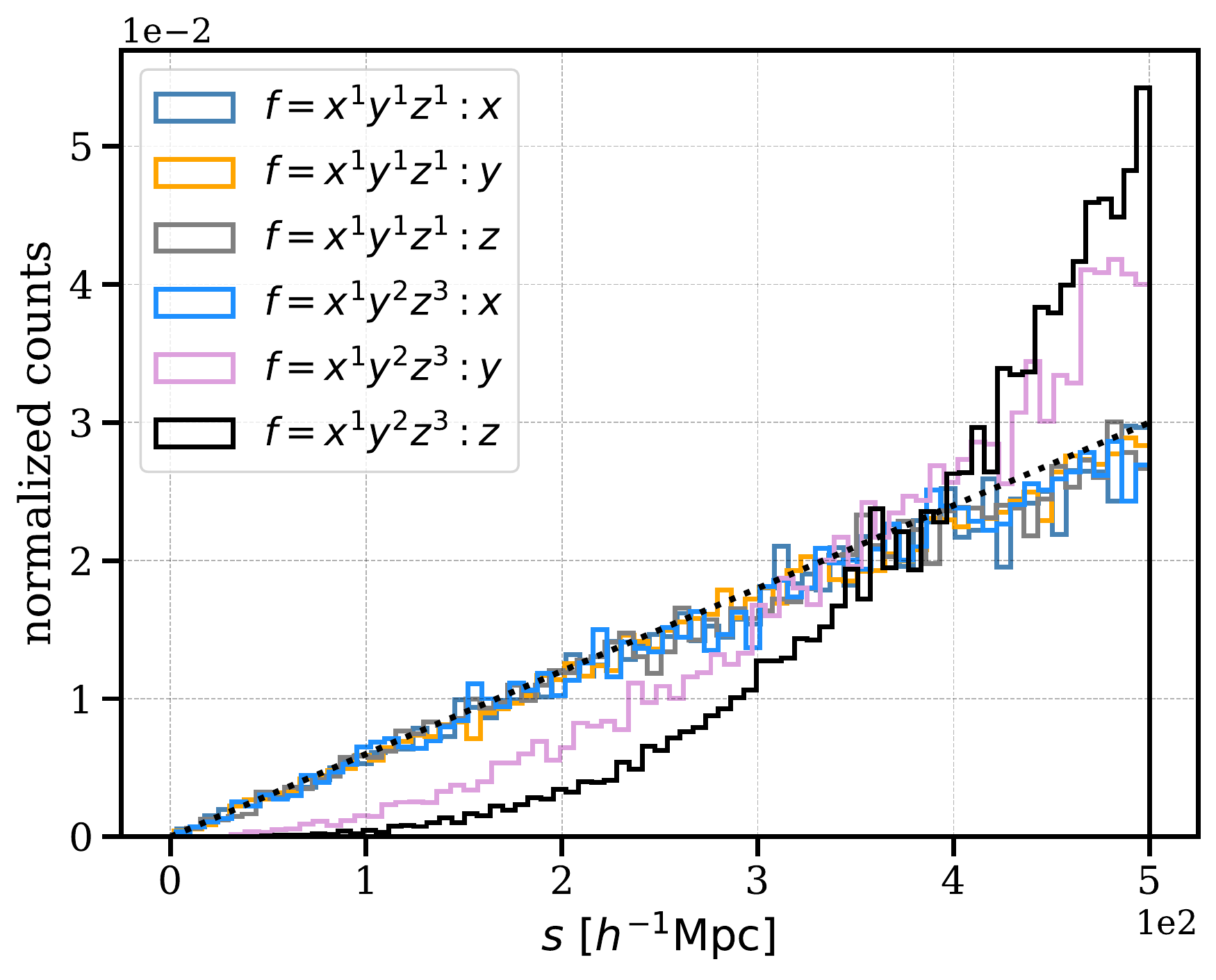}
    \caption{Normalized number count distribution along the $x$-, $y$-, and $z$-axes of mocks with a spatially-random underlying distribution but modulated by a power-law in each direction (\S\ref{subsub:toy_model}, Eq. \ref{eqn:pl_mod}), as indicated in the legend; the function $f$ shows the modulation along each axis, and the variable after the colon which axis is plotted. We show two cases along each of the three axes; powers $p_x = p_y = p_z =1$ and then $p_x=1,\; p_y=2,\; p_z=3$.}
    \label{fig:nocluster500mpc_sel_xyz_111_123}
\end{figure}

Though the above shows that on average (in an infinite volume), a power-law modulation would produce no spurious parity-odd signal, such modulation could increase the fluctuations observed. If such a systematic were present in the data but not the mocks, then the covariance used (from mocks or calibrated by mocks) would be an underestimate, and this could produce an apparent detection. Hence, we explore this power-law toy model numerically to assess if such an effect might occur. We construct $100$ mocks whose points have a spatially random distribution, each in a cubic box with side length $L_{\rm box} = 500 \,\mpch$ and number density $\bar{n} = 3\times 10^{-4} \left[\mpch\right]^{-3}$. The distribution of the mock number density along the $x$-, $y$-, and $z$-axes is shown in Fig.~\ref{fig:nocluster500mpc_sel_xyz_111_123}. We investigate two cases, $p_x=1$, $p_y=1$, $p_z=1$ and $p_x=1$, $p_y=2$, $p_z=3$. We also create a set of $100$ mocks with the same box size and number density but with a uniform distribution along the three axes. For all cases, we use the same random catalog with a uniform distribution at all spatial positions to convert the density field into a density fluctuation field (the only relevant part of edge correction for a periodic box).

\begin{figure}
     \centering
     \begin{subfigure}[b]{0.3\textwidth}
         \centering
         \includegraphics[width=1\textwidth]{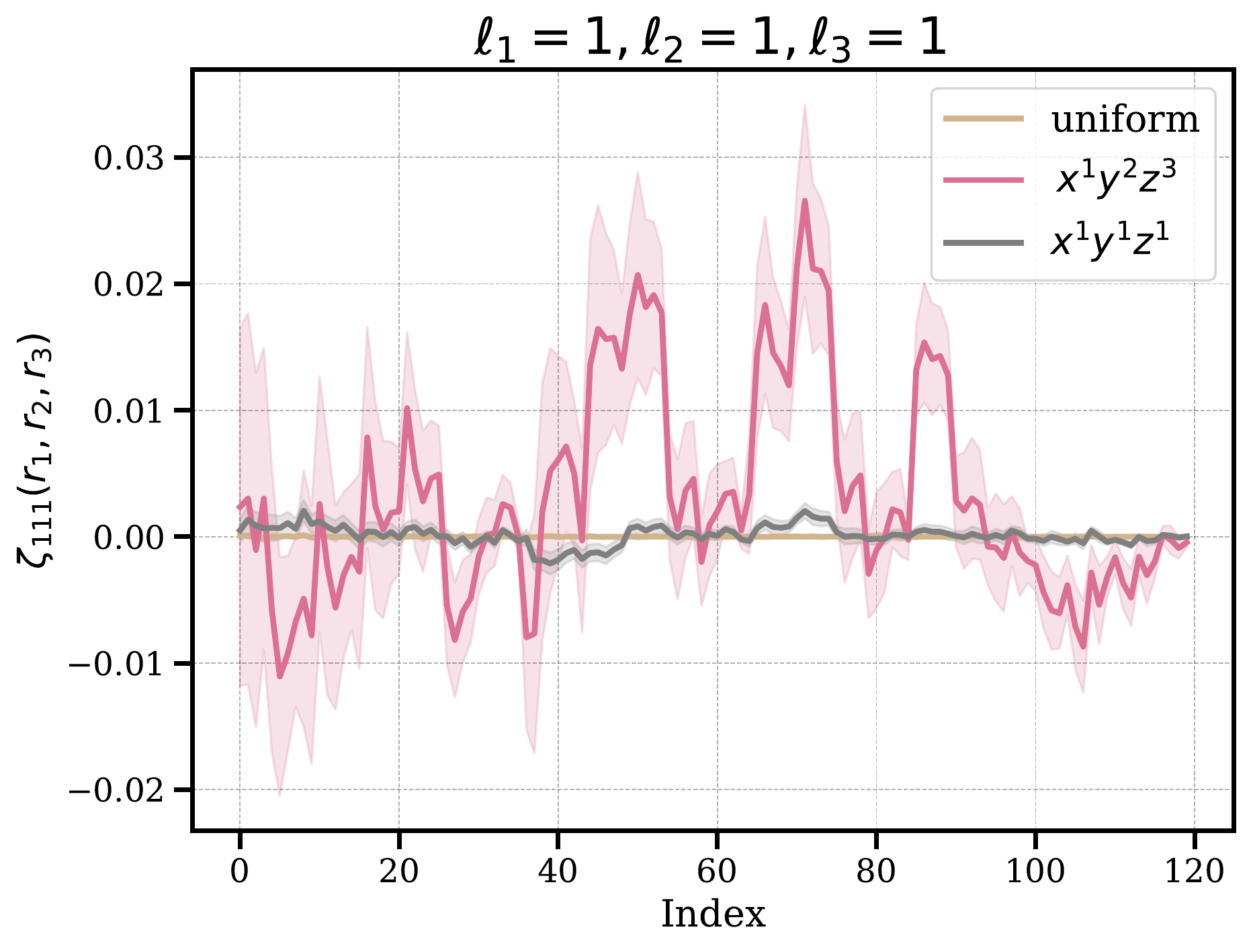}
         \label{fig:zeta_nonclus_99mocks_500mpch_box_ell111}
     \end{subfigure}
     \begin{subfigure}[b]{0.3\textwidth}
         \centering
         \includegraphics[width=1\textwidth]{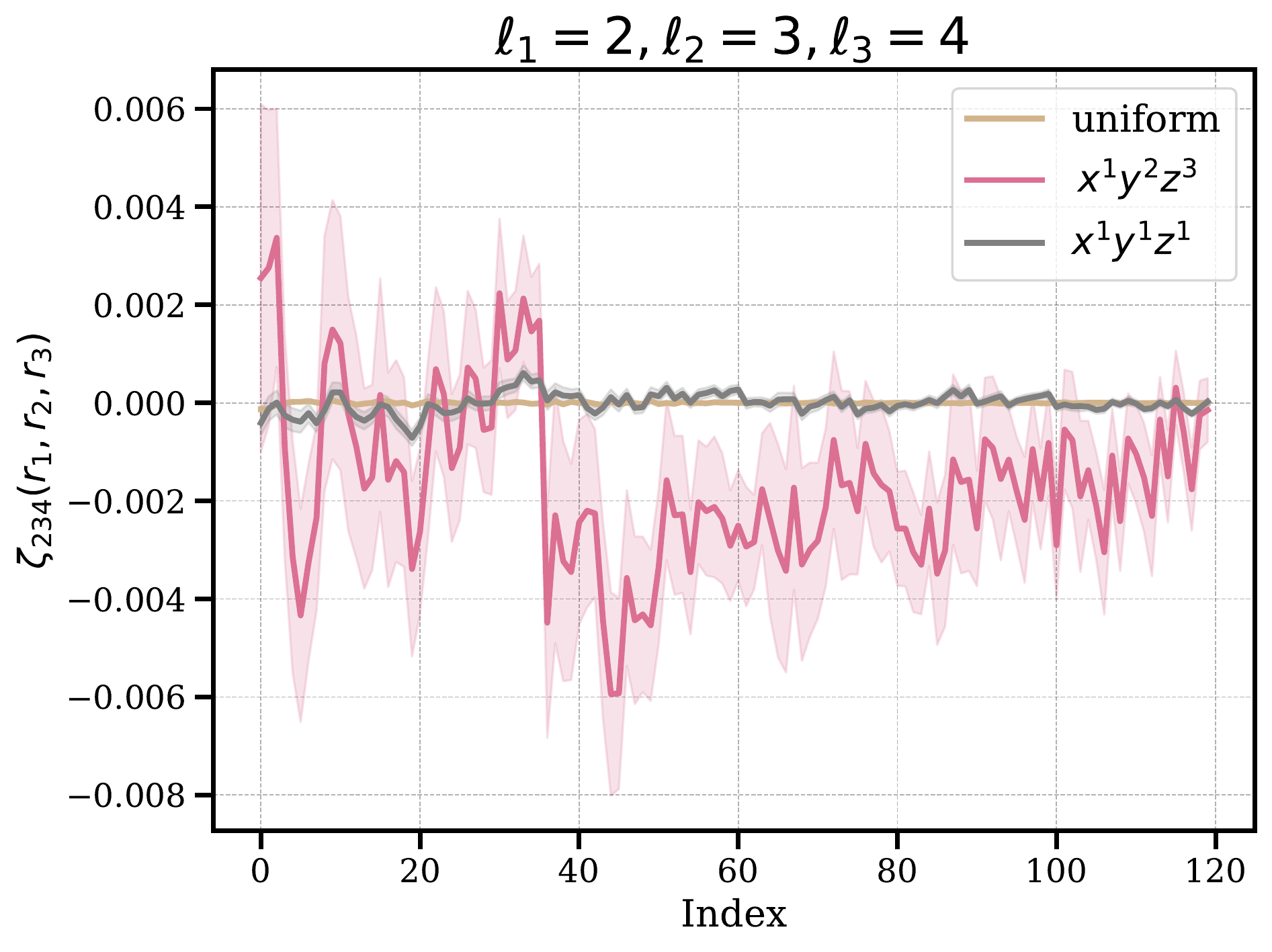}
         \label{fig:zeta_nonclus_99mocks_500mpch_box_ell234}
     \end{subfigure}
     \begin{subfigure}[b]{0.3\textwidth}
         \centering
         \includegraphics[width=1\textwidth]{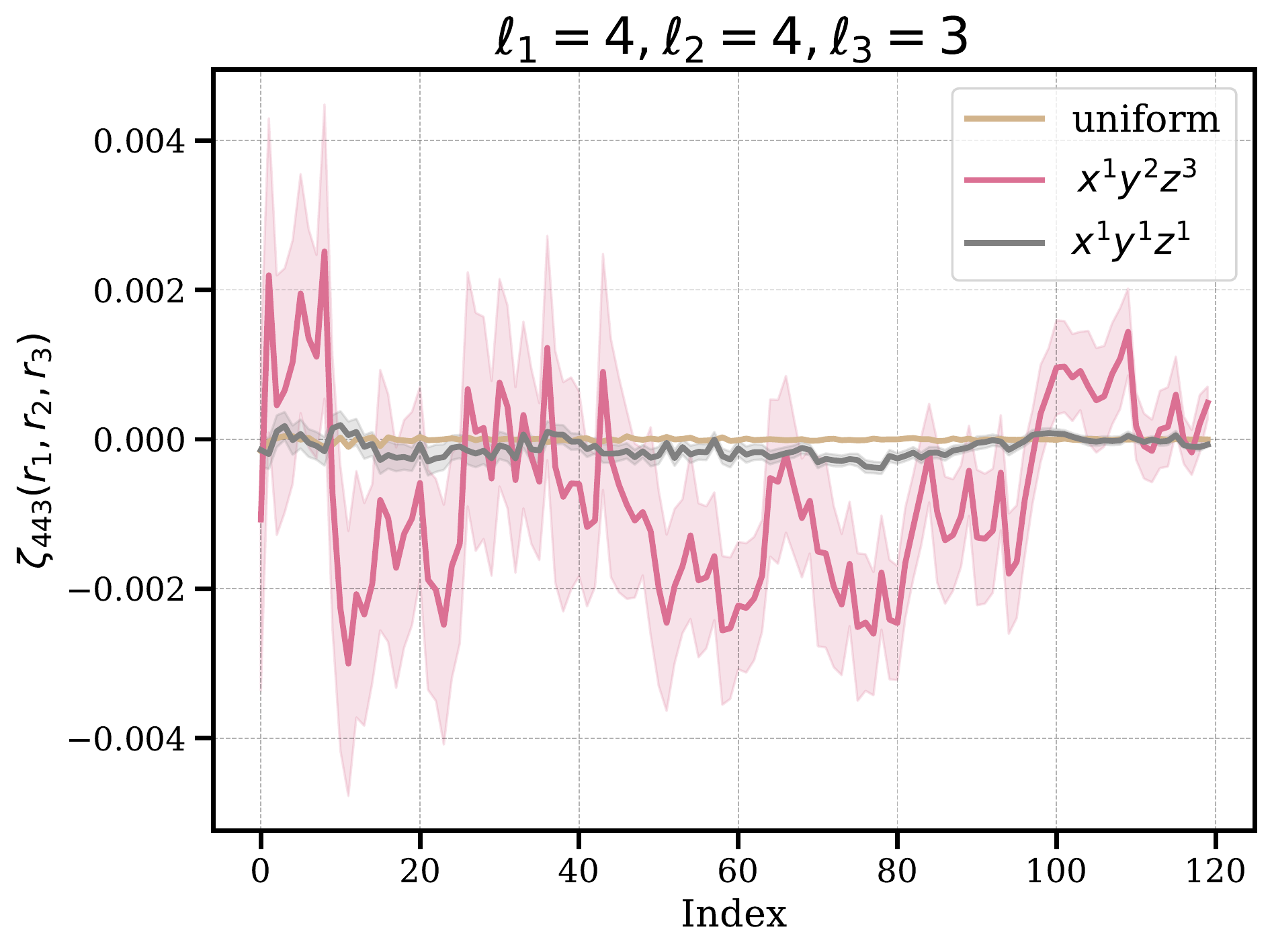}
         \label{fig:zeta_nonclus_99mocks_500mpch_box_ell443}
     \end{subfigure}
     \caption{Several parity-odd 4PCF coefficients as measured from our spatially-random power-law-modulated mocks (\S\ref{subsub:toy_model}). We show results from $100$ mocks  with number density $\bar{n} = 3\times 10^{-4}\, [\mpch]^{-3}$ for no modulation (tan), $x^1 y^2 z^3$ (red), and $x y z$ (grey). We find that the $\chi^2$ for the power-law-modulated mocks is smaller than that for the unmodulated ones, implying that this modulation does not produce a significant parity-odd 4PCF at the signal level.}
     \label{fig:zeta_nonclus_99mocks_500mpch_box}
\end{figure}

Fig.~\ref{fig:zeta_nonclus_99mocks_500mpch_box} shows the 4PCF coefficients measured from the no-clustering mocks. From left to right, we plot the channels  $\{\ell_1=1,\ell_2=1,\ell_2=1\}$, $\{\ell_1=2,\ell_2=3,\ell_2=4\}$, and $\{\ell_1=4,\ell_2=4,\ell_2=3\}$. In each panel, we compare a parity-odd 4PCF coefficient for an unmodulated sample and two differently-modulated samples.  As expected, the 4PCFs measured from the uniform sample have very small amplitudes. The $\chi^2$ for the uniform sample is comparable to those modulated samples. Overall, this means that the power-law modulation would not produce a spurious detection of parity-odd modes at the signal level.

\subsubsection{Density Distortions, Selection Function, and Fiber Collisions}
\label{subsub:density_distort}
In theory, one can expect parity-odd signals to be induced solely by 3D modulations of the density field because such modulation could convert a parity-even mode into a parity-odd mode. A 2D screen uncorrelated with the underlying density field can be separately averaged over rotations, and the rotation will force any parity-odd piece of the screen to vanish. An example of an uncorrelated screen is that given by the veto mask, which was designed to remove angular regions contaminated by foreground bright stars, plate holes, \textit{etc.} and is not correlated with the underlying 3D galaxy density.

Another 2D effect is ``fiber collisions'' (FC). BOSS uses fibers to guide the light of the observed objects from the focal plane to the spectrograph. Each fiber has a limited angular diameter ($62^{''}$) and multiple galaxies that fall within this radius cannot be resolved within one exposure. In the data, fiber collisions are corrected by upweighting the nearest angular neighbor ~\citep{ROSS2012BOSSsys,Reid2016boss}. This approach can potentially introduce long wavelength mode-coupling because it identifies neighbors only on an angular basis.  Although one can model the fiber-collision effect as a top-hat function~\citep{Hahn2017FiberCollision}, the exact physical scale could still be redshift-dependent (see~\citealt{Hou2021eBOSS} for quasars, which span a wide redshift range).

To demonstrate the impact of these angular effects on the parity-odd modes, we turn off the veto mask weights implemented in the \patchy mocks as well as the fiber collision weights. By switching on and off the fiber collision weight we can see the impact of the nearest-neighbor upweighting on the parity-odd detection significance. 
The left panel of Fig.~\ref{fig:chi2_patchy_ngc_weights} shows the distribution of the resulting $T^2$ values compared to that expected. When the covariance is inferred from the contaminated mocks themselves, there is good agreement between the contaminated mocks' $T^2$ distribution and that for the standard case. We, therefore, conclude that these effects do not induce parity-odd modes at the signal level. 

{\color{black} We also explored the effect of the fiber collision weights on the signal as found in the real CMASS data. We set these weights to be identically unity on the data, thus undoing the upweighting of the nearest angular neighbour of a galaxy lost to fiber collision. We also make this same change to the fiber collision weights on the mocks, and then remeasure their 4PCF and refit our analytic covariance matrix template using it.

We found that with this setup, the ten-bin parity-odd detection significance increased by $6\sigma$. We also computed the parity-even connected 4PCF, to assess if an unaccounted-for error of this type in fiber collision weights on the data would create tension between data and mocks. Indeed, we found that with all fiber-collision weights on the data set to unity (as if one had made a gross error in the data weights), but with standard weights on the mocks (and also in fitting the covariance template), there would be a $5\sigma$ disagreement in the parity-even sector between mocks and data. This argues that any serious error in the fiber-collision weights that impacts the parity-odd sector can be controlled by enforcing consistency in the parity-even sector.

We now briefly discuss what we regard as the most likely reason the parity-odd significance in the data increased in the test above. We know from the tests on the mocks alone that an error in the fiber-collision weights cannot induce a true parity-odd signal where there was initially none (left panel of Fig.~\ref{fig:chi2_patchy_ngc_weights}). Thus, the increased detection significance in the data is presumed to result from the change in the fiber collision weights' increasing the \textit{variance} in the data more than it does in the mocks used to estimate the covariance.

We suspect this disproportionate increase is due to the following. In the real survey, each observational tile is allowed to be overlapped with others to maximize the fraction of targets that can be assigned fibers. Thus, uncorrected fiber collisions likely have a lower impact on the data than on the mocks. In particular, even with the fiber collision weights set to unity, there is likely better sampling of the densest regions on the sky in the data than in the mocks. This would tend to increase the data's variance relative to that of the mocks in the situation where neither are corrected for fiber collision.}

\begin{figure}
    \centering
    \includegraphics[width=.5\textwidth]{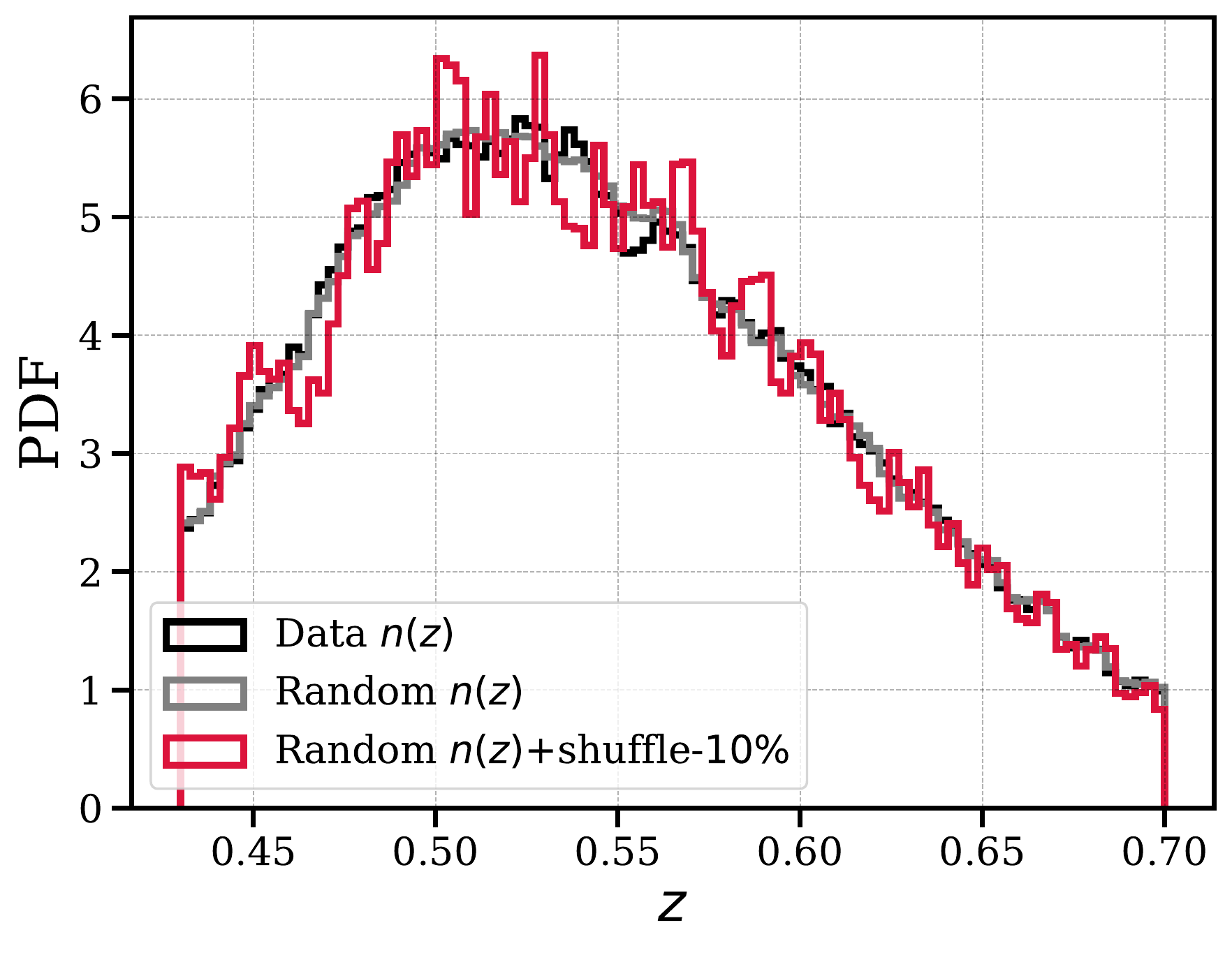}
    \caption{PDF of the radial selection function for \patchy NGC. The selection function applied to the \patchy mocks is designed to match that from the BOSS data for both the mock data catalog (black histogram) and the random catalog (grey histogram). To test whether an imperfect inference of the radial selection function can lead to a spurious parity-odd signal, we deliberately distorted the $n(z)$ for the random catalog (red histogram), by shifting it by 10\% on average, where the percentage shift is drawn from a Gaussian with that mean and width of 5\%. We show the distribution of $T^2$ values for \patchy mocks analyzed using these ``shuffled'' randoms in Fig. \ref{fig:chi2_patchy_ngc_weights}; it is consistent with no detection of a parity-odd 4PCF.}
    \label{fig:nz_shuffle10_patchy_ngc}
\end{figure}

\begin{figure}
\begin{subfigure}[b]{0.48\textwidth}
    \centering
    \includegraphics[width=0.95\textwidth]{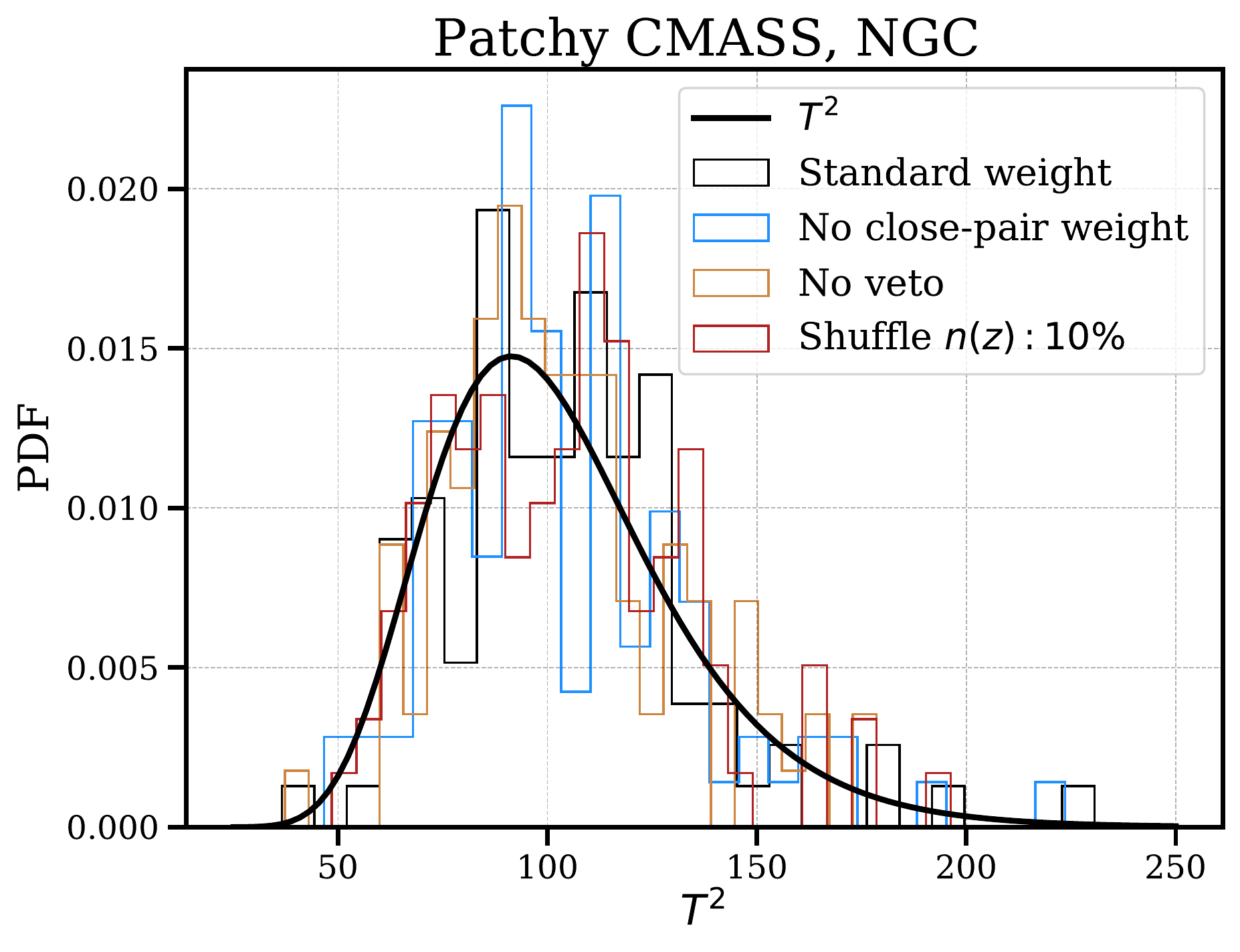}
\end{subfigure}\hfill
\begin{subfigure}[b]{0.48\textwidth}
    \centering
    \includegraphics[width=0.95\textwidth]{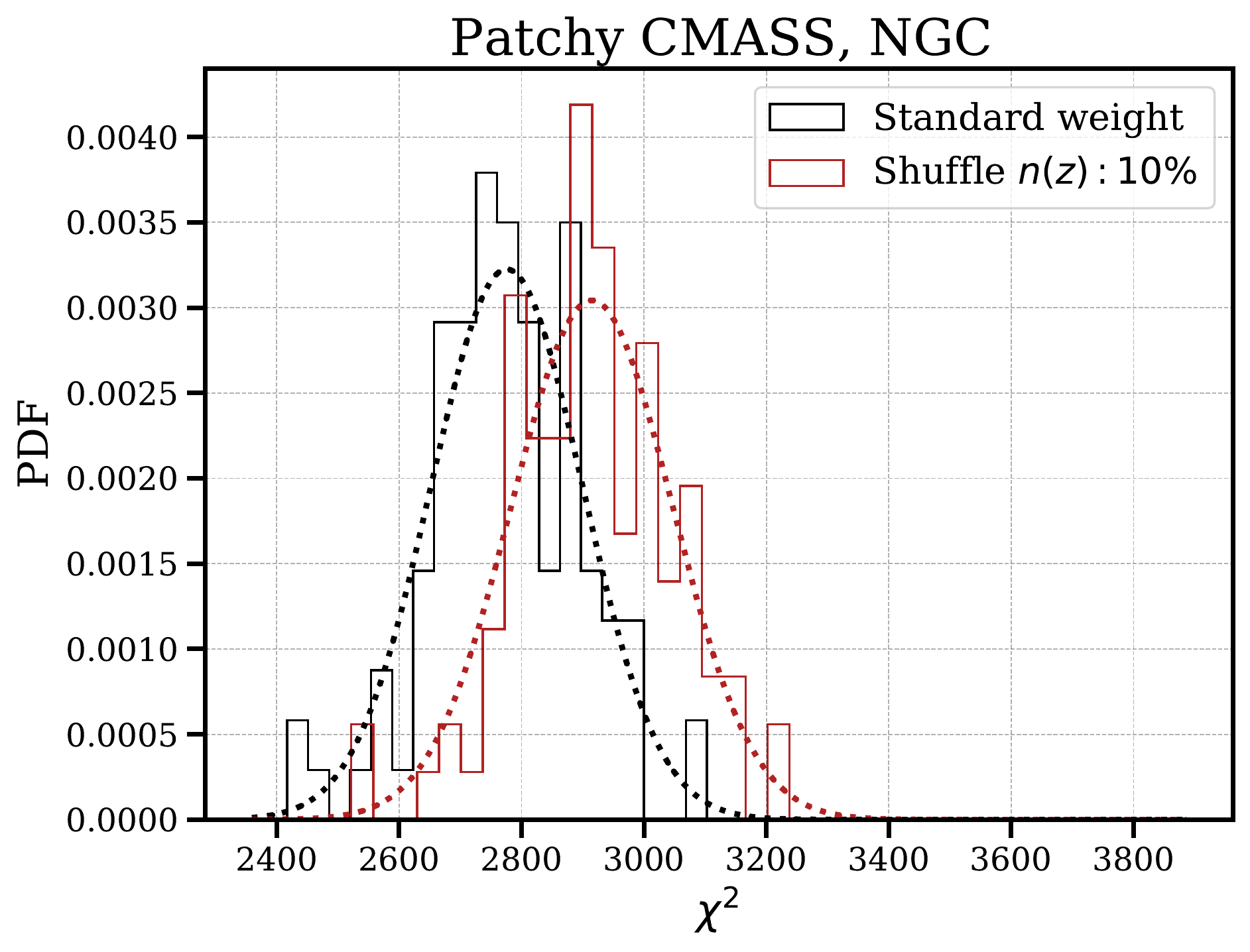}
\end{subfigure}
    \caption{Histograms of statistics for \patchy mocks' 4PCFs analyzed with different weighting schemes. {\it Left:} $T^2$ values calculated with 50 eigenvalues using our data compression scheme (\S\ref{subsub:comp}) with the covariance matrix inferred from the systematics-integrated mocks themselves. The black curve is the expected $T^2$ distribution. The histogram for the \patchy mocks with the standard weighting scheme is in black, with  the close pair weight turned off to assess the impact of fiber collisions in blue, with the veto mask turned off to assess the impact of the angular mask in orange, and finally, in red, the results of shuffling the randoms' redshift distribution by 10\% as described in Fig. \ref{fig:nz_shuffle10_patchy_ngc}. None of these alterations to the 4PCF analysis produce a distribution of $T^2$ values that are inconsistent with that expected under the null hypothesis of no-parity-odd 4PCF (black curve). {\it Right:} $\chi^2$ values calculated using the analytic covariance calibrated to mocks analyzed with the standard weights. Here we can see that the $\chi^2$ distribution has a higher mean for the shuffled mocks than for the standard mocks, suggesting that although such a systematic does not induce a parity-odd \textit{signal}, it might lead to underestimation of the covariance.}
    \label{fig:chi2_patchy_ngc_weights}
\end{figure}

In addition to the angular effects, we also distort the radial selection function by introducing a $10\%$ scatter in the $n(z)$ of the randoms. At each redshift in Fig.~\ref{fig:nz_shuffle10_patchy_ngc}, we draw a value from a Gaussian with mean $10\%$ and width $5\%$, and then decide randomly and with equal probability whether to give it a plus or a minus sign. We then use this value to set a new probability of a random particle at that redshift being selected. This procedure results in the red curve in Fig.~\ref{fig:nz_shuffle10_patchy_ngc} for the shuffled randoms' $n(z)$. The FKP weights for the randoms are also recomputed after this. 

The distribution of $T^2$ values for the \patchy mocks with shuffled $n(z)$  when we use the compression scheme with the covariance inferred from the contaminated mocks agrees well with that obtained using the standard weights (black histogram) and also with the predicted $T^2$ distribution (black curve) as shown in the left panel of Fig.~\ref{fig:chi2_patchy_ngc_weights}. However, the radially-distorted randoms lead to an increase in the covariance, which is interestingly in contrast to mocks with angular contamination. As a result, the mock distribution shifts towards higher $\chi^2$ when we use an analytic covariance matrix calibrated with respect to the standard mocks (right panel of Fig.~\ref{fig:chi2_patchy_ngc_weights}). In other words, had this systematic been present in our BOSS data, the covariance matrix inferred from the standard \patchy mocks would have been underestimated. In table~\ref{tab:syst_budget} we estimate how much this effect could have been affecting our detection significance for {\bf ten} and {\bf eighteen} bin tests, respectively.

\subsubsection{Magnification Bias}
We consider magnification bias qualitatively. If the underlying density field produces non-zero amplitudes only in even-parity 4PCF channels, it is sufficient to ask whether magnification bias will preferentially affect tetrahedrons of one handedness. Much like the method of image charges, we may consider a notional pair: a tetrahedron and its mirror image. As long as we can prove that for any such pair, there is no effect, no parity-breaking will be introduced. 

Consider for simplicity a tetrahedron for which the three galaxies defining a triangular base are at the same redshift, and the fourth point is at a higher redshift. If the three ``base'' points are closely enough concentrated, they may magnify the fourth galaxy behind them and on average make it more likely to be selected than otherwise. However, this will be the case also for the mirror image of this tetrahedron. Indeed, magnification will be invariant to the sign-flip of the two plane-of-sky coordinates; at most, it can be parity-odd in $z$, but any such contribution from the $z$ behavior alone will vanish after rotation-averaging. 

\subsubsection{Splitting the Sky into Angular Patches}
\label{subsub:split_sky}

To assess if our detection is coming from one part of the sky preferentially, which might indicate a systematic, we split the sky into eight angular patches, each of roughly $40 \times 35\, {\rm deg}^2$ (right ascension, RA, times declination, DEC), and run our analysis on each independently. Each angular patch  implies a subvolume (hence SV) as we extend it out in the redshift direction. 
The right panel shows some examples of the 4PCF measurement and compares them to the standard deviation of each subvolume. We estimate the covariance of each subvolume by splitting the \patchy mocks in the same way. Since the subvolumes receive modulations from Fourier modes larger than them, ``beat-coupling'' leads to additional corrections to the covariance matrix~\citep{li_ssc, dePutter2012box}. Thus a naive rescaling of the covariance by its volume relative to the total survey volume would likely give an underestimate and thence produce artificially high $T^2$ values.

Fig.~\ref{fig:chi2_boss_ngc_sv8} shows the $T^2$ computed from the entire BOSS CMASS sky (thick dashed vertical black line) and the eight subvolumes (thin colored vertical lines). The expected $T^2$ distribution is the black curve. We combine the eight subvolumes by taking an inverse-variance weighted sum of the $T^2$ values. Using the diagonals of the covariance matrix (see the right panel of Fig.~\ref{fig:chi2_boss_ngc_sv8}) we find the combined $T^2=128$, which differs from the one from the entire BOSS CMASS sky $T^2=138$. We repeat this using the full covariance of the subvolumes and find $T^2 = 129$, quite similar to the result using just the diagonal. We see that subvolume 2 and 3 have the highest detection significance compared to the average; given the errorbar of the $\chi^2$ shown in the right panel of Fig.~\ref{fig:chi2_boss_ngc_sv8}, this corresponds to a $1.8\sigma$ difference. We also note that combining the signal at the estimator level is not a simple linear operation due to the edge correction (see also Eq.~\ref{eqn:4pcf_estimator_edge_corrected}). Moreover, since we use the same analytic covariance for different subvolumes, the amount of information picked out by the ranked eigenvalues could differ for each subvolume.  We, therefore, expect that there is a difference between analyzing the data on the whole sky or splitting it and then combining it. 

We also repeat this same test on the SGC and on the LOWZ sample. Overall, we do not find any subvolume that has a particularly high or low detection significance compared to the average.

\begin{figure}
     \centering
\begin{subfigure}[b]{0.45\textwidth}
    \includegraphics[height=6cm, width=\textwidth]{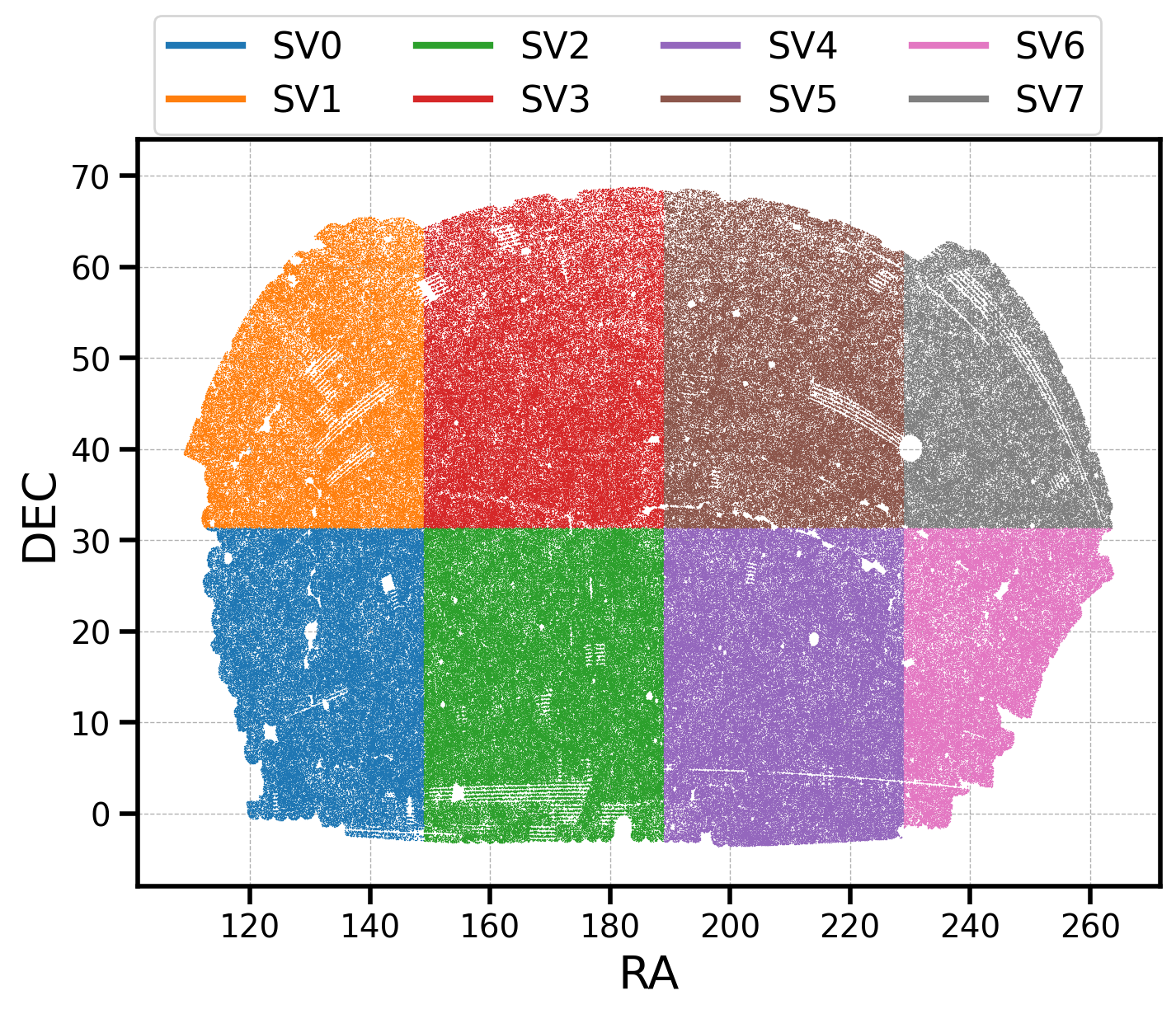}
\end{subfigure}
     \begin{subfigure}[b]{0.45\textwidth}
         \includegraphics[height=6cm, width=\textwidth]{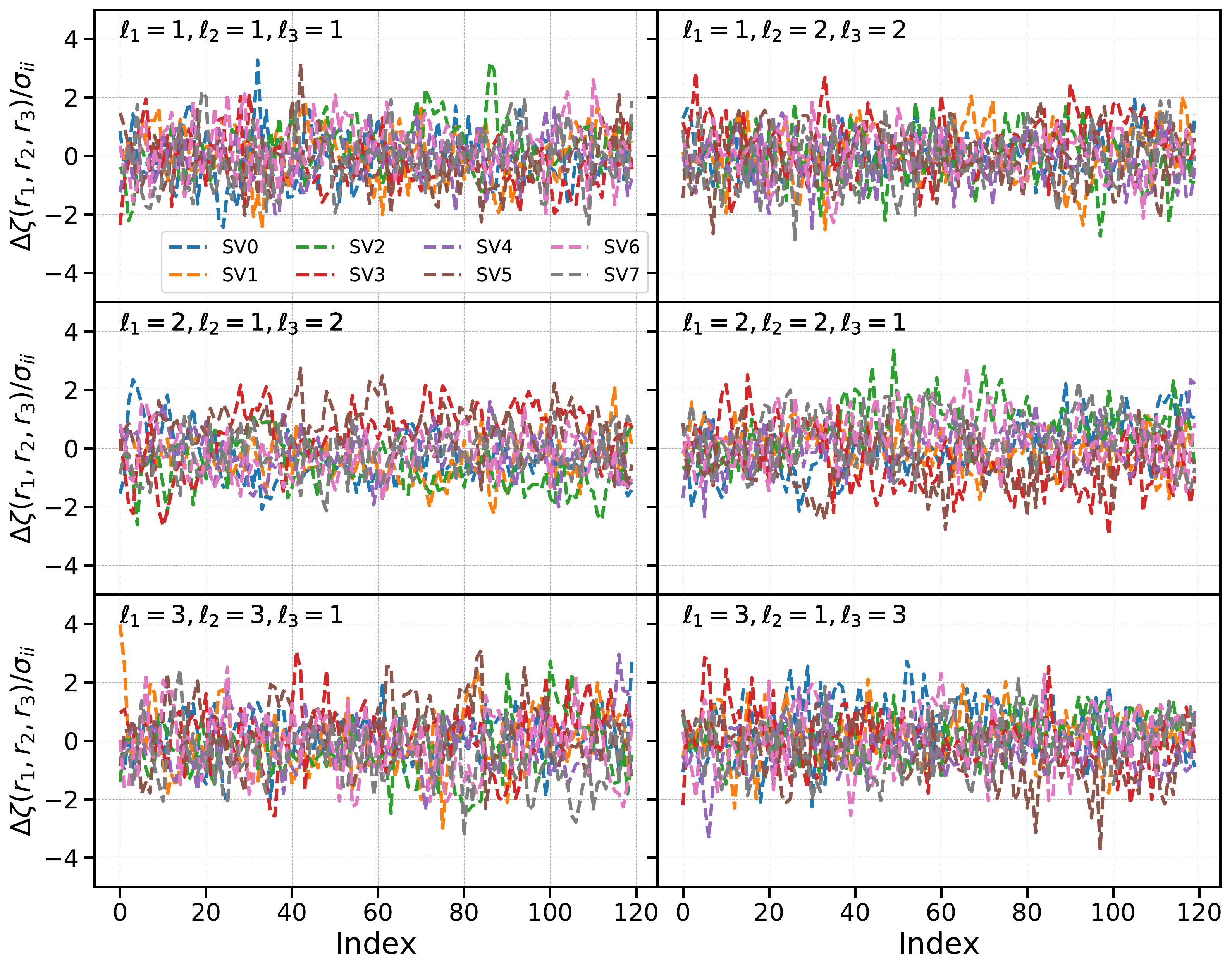}
    \end{subfigure}
    \caption{{\it Left:} Footprint of CMASS NGC showing how we split the data into eight patches, each roughly $40^{\circ}$ in right ascension and $35^{\circ}$ in declination. ``SV'' stands for sub-volume, as each patch extends into the redshift direction. {\it Right}: For six parity-odd channels, we display the difference between the 4PCF coefficients of the eight sub-volumes created by extending these angular patches in the redshift direction, and those from the full NGC sky. We divide these differences by the standard deviation of the \patchy results. The colors for the eight sub-volumes match in the left and right panels. Overall, we see that no one angular region stands out as producing a great deal more parity-odd signal than another. We offer a more rigorous demonstration of this claim in Fig. \ref{fig:chi2_boss_ngc_sv8}, which shows the $\chi^2$ for each sub-volume.}
    \label{fig:boss_ngc_sv8}
\end{figure}

\begin{figure}
    \centering
    \begin{subfigure}[b]{0.46\textwidth}
    \includegraphics[width=.9\textwidth]{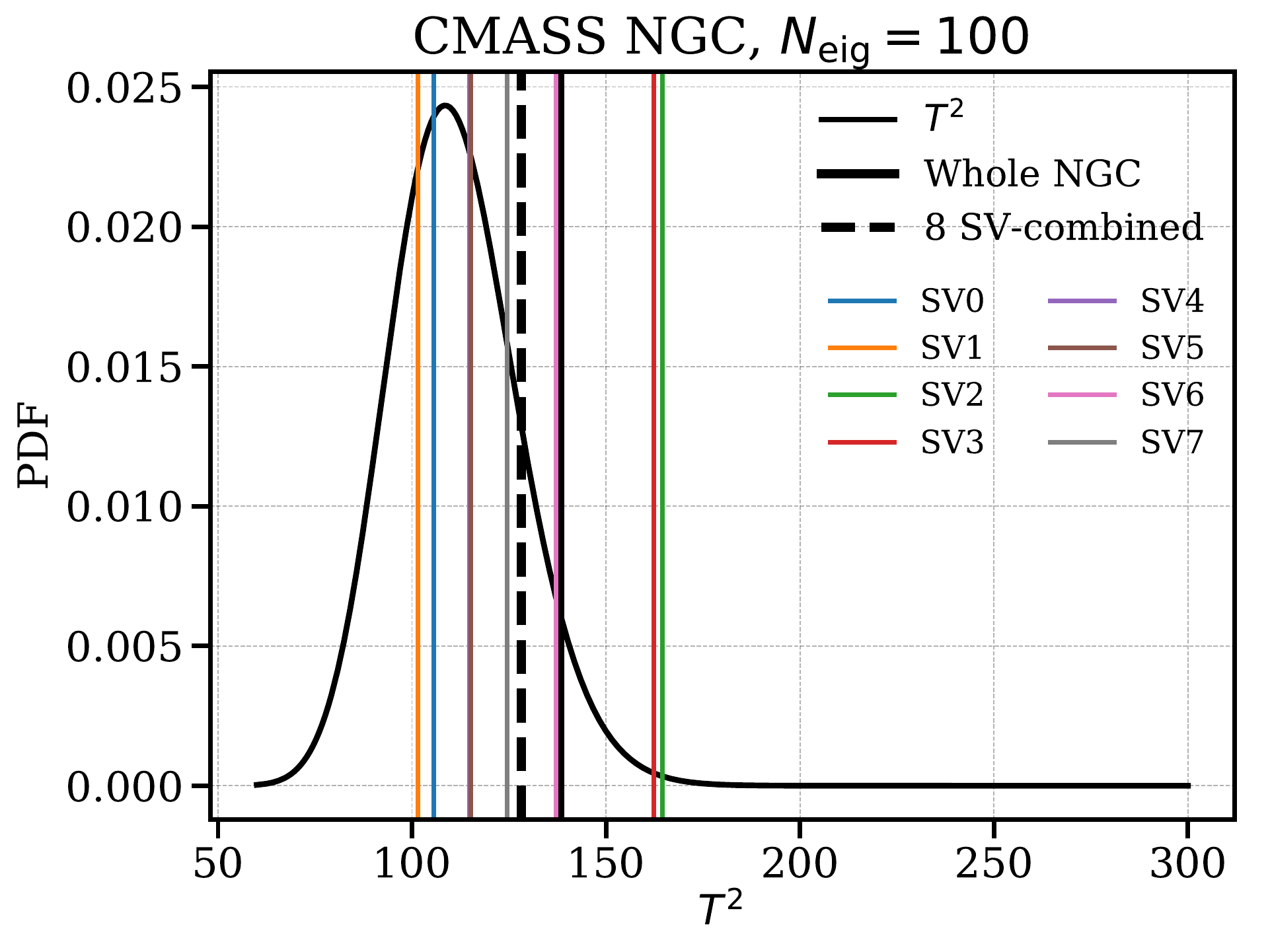}
    \end{subfigure}
    \begin{subfigure}[b]{0.44\textwidth}
    \includegraphics[width=.9\textwidth]{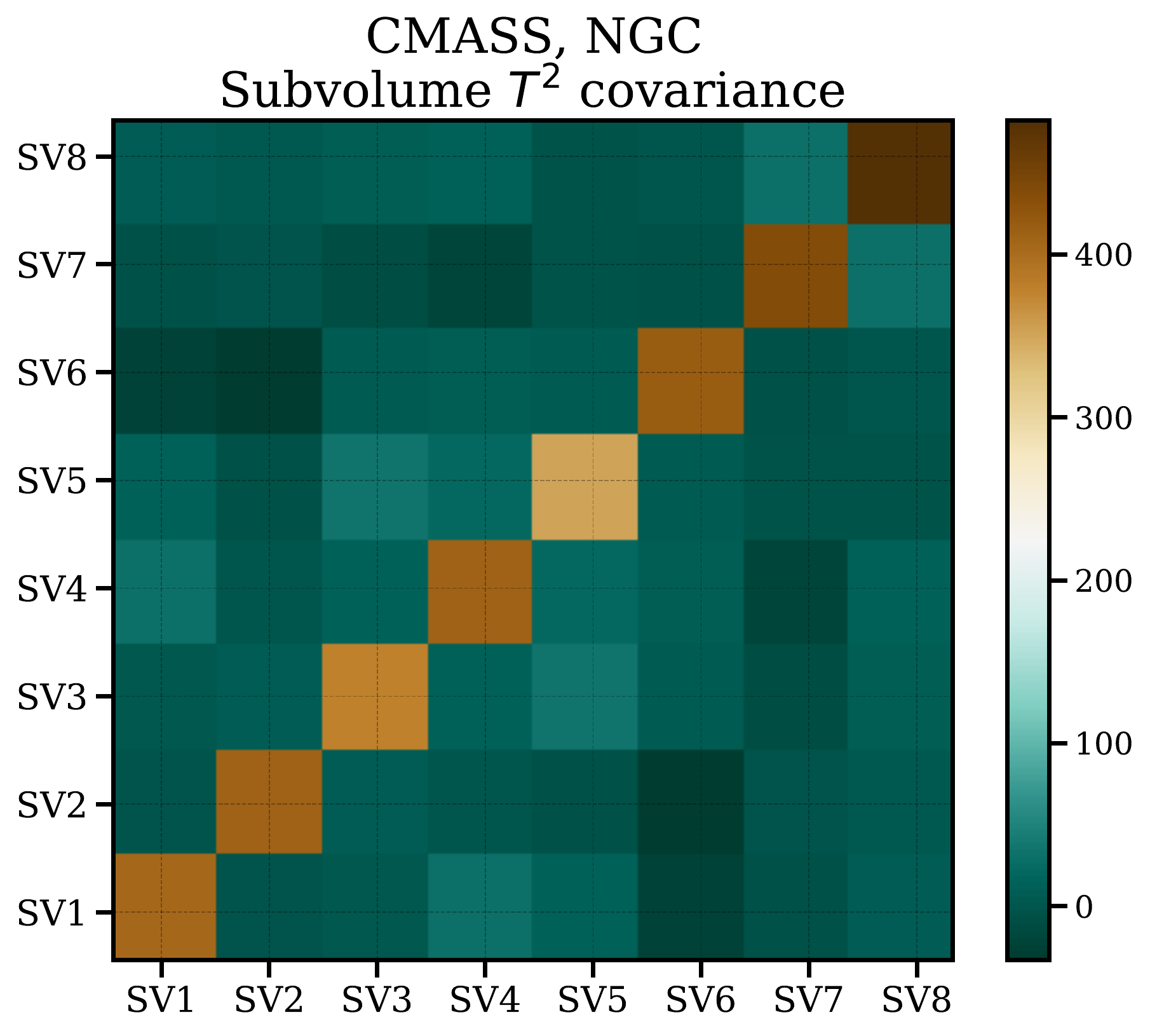}
    \end{subfigure}
    \caption{{\it Left:} Here we show the $T^2$ values from our angular patch test as described in Fig. \ref{fig:boss_ngc_sv8}. The  $T^2$ value for the entire CMASS NGC is the thick black line, the eight subvolumes' $T^2$ values are in thin, colored lines, and the combined $T^2$ from the eight subvolumes are in dashed black. The predicted $T^2$ distribution (under the null hypothesis of no parity-odd signal) with $100$ degrees of freedom is the black curve; we have used the compressed analysis described in \S\ref{subsub:comp}. {\it Right}: Covariance for the $T^2$ values of the eight sub-volumes, estimated using the \patchy mocks. This covariance is needed to properly co-add the $T^2$ values for the eight sub-volumes to report the combined $T^2$ given in the lefthand panel (dashed line there). The slight non-vanishing off-diagonal indicates correlations between the eight sub-volumes, but this turns out not to affect substantially the combined $T^2$ value. We may observe that the combined $T^2$ value differs from that for the whole NGC by about 10\%; we believe this is due to missing the produced by quartets that are split across more than one sub-volume, which go uncounted in our sub-volume analysis. We would expect that adding them in (as the whole NGC analysis does) would indeed raise the $\chi^2$.}
    \label{fig:chi2_boss_ngc_sv8}
\end{figure}

\subsubsection{Redshift Failure}
\label{subsubsec:zfail}

The BOSS fibers are not completely randomly positioned in an observing tile. The resolving power of the fibers is impacted by their positions on the CCD cameras. Those fibers near the edge of the spectrograph slit are less likely to obtain a high-quality spectrum than are more central fibers (see \S4.2.2 of \citealt{Smee2013BOSS}). Consequently, redshift failure occurs in a way that depends on 2D position on the telescope plate. This dependence is not wholly symmetric under a parity transformation of the $x$ and $y$ axes on the plate. 

Now, when a redshift failure occurs, the nearest angular neighbor of the failed galaxy is upweighted. The neighbor's redshift should be a fair draw from the redshift distribution of the survey. However, redshift failure is statistically slightly more likely for a galaxy at higher redshift, as it will on average be fainter, and hence more affected by the reduction in spectrum quality from being at the edge of a CCD. Thus, replacing it by upweighting a galaxy fairly drawn from the survey's redshift distribution can actually produce a \textit{bias}. Effectively, we will systematically be pulling galaxies towards lower redshift  as a function of 2D position on the plate. This 2D function, as already noted, is not fully symmetric under $(x, y) \to (-x, -y)$ and hence, in combination with the coupling to $z$ produced by the above, can in principle imprint a parity-odd signal on the density.

We now seek to explore this issue. CMASS has $1.8\%$ of its objects undergo redshift failures, while LOWZ has only $0.3\%$ do so~\citep{ROSS2012BOSSsys}. Thus we mainly focus on CMASS. Since the redshift failure galaxies are already removed from the catalog, we explore this effect by creating \textit{more} redshift failures and seeing whether it appreciably alters our parity-odd signal.

We first infer the redshift efficiency according to the inverse of redshift failure weight, then we apply a fit and assign each fiber a redshift efficiency rate accordingly. We apply a magnitude cut with ${{\rm mag}_{\rm cut}} = 19.20$ and from which a galaxy subsample is selected weighted by the fiber efficiency rate such that the subsample is $1.8\%$ of the total galaxies. We then identify the faint and bad fiber galaxies’ neighbors and select the ones that are closest in angular distance. Finally, we upweight those galaxies’ redshift failure weights by one and renormalize the random catalog’s weights such that it matches the total weights of the new data catalog. We repeat the above steps for both NGC and SGC.

Here we do not transfer the total systematic weights, as they cannot be simply propagated to the nearest neighbours. Assuming the total systematic weights to be unity for those galaxies has an impact on the total weights of only $~0.1\%$, so we do not expect an issue to arise due to this. We also retain the current estimation of the sector completeness, as this cannot be easily redone at the catalog level.

We found that doubling the redshift failures increased the total parity-odd detection on the joint sky by $2.5\sigma$ when using eighteen radial bins and $1.5\sigma$ when using ten radial bins, both with the analytic covariance; these results are summarized in Table \ref{tab:syst_budget}. This increase could be due to the 3D modulation of the density field producing a parity-odd mode as discussed above, but it could also simply be due to increased \textit{variance} in the data (\textit{i.e.} that any parity-odd \textit{signal} so produced would average to zero in the infinite limit). This increased variance could stem from our inability to correct the sector completeness to account for the removal of more objects. 

Given that the angular separation between galaxies and their nearest neighbours is typically much smaller than the sky coverage of a sector, and most of the upweighted galaxies and the removed galaxies are still within the same sector and this, we might expect that the impact of uncorrected sector completeness in our test setup is small. However, we need to quantify its impact. In order to do so, we applied a uniform selection in redshift and selected $1.8\%$ of those galaxies that are close to bad fibers. We proceed by upweighting their neighbours. There we observed $2.2\sigma$ and $0.5\sigma$ increase in detection significance for the joint sky using eighteen bins and ten bins, respectively. Hence the increase in the detection significance is actually dominated by the imperfect compensation between the random catalog and the real data produced by the uncorrected sector completeness in our test setup, which leads to a spurious variation in the density field. This is qualitatively similar to what we find with the power-law toy model in \S\ref{subsub:toy_model}.) Furthermore, redshift failures also impact the parity-even modes; this impact (for our ``doubling'' test) is given in Table \ref{tab:syst_budget}. The good agreement between this mode in the true BOSS data and in the distribution of the mocks~\citep{Philcox2021detection} demonstrates that redshift failure at the level it is actually occurring in BOSS should not have a strong impact on the detection significance.

\subsection{Observer-Related Effects}
\label{subsec:observe_sys}

\subsubsection{Redshift-Space Distortions (RSD)}
\label{subsub:rsd}
Galaxies' radial distances are inferred from their redshifts, however, their peculiar velocities caused by the local gravitational environment add a component in addition to their cosmological redshifts. This distortion in galaxies' positions is known as redshift-space distortions (RSD). 

In order to understand the impact of the RSD on parity-odd modes, we first consider the ``global'' line of sight approximation and take the $\hat{\bf z}$ as the line of sight. RSD modulates the density by the usual \citet{kaiser} factor, which is even-parity with respect to the line of sight. The primary galaxy itself is at the origin and so does not carry any angular momentum; multiplying it with three density fields, with each of them being parity-even, will be parity-even. Given that rotation averaging does not alter the parity, RSD considered in the ``global'' line of sight approximation cannot produce parity-odd modes. We can easily extend this logic to a ``local'' line of sight by taking the line of sight to a quartet to be that to the primary. The modulation in the overdensity field due to the RSD will be proportional to $(\hat{\bf n} \cdot \bfkhat)^2$, but again results only in parity-even combinations. Given that rotation averaging does not alter the parity, certainly, RSD considered in the ``local'' line of sight approximation cannot produce parity-odd modes.\footnote{\citet{beutler_wa} find parity-odd modes in the power spectrum multipoles (\textit{i.e.} odd $\ell$; with respect to the line of sight) but that is an artificial consequence of the way that the Yamamoto estimator, which selects one of the two galaxies to define the line of sight, is implemented in Fourier space. Even in configuration space, the Yamamoto estimator does not produce parity-odd modes with respect to the line of sight, as shown in \cite{SE_wa}}. We also note~\citet{Jeong2019BispectrumOdd} have considered parity-odd bispectrum induced by galaxies' radial velocity by introducing explicit coupling to the line of sight degree of freedom, by doing so the total rotational invariant system includes the additional line of sight vector, which by itself has no constraint on parity as long as the combined system has total zero angular momentum. In our work, instead, the line of sight dependencies are already averaged over and we are not sensitive to the parity-odd mode due to the RSD as in~\citet{Jeong2019BispectrumOdd}.\footnote{We have calculated that each term in their galaxy bias model that sources an odd-parity bispectrum will not lead to an odd-parity 4PCF after averaging with respect to the line of sight.}

\begin{figure}
     \centering
     \begin{subfigure}[b]{0.3\textwidth}
         \centering
         \includegraphics[width=1\textwidth]{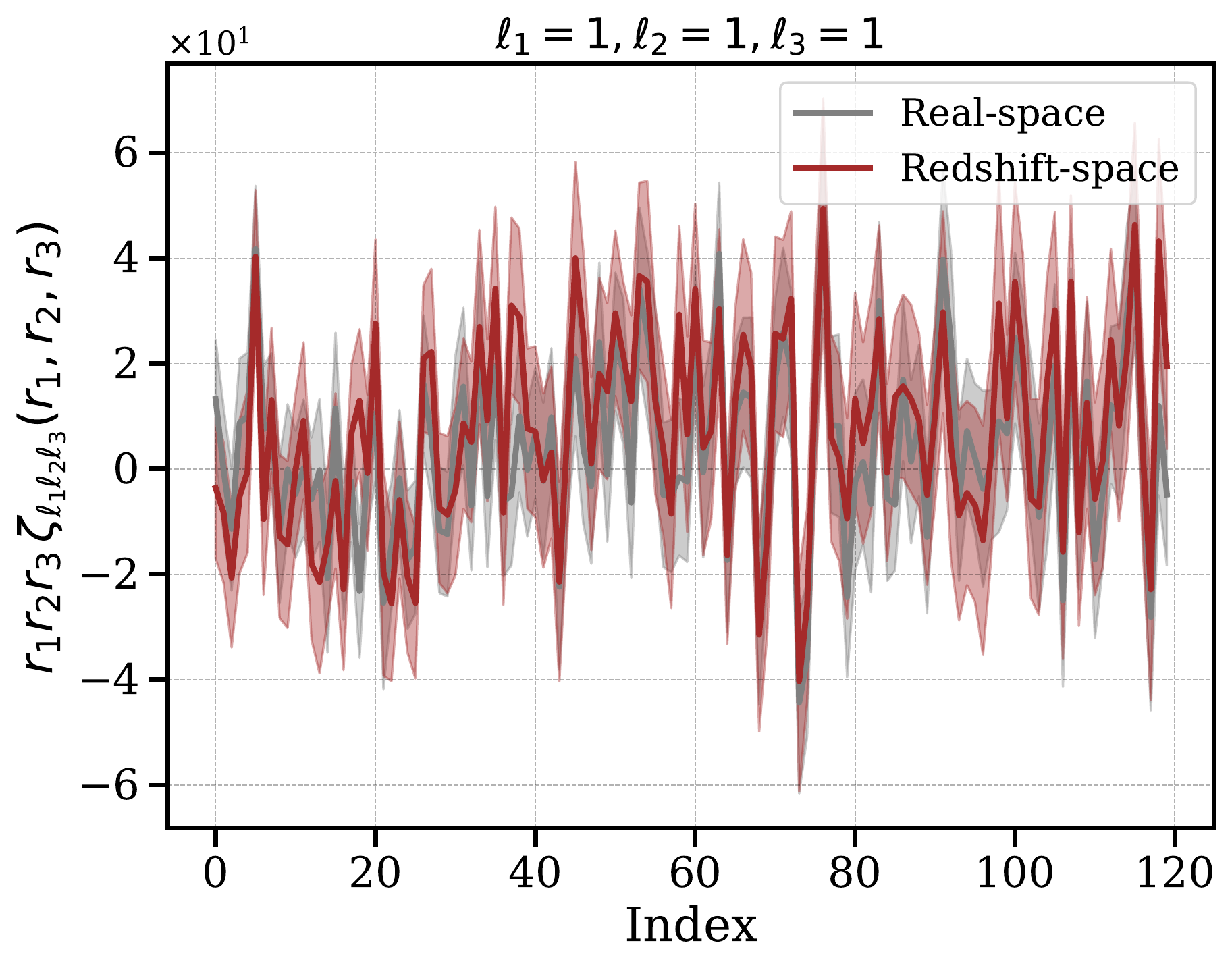}
         \label{fig:4PCF_measurement_odd_patchy_box4_rsd_ell111}
     \end{subfigure}
     \begin{subfigure}[b]{0.3\textwidth}
         \centering
         \includegraphics[width=1\textwidth]{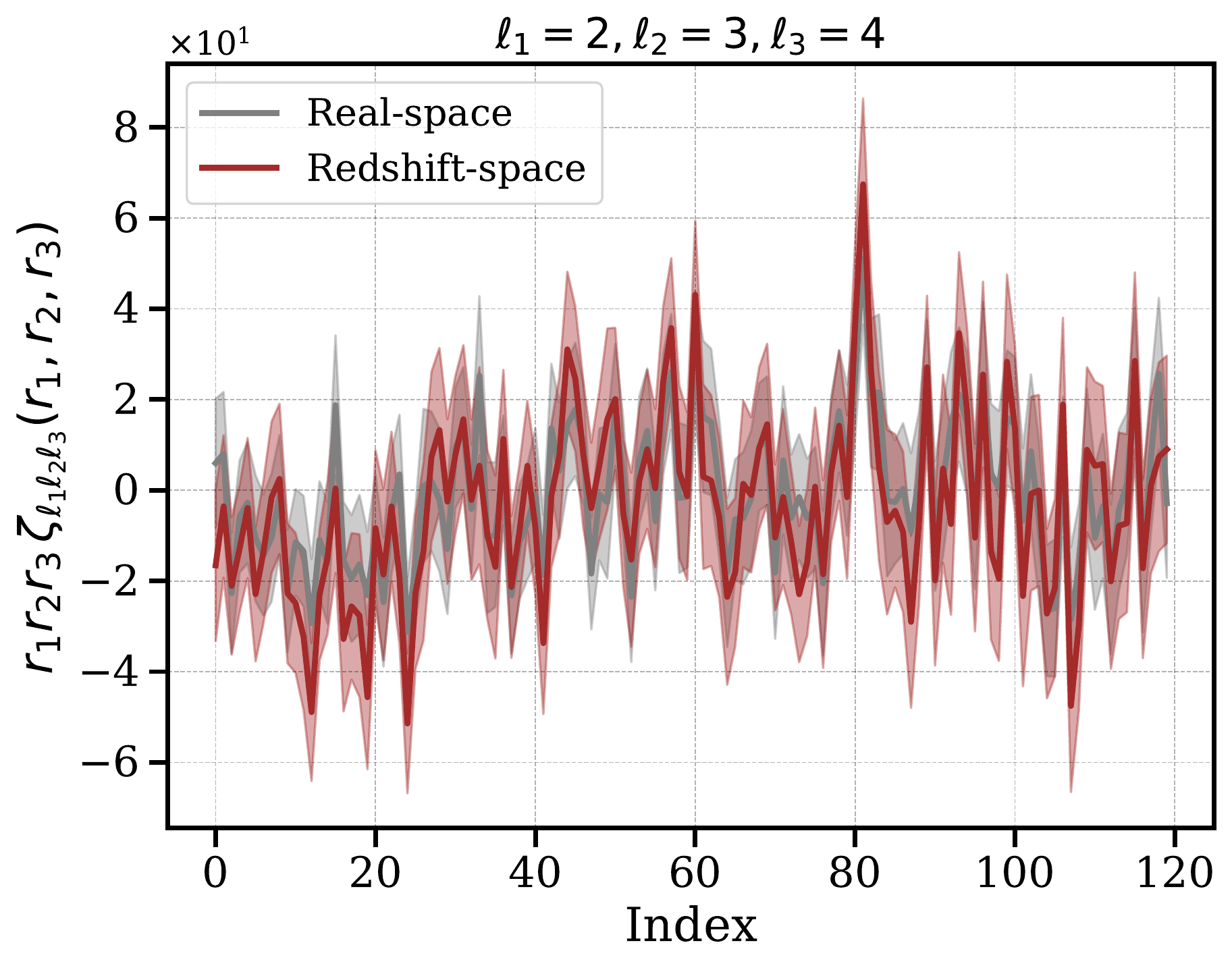}
         \label{fig:4PCF_measurement_odd_patchy_box4_rsd_ell234}
     \end{subfigure}
     \begin{subfigure}[b]{0.3\textwidth}
         \centering
         \includegraphics[width=1\textwidth]{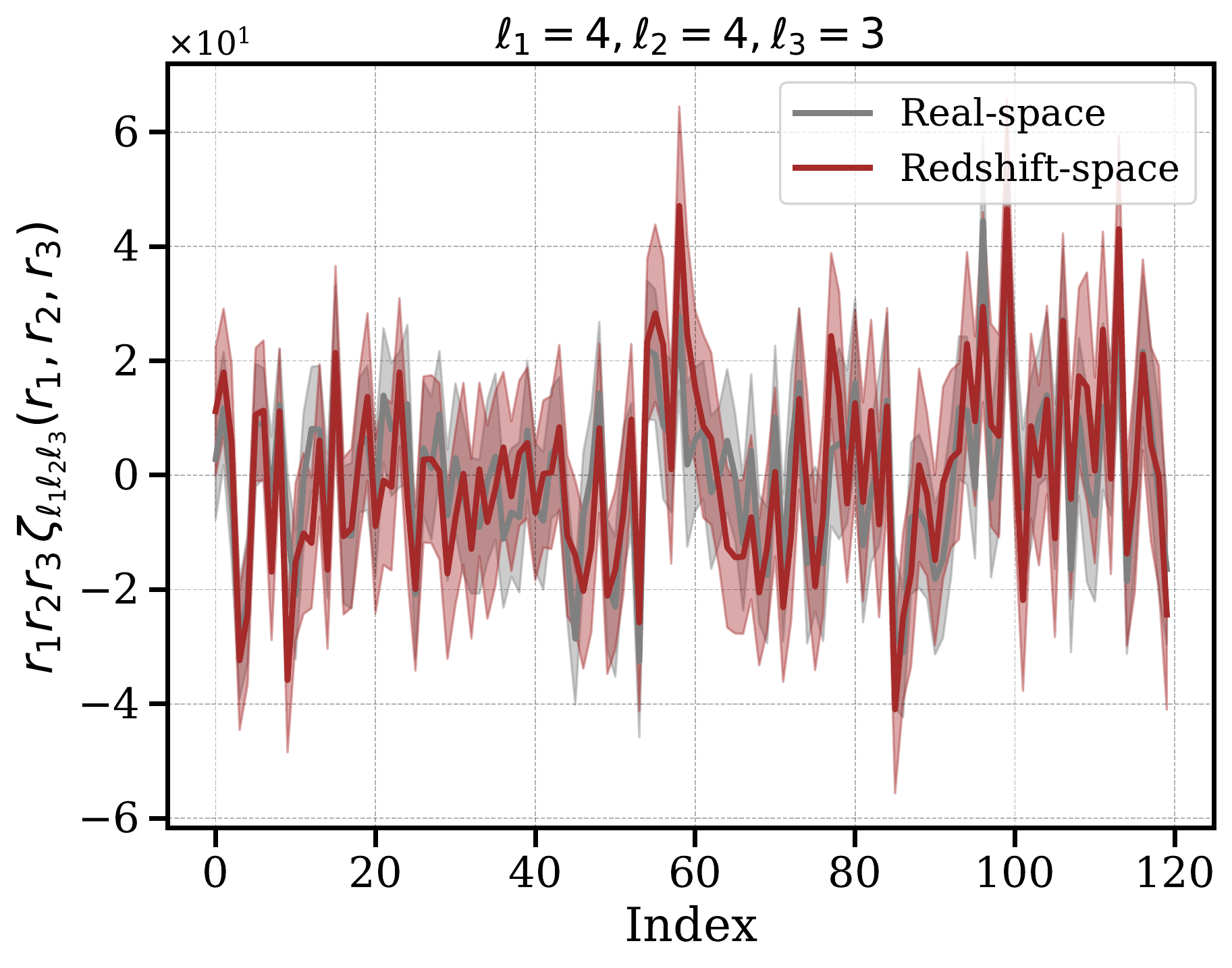}
         \label{fig:4PCF_measurement_odd_patchy_box4_rsd_ell443}
     \end{subfigure}
     \caption{parity-odd 4PCF channels $\{\ell_1=1, \ell_2=1,\ell_3=1\}$, $\{\ell_1=2, \ell_2=3,\ell_3=4\}$, and $\{\ell_1=4, \ell_2=4,\ell_3=3\}$. We compare the measured 4PCF coefficients from $99$ \patchy mocks in real space (grey) and in redshift space (red) with uniform geometry. The error bar is inferred from the standard deviation of the 99 mocks.}
     \label{fig:patchy_4pcf_rsd_real}
\end{figure}

To investigate the impact of RSD, we also use the \patchy mocks in real and redshift space.  For this test, the \patchy mocks used are in a cubic box of side length $L_{\rm box} = 2,500 \mpch$ and with number density $\bar{n} = 6.6\, [\mpch]^{-3}$ at redshift $z=0.57$. The galaxies are shifted using the exact 3D velocities of each galaxy in the mock.
A comparison of the 4PCF in real- and in redshift-space is shown in Fig.~\ref{fig:patchy_4pcf_rsd_real}. We do not observe an overall enhancement in the amplitude for the 4PCF.

\subsubsection{Tidal Alignment}
\label{subsub:tidal}
The large-scale tidal field impacts galaxies' orientations, which might lead to an anisotropic galaxy selection. For instance, in the case of spirals, an imaging survey might be more likely to detect a face-on rather than edge-on disk. Even LRGs are stretched ellipsoids, with aspect ratios of order 2:1, so orientation could impact the selection, though the effect would likely be smaller. For LOWZ, the sample is quite pure (LRGs), but for CMASS, the sample is roughly 26\% disky blue galaxies.  We might thus expect any anisotropic selection would manifest more strongly in CMASS. 

However, given that the observer's orientation is random with respect to the tidal field (and thus the orientation of the galaxies being selected), we suspect that such an effect could not produce parity-odd modes. If it did, we might expect a stronger detection of them in CMASS for the reason noted above. 

However, for completeness, we here work out what a tidal model might do if the selection depended upon it. Following~\citet{Hirata2009TidalAlignment}, the observed galaxy density fluctuation can be expressed as
\begin{eqnarray}
\delta_{\rm g}^{\rm obs} + 1 = \left(\delta_{\rm g}^{\rm true} + 1\right)\left(1+\epsilon(\bfnhat|\bfx)\right),
\end{eqnarray}
where $\epsilon$ is the selection function at position $\bfx$, and it depends on the viewing direction $\bfnhat$. The average of the selection function $\epsilon$ over the unit sphere is zero. By treating the large-scale tidal field as a perturbation, the lowest-order Fourier-space contribution (after Taylor-expanding) is (\citealt{Hirata2009TidalAlignment} equation (30)):
\begin{eqnarray}
\epsilon(\bfnhat|\bfk) = A \left[(\bfnhat\cdot {\bf \hat k})^2-\frac{1}{3} \right] \delta_{\rm m}(\bfk)
\end{eqnarray}
where $A$ is an amplitude and $\delta_{\rm m}$ is the matter overdensity.
We see that the selection function $\epsilon$ is parity-even, and so multiplying it onto a single density does not alter the parity of any spherical harmonic expansion coefficient associated with that density field. Hence, if there is on average signal in only parity-even channels in the underlying galaxy 4PCF, modulating each density point by $\epsilon(\bfnhat|\bfk)$ above will not convert them into a signal in parity-odd channels.

\subsection{Procedure-Related Effects}
\label{subsec:procedure}
\subsubsection{Fiducial Cosmology}
\label{subsub:fidcosmo}
In order to convert the sky coordinates to Cartesian coordinates, we need to assume a fiducial cosmology. Using a fiducial cosmology that differs from the true cosmology may lead to distortions in the clustering. 

At the signal level, if any galaxy quadruplets have their parity flipped by such an error, it is equally likely to happen for a given tetrahedron and for its mirror image. Therefore, on average, even if a ``parity-flip'' occurs, it will be canceled out after averaging over all galaxies in the survey. The radial bin separation $\Delta r$ we choose is $14\;\mpch$ when we use ten radial bins and $8\mpch$ when we use eighteen radial bins. Beginning with the cosmology of ~\citet{PlanckCollaboration2020} and moving $\pm 3\sigma$, the change in the angular diameter distance is less than $1\%$, which will not be resolved by our radial bin choice. At the covariance level, a different cosmology could lead to a different estimate of the best-fit number density and volume (for the analytic covariance) and would also change the power spectrum shape. Even though we use the same cosmology to convert the galaxies in the real data and the \patchy mocks, the statistical fluctuations do not necessarily scale the same way as we assume a fiducial cosmology that differs from the true one. 
We can see a $6\%$ change in the volume when distorting our fiducial cosmology by $3\times$ the constraints on $\Omega_{\rm m}$ obtained in \citet{PlanckCollaboration2020}. Such a difference corresponds to an overestimation of the detection significance by $\sim\! 2.5\sigma$ (see Table \ref{tab:syst_budget}). 
However, if there is an obvious difference between our choice of fiducial cosmology and the underlying one, it will also be visible in the comparison of the parity-even 4PCF measurement. The fact that we saw good agreement between the observed parity-even 4PCF and the mock distribution suggests wrong fiducial cosmology is not relevant here.

\subsubsection{Survey Geometry and Edge Correction}
\begin{figure}
     \centering
     \begin{subfigure}[b]{0.45\textwidth}
         \centering
         \includegraphics[width=1\textwidth]{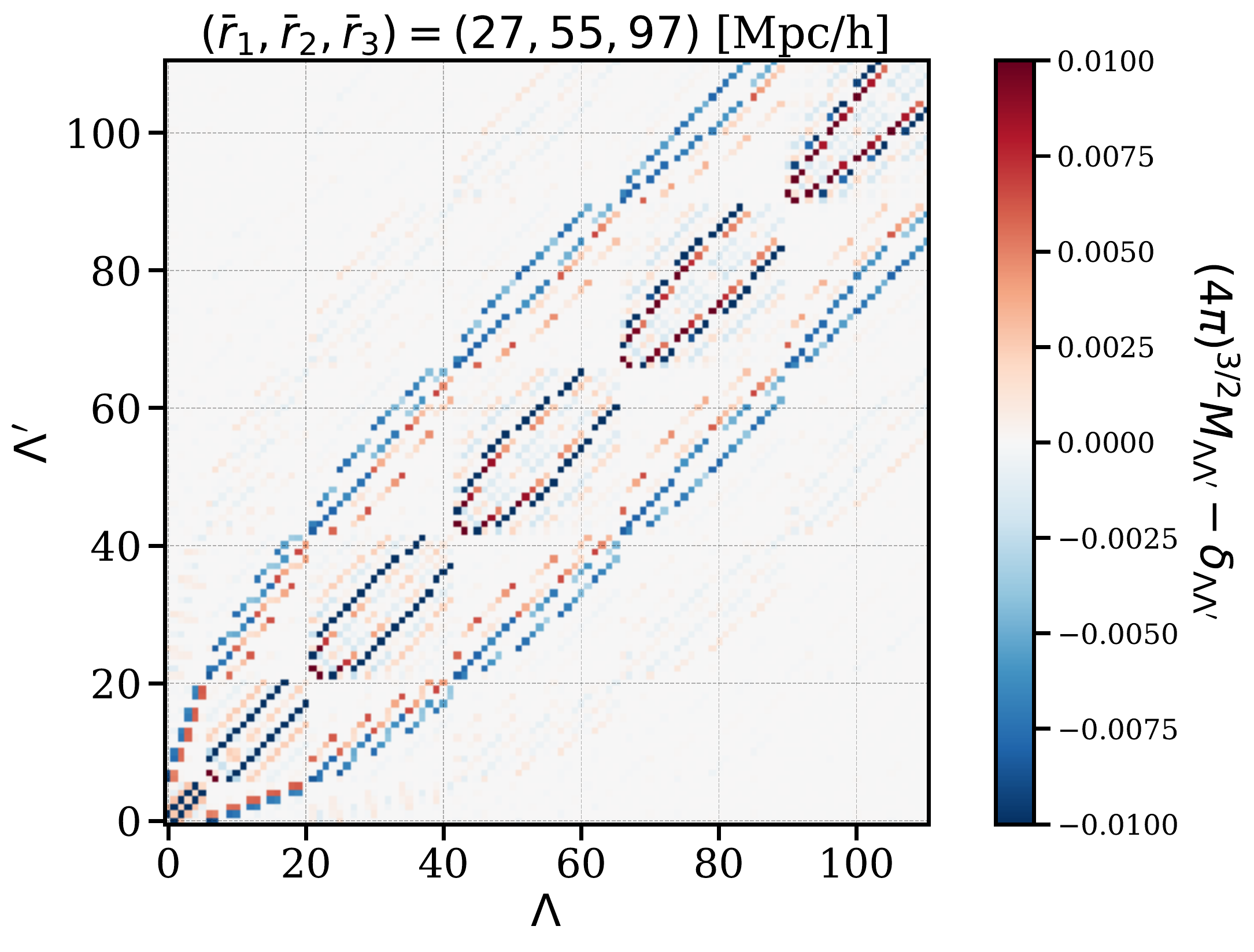}
         \label{fig:edge_correction_matrix_patchy}
     \end{subfigure}
     \begin{subfigure}[b]{0.43\textwidth}
         \centering
         \includegraphics[width=1\textwidth]{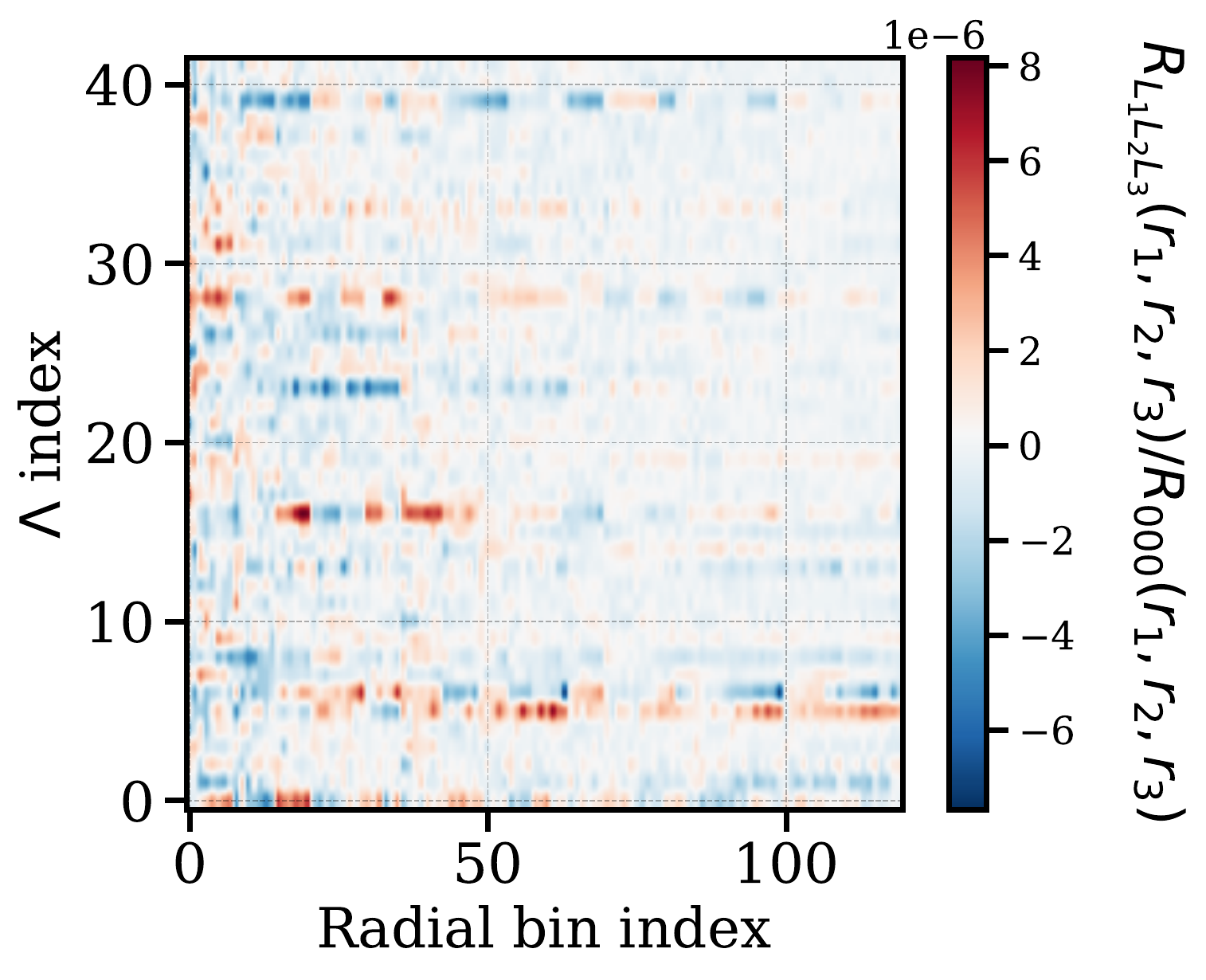}
         \label{fig:ratio_Rl123_R000_patchy_cmass}
     \end{subfigure}
     \caption{{\it Left:} Edge-correction matrix from \patchy mocks with CMASS NGC geometry at a given radial bin $(\bar{r}_1, \bar{r}_2, \bar{r}_3)=(27, 55, 97)\; [h^{-1}\;{\rm Mpc}]$. Each $\Lambda=\{\ell_1,\ell_2,\ell_3\}$ denotes a specific angular channel for both even and odd parity. The non-vanishing off-diagonal features indicate that the survey geometry couples even and odd modes. {\it Right}: The odd coefficients of the randoms normalized by $\mathcal{R}_{000}$. We have mapped the odd angular momentum triplets given by $\Lambda$ to a 1D index (vertical axis), and show the radial bin also so mapped on the horizontal axis. Here we see the dimensionless parity-even-to-parity-odd mixing is of order $10^{-6}$, showing that the dominant change to the odd-parity modes from edge-correction is due to $\mathcal{R}_{000}$, \textit{i.e.} overall renormalization from dividing by the mean number density.}
     \label{fig:edge_correction}
\end{figure}

As discussed in \S\ref{sec:basis}, the 4PCF measurement requires correction for the survey geometry as given in Eq.~\eqref{eqn:4pcf_estimator_edge_corrected}. In principle, there can be a mixing between parity-even and parity-odd channels due to the survey geometry. The left panel of Fig.~\ref{fig:edge_correction} shows the coupling matrix ${\rm \bf{M}}_{\ell_1\ell_2\ell_3, \ell'_1\ell'_2\ell'_3}$ as in Eq.~\eqref{eqn:4pcf_coupling_matrix} from \patchy mocks with CMASS NGC geometry at a radial bin combination $(r_1, r_2, r_3) = (27, 55, 97)\; [h^{-1}\;{\rm Mpc}]$. This coupling matrix has off-diagonal elements, which means that the parity-even mode can couple to the parity-odd mode. To assess the correction of the parity-odd modes due to the survey geometry, we plot the ratio of the random coefficients to the lowest-order one, $\calR_{\ell_1\ell_2\ell_3}/\calR_{000}$ (right panel of Fig. \ref{fig:edge_correction}). These ratios are of order $10^{-6}$; the dominant edge correction is thus simply $\mathcal{R}_{000}$, which does not mix parity. In addition, the fact that the \patchy mocks (which also get edge-corrected) are consistent with the predicted distribution indicates that survey geometry correction is not inducing spurious parity-odd modes. 

We also may develop a toy model for the edge-correction factors based on \citet{SE_3pt} \S4.3, where the impact of a planar edge on a uniform density sphere was considered. We consider a sphere of radius $R$ some distance away from a planar edge of the survey, and orient the $z$-axis perpendicular to that plane. We may define a critical angle cosine $\mu_{\rm c}$; for $\mu < \mu_{\rm c}$, a point will fall outside the survey. We may compute the $a_{\ell m}$ as a function of $\mu_{\rm c}$, then form the 4PCF estimate, and finally average over all possible $\mu_{\rm c}$, which corresponds to averaging over how far away the sphere's center is from the edge. The choice of $z$-axis means that only $m=0$ $a_{\ell m}$ are non-zero. The setup is described more extensively in \citet{SE_3pt} \S4.3.

Proceeding in this way, we may find a general form for $a_{\ell 0}$, which is given in \citet{SE_3pt}. Here, we will only compute the first few leading edge correction factors, so we find it more useful to simply quote the first few $a_{\ell 0}$ explicitly. We have $a_{00} = \sqrt{\pi}\;[\mu_{\rm c} + 1]$, $a_{10} = \sqrt{3\pi} \;[\mu_{\rm c}^2-1]/2$, and $a_{20} = \sqrt{5\pi}\; [\mu_{\rm c}^3 - \mu_{\rm c}]/2$. Cubing $a_{00}$, we find that $\mathcal{R}_{000} = \pi^{3/2} [15/4]$; the 3-$j$ symbol involved in $\mathcal{R}_{000}$ is unity. Similarly, we obtain $\mathcal{R}_{011} = -\pi^{3/2} \sqrt{3} [7/40]$, $\mathcal{R}_{112} = -\pi^{3/2} \sqrt{3/2}/32$, and $\mathcal{R}_{022} = -\pi^{3/2} \sqrt{5} (33/1120)$. These lead to, respectively, $f_{011} = 8\%$, $f_{112} = -1\%$, and $f_{022} = 2\%$, where these are the ratios of the $\mathcal{R}$ with the indicated angular momenta to $\mathcal{R}_{000}$. Furthermore, we note that only of order 20\% of spheres in the survey impinge on a boundary at all for the BOSS footprint; hence these factors would be expected to scale down by roughly a factor of five. These estimates show that the edge correction is not a substantial alteration to the measured 4PCF, and indeed that the dominant effect is that of $\mathcal{R}_{000}$, with the higher-order edge-correction factors contributing of order a few percent at most (after division by the factor of five).

\section{Conclusion}
\label{sec:conclusion}

\begin{table}
\centering
\caption{Here we summarize the shifts in detection significance, in units of $\sigma$, due to various systematics for both the parity-odd (top two rows) and parity-even (bottom two rows) modes, including redshift failure (\S\ref{subsubsec:zfail}), the impact of the fiducial cosmology (\S\ref{subsub:fidcosmo}), and a distorted selection function (\S\ref{subsub:density_distort}). We used the N+S of CMASS and the analytic covariance for these calculations.}
\label{tab:syst_budget}
\begin{tabular}{lcccc} \toprule
Parity & \begin{tabular}[c]{@{}c@{}}Number of\\ Bins\end{tabular} & Redshift Failure & \begin{tabular}[c]{@{}c@{}}Fiducial \\ Cosmology\end{tabular} & \begin{tabular}[c]{@{}c@{}}Distortion in\\ Selection Function\end{tabular} \\ \midrule
\multirow{2}{*}{Odd} & 10 & $1.5 $ & $\sim\! 1.5 $ & $\sim\! 1.0 $ \\
 & 18 & $2.5 $ & $\sim\! 2.5 $ & $\sim\!1.7 $ \\ \cmidrule{1-5}
\multirow{2}{*}{Even} & 10 & $0.9 $ & $\sim\! 0.7 $ & $\sim\! 0.6 $ \\
 & 18 & $1.8 $ & $\sim\! 1.9 $ & $\sim\! 2.4 $ \\ \bottomrule
\end{tabular}
\end{table}

In this work, we have used a novel, recently-suggested method \citep{Cahn2021parity} to search for parity-violation in 3D large-scale structure. This method exploits the fact that the lowest-order shape that cannot be rotated into its mirror image in 3D is a tetrahedron, and thus counting tetrahedra and looking for excesses of one type over another can reveal parity violation. We have detected strong evidence for it in both the CMASS and LOWZ samples of BOSS, and in both the NGC and SGC of each, summarized in Table \ref{tab:sigs}. We have explored numerous systematics to assess whether any could produce either a spurious parity-odd signal, or a substantial change to the covariance matrix used such that we would underestimate the statistical error bars so as to make a spurious detection. We have not found any systematics that seems to do either. We have also performed a number of tests of our covariance matrix and pursued a number of analysis variations both as regards covariance matrix and radial binning, scales, and angular momenta used. 

Overall, the compact takeaway from this work is that CMASS with eighteen radial bins has a $7\sigma$ detection of parity-odd modes when using the analytic covariance with all degrees of freedom, and a roughly $4\sigma$ detection when using an order of 500 eigenvalues plus a covariance estimated directly from the mocks. Given a large number of degrees of freedom in our analytic-covariance analysis, underestimating the noise by 20\% in every degree of freedom would be enough to produce this detection spuriously. In the ``compressed'' analysis with 500 eigenvalues, we would have had to underestimate the covariance by 30\% in every degree of freedom to spuriously produce our $4\sigma$ result. Such a substantial underestimate seems unlikely but given that the result, if cosmological in origin, would be significant, caution remains warranted.\footnote{We chose 500 eigenvalues for discussion here because that strikes a balance between detection significance yet has several mocks per element in the covariance. This has 1,000 degrees of freedom because we treat eigenvalues from North and South as separate degrees of freedom.}

We now briefly revisit the key technical aspects of our analysis and of systematics search.

\paragraph*{Radial Binning}\mbox{}\\
In this work, we tried three different radial binning schemes and found different detection significances. We hypothesize that the increasing detection significance is due to the reduction of internal cancellation. The finest binning we use in this work has $\Delta s=8\;\mpch$, which is roughly half of the average inter-particle separation for BOSS. In this work, our main goal is to explore a possible parity-odd signal; we leave the exploration of the optimal radial binning to future work. 
\paragraph*{CMASS vs. LOWZ Detections}\mbox{}\\
CMASS and LOWZ are both mainly composed of LRGs, but the two samples behave differently with regard to i) sensitivity of the detection significance to varying radial binning and  ii) sensitivity of the detection significance to varying $\ell_{\rm max}$. These different behaviours may be due to differences in the number density, mean redshift, galaxy population, and redshift failure rates of the two samples. 
\paragraph*{Correlation Across the Sky}\mbox{}\\
We do not observe a statistically significant positive cross-correlation between the NGC and SGC in each dataset, and we find that this test is sensitive to radial binning. We explored a simple phenomenological model to understand if high detection significance but low cross-correlation is generically possible, and it seems so. These two measures are algebraically independent so this finding was not outside the realm of expectation. 
\paragraph*{Systematics}\mbox{}\\
While we have shown that nearly all the systematics considered do not produce parity-odd modes at the \textit{signal} level (\textit{i.e.} they would contribute vanishingly in the infinite-volume limit), some can enhance the variance of the observed data relative to the covariance from the mocks if they are present in the data but not the mocks. If substantial enough, this in principle could produce a spurious detection. In practice, we have not found any systematic that, even if grossly larger than what we believe to be present in the data, could increase the variance enough to explain our results. We have also looked for consistency in the even-parity sector as a way to guard against any systematics' unbeknownst to us increasing the variance. 
\paragraph*{Implications}\mbox{}\\
Given the unexpected nature of the detected signal, it is premature to speculate on the implications. Following the advice of Newton: \textit{hypotheses non fingo.}\footnote{``I contrive no hypotheses,'' from Newton's ``General Scholium'' added to the \textit{Principia} of 1713.} Nonetheless, it is worth placing our results in the context of possible theoretical models that could explain them, should the signal here be of genuinely cosmological origin. Besides Chern-Simons-like interactions~\citep{Deser1982TopGauge,Deser1982Gauge3D,Witten1989JonesPoly} with couplings of different forms and various extensions~\citep{Freidel2005spinor,Alexander2006Leptogenesis,Liu2020CosmoCollider,Li2022teleparallel}, string-sourced perturbations~\citep{Pogosian2008CosmicString}, primordial vortices~\citep{Vilenkin1978RotatingBH,Vilenkin1982PVmagnetic,Brizard2000magnetic}, broken symmetry during phase transition~\citep{Hooft1974Higgs,Quashnock1989QCDPT,Baym1996EWPT}, anomalies in gravity~\citep{AlvarezGaume1983GravAnomaly,Contaldi2008CMBgravityAnomaly,Kuntz2019PVGravityAnomaly} can all produce parity-violation.
Among these mechanisms, the simplest relevant for us is where there is a non-vanishing parity-odd trispectrum (Fourier-space analog of the 4PCF) in the scalar sector~\citep{Shiraishi2014TrispectrumEven,Shiraishi2016PVtrispectrum}, as curvature perturbations directly seed the galaxy density perturbations. Extending such models to include the full wave-vector dependence as well as ``late-time'' effects such as galaxy biasing and redshift-space distortions will be an important step on the road to comparing models to data and determining if the signal here observed is consistent with any such models. Especially given that the detection significance we report here is a cumulative result from hundreds to thousands of degrees of freedom, a model comparison will provide much more constraining power as it will enable the use of shape information. 
\paragraph*{Outlook}\mbox{}\\
There are two main directions we suggest on the data side to further pursue the parity-odd sector. First, it is worth validating the measurement on different surveys with different designs. Upcoming surveys such as Dark Energy Spectroscopic Instrument~\citep[DESI;][]{DesiCollaboration2016}, Euclid ~\citep{Laureijs2011}, and Rubin~\citep{collaboration2009lsst}, which will provide much-larger samples and with different systematics,   will offer expanded insight into the signal's true origin. In particular, if our signal is of truly cosmological origin, it would likely be detectable in DESI at much higher statistical significance due to the larger volume. Thus, while in the present paper, making an error in our covariance matrix at the 20-30\% level would produce a spurious detection, in DESI, we expect one would have to make a much larger error in the covariance. Since this is unlikely, finding evidence for this signal in DESI would be a strong indicator that either the signal is real, or that it is due to some as yet-undiscovered parity-breaking systematic shared between BOSS and DESI. Yet, given that DESI has a different imaging survey, a different instrument, and a method for fibering objects in taking spectra, and is on a different telescope than BOSS, this latter possibility also seems unlikely.
Second, it is worth pursuing further approaches to search out systematics. For example, assuming a linear relationship between the systematics and the observed galaxy density field, one can study cross-correlations between them to reveal residual systematics~\citep{Ho2012LRGsys}. Alternatively, likelihood-based forward modelling can also be applied to search for unknown systematics~\citep{Lavaux2019BorgSys}. The first approach would require a non-trivial extension of our current NPCF algorithm in order to compute cross-correlations between fields, and the second requires injecting inferred systematics maps into mock catalogs. We leave this for future work.

Whether the detection of the current work ultimately turns out to be cosmological in origin or not, we hope that the careful and extensive treatment of systematics here offers a well-delineated avenue forward for future studies of parity-breaking early-Universe physics using 3D large-scale structure.

\section*{Acknowledgements}
We thank K. Dawson, H. Ding, D.P. Finkbeiner, A. Ginsburg, R. Guzm\'an,  D. Jeong, A. Krolewski, E. Lada, W. Percival, O. Philcox, S. Portillo, D. Richardson, A.J. Ross, U. Seljak, D.J. Schlegel, M. Slepian, D. Slepian, D. Spergel, C. Steinhardt, C. Telesco, and W. Xue for useful conversations. We especially thank D.J. Eisenstein, A. de Mattia, S. Saito, and A. G. S\'{a}nchez for extensive comments on the paper.
We also thank A. Chuang, F.S. Kitaura, and C. Zhao for providing access to the \patchy mocks. We thank E. Deumens of UF Research Computing for assistance using HiPerGator. Finally, we thank the members of the Slepian group for thoughtful discussions throughout the course of this work: N. Brown, J. Chellino, M. Hansen, F. Kamalinejad, K. Meigs, W. Ortol\`a Leonard, J. Sunseri, and C. Zhang. JH is in indebted to G. Hou for all their unconditional support in science. JH has received funding from the European Union’s Horizon 2020 research and innovation program under the Marie Sk\l{}odowska-Curie grant agreement No 101025187.

\section{Data Availability Statement}
The datasets underlying this article that are used to measure the 4-Point Correlation Functions and the covariance matrices are available via \url{https://data.sdss.org/sas/dr12/boss/lss/}.



\bibliographystyle{mnras}
\bibliography{ref.bib} 


\appendix

\section{Parity-odd Basis Functions in Cartesian Representation}
\label{app:fns}
Below we list parity-odd basis Functions in Cartesian representation for angular momentum up to $\ell_{\rm max} = 4$.
\begin{eqnarray}
\mathcal{P}_{111}(\hat{\mathbf{r}}_{1}, \hat{\mathbf{r}}_{2},\hat{\mathbf{r}}_{3})&=&-i\frac{3}{\sqrt{2}}(4 \pi)^{-3 / 2}\; \hat{\mathbf{r}}_{1} \cdot\left(\hat{\mathbf{r}}_{2} \times \hat{\mathbf{r}}_{3}\right),\nonumber\\
\mathcal{P}_{122}(\hat{\mathbf{r}}_{1}, \hat{\mathbf{r}}_{2},\hat{\mathbf{r}}_{3})&=&i\sqrt{\frac{45}{2}}(4 \pi)^{-3 / 2}\; \hat{\mathbf{r}}_{1} \cdot\left(\hat{\mathbf{r}}_{2} \times \hat{\mathbf{r}}_{3}\right)\left(\hat{\mathbf{r}}_{2} \cdot \hat{\mathbf{r}}_{3}\right),\nonumber\\
\mathcal{P}_{133}(\hat{\mathbf{r}}_{1}, \hat{\mathbf{r}}_{2},\hat{\mathbf{r}}_{3})&=&-i\frac{15}{4} \sqrt{7}(4 \pi)^{-3 / 2}\;\hat{\mathbf{r}}_{1} \cdot\left(\hat{\mathbf{r}}_{2} \times \hat{\mathbf{r}}_{3}\right)\left[\left(\hat{\mathbf{r}}_{2} \cdot \hat{\mathbf{r}}_{3}\right)^{2}-\frac{1}{5}\right],\nonumber\\
\mathcal{P}_{144}(\hat{\mathbf{r}}_{1}, \hat{\mathbf{r}}_{2},\hat{\mathbf{r}}_{3})&=&+i\frac {21\sqrt{15}}{4}(4\pi)^{-3/2}\, \bfrhat_1\cdot(\bfrhat_2\times\bfrhat_3)\left[(\bfrhat_2\cdot\bfrhat_3)^3-\frac 37(\bfrhat_2\cdot\bfrhat_3)\right],\nonumber\\
\mathcal{P}_{223}(\hat{\mathbf{r}}_{1}, \hat{\mathbf{r}}_{2},\hat{\mathbf{r}}_{3})&=&-i{15}\sqrt{\frac 58}(4\pi)^{-3/2}\;\bfrhat_1\cdot(\bfrhat_2\times\bfrhat_3)\left[(\bfrhat_1\cdot\bfrhat_3)\,(\bfrhat_2\cdot\bfrhat_3) -\frac 15\bfrhat_1\cdot\bfrhat_2\right],\nonumber\\
\mathcal{P}_{234}(\hat{\mathbf{r}}_{1}, \hat{\mathbf{r}}_{2},\hat{\mathbf{r}}_{3})&=&+i\frac 4{7\cdot 15}(4\pi)^{-3/2}\;\triple\left[(\dprod 13)(\dprod 23)^2-\frac 17\dprod 13-\frac 27(\dprod12)\,(\dprod 23)\right],\nonumber\\
\mathcal{P}_{333}(\hat{\mathbf{r}}_{1}, \hat{\mathbf{r}}_{2},\hat{\mathbf{r}}_{3})&=&i\frac{5\cdot 105}{6\sqrt 6}(4\pi)^{-3/2}\;\triple\bigg[(\dprod 12)(\dprod 13)(\dprod 23)\nonumber \\ &&\shoveright\qquad\qquad-\frac 15 \left[(\dprod 12)^2+(\dprod 13)^2+(\dprod 23)^2\right]+\frac 2{25}\bigg],\nonumber\\
\mathcal{P}_{344}(\bfrhat_1,\bfrhat_2,\bfrhat_3)&=&-i\frac{15\cdot49}4\sqrt{\frac5{11}}(4\pi)^{-3/2}
\triple\bigg[(\dprod 12)(\dprod13)(\dprod23)^2\nonumber\\ &&\;\;-\frac 15(\dprod23)^3+\frac{31}{245}(\dprod23) -\frac3{49}(\dprod12)(\dprod13)
 -\frac 27(\dprod13)^2(\dprod23)-\frac 27(\dprod12)^2(\dprod23)\bigg].
\end{eqnarray}
The eight functions above are the fundamental set and the remaining fifteen are generated by interchanging the arguments.

\section{Discussion of Pearson correlation Coefficient between two datasets}
\label{app:pears}
Consider two datasets $\vd_1 = \vs_1 + \vn_1 - \av{\vs_1}_{\rm all}\equiv \Delta \vs_1 + \vn_1$ and $\vd_2 = \vs_2 + \vn_2-\av{\vs_2}_{\rm all}\equiv \Delta \vs_2 + \vn_2$, with $\vs$ the signal-to-noise ratio and $\vn$ the noise. These are data vectors---in our case, over all channels and radial bins. $\av{\vs_j}_{\rm all}$ denotes an average over all channels and radial bins.

Given two datasets, each with a certain intrinsic statistical noise drawn from Gaussian distribution, what is the maximum allowed correlation between them? In the decorrelated basis, we can treat each data point independently. 
We recall from Eq.~\eqref{eqn:corr_rs} the definition of the Pearson correlation coefficient and insert in it the data vectors as defined above, finding
\begin{eqnarray}
{r_{\rm p}} &=& {\frac{\sum_{i=1}^{N_{\rm bin}} \left(d_1^{(i)}-\av{\vd_1}_{\rm all}\right)\left(d_2^{(i)}-\av{\vd_2}_{\rm all}\right)}{\sqrt{\sum_{i=1}^{N_{\rm bin}} \left(d_1^{(i)}-\av{\vd_1}_{\rm all}\right)^2 \sum_{i=1}^{N_{\rm bin}} \left(d_2^{(i)}-\av{\vd_2}_{\rm all}\right)^2 }}}.
\end{eqnarray}
Assuming a linear relation between the signals from the two samples, \textit{i.e} $\vs_2 = b \vs_1$, that the average observed signal is given by $\av{\vd_j}_{\rm all} = N_{\rm bin}^{-1} \sum_{i=1}^{N_{\rm bin}} s_j^{(i)}$ for $j=1, 2$, and that the noise vanishes after averaging, we have for the numerator
\begin{eqnarray}
&&\sum_{i=1}^{N_{\rm bin}}\left(d_1^{(i)}-\av{\vd_1}_{\rm all}\right)\left(d_2^{(i)}-\av{\vd_2}_{\rm all}\right)
= \av{\vs_1 \vs_2}_{\rm all} + \av{\vs_1 \vn_2}_{\rm all} + \av{\vs_2 \vn_1}_{\rm all} + \av{\vn_1 \vn_2}_{\rm all} - \av{\vs_1}_{\rm all}\av{\vs_2}_{\rm all}.
\end{eqnarray}
For the denominator, we use that
\begin{eqnarray}
\sum_{i=1}^{N_{\rm bin}}\left(d_j^{(i)}-\av{\vd_j}_{\rm all}\right)^2
&=&  \av{\vs_j^2}_{\rm all} - \av{\vs_j}_{\rm all}^2 + \av{\vn_j^2}_{\rm all} + 2\av{\vs_j \vn_j}_{\rm all}\nonumber\\
&\equiv& \sigma^2_{\vs_j} + \sigma^2_{\vn_j} + 2\av{\vs_j \vn_j}_{\rm all}.
\end{eqnarray}
The correlation coefficient is then
\begin{eqnarray}
{r_{\rm p}} = {\frac{\av{\vs_1 \vs_2}_{\rm all} -\av{\vs_1}_{\rm all}\av{\vs_2}_{\rm all} + \av{\vs_1 \vn_2}_{\rm all} + \av{\vn_1 \vs_2}_{\rm all} + \av{\vn_1 \vn_2}_{\rm all}}{\sqrt{(\sigma^2_{\vs_1} +2\av{\vs_1 \vn_1}_{\rm all} +\sigma^2_{\vn_1}) (\sigma^2_{\vs_2} +2\av{\vs_2 \vn_2}_{\rm all} +\sigma^2_{\vn_2})}}}.
\end{eqnarray}
First we discuss the low signal-to-noise limit for $\vs_1\ll \vn_1$ and $\vs_2\ll \vn_2$. We assume the correlation between the signal and noise is zero, and that the signal covariance is much smaller than that of the noise, \textit{i.e.} $\av{\vs_1 \vs_2}_{\rm all} \ll \av{\vn_1 \vn_2}_{\rm all}$. We find in this limit that
\begin{eqnarray}
{r_{\rm p}} = {\frac{\av{\vn_1 \vn_2}_{\rm all}}{\sigma_{\vn_1} \sigma_{\vn_2}}} = 0.
\end{eqnarray}
where to obtain the last equality we assumed that the noise is independent in each bin (as is assumed for a Pearson analysis, and as is the case once we have rotated to the decorrelated basis as discussed in the main text, \S\ref{subsec:cross}).

Second, in the high signal-to-noise limit that $\vn_1\ll \vs_1$ and $\vn_2\ll \vs_2$, and now with $\av{\vn_1 \vn_2}_{\rm all} \ll \av{\vs_1 \vs_2}_{\rm all}$, we find
\begin{eqnarray}
{r_{\rm p}} = {\frac{\av{\vs_1 \vs_2}_{\rm all}-\av{\vs_1}_{\rm all}\av{\vs_2}_{\rm all}}{\sigma_{\vs_1} \sigma_{\vs_2}}}.
\end{eqnarray}

We note that $\sigma$ here is the rms of the relevant vector \textit{over radial bins and channels}, not any kind of statistical error. It is simply a measure of how much that vector varies as the angular momenta and the radial bins are changed. This equation thus shows that, even if one had a signal of very high amplitude (and hence high S/N and high detection significance), if one had little variation over channels and radial bins, one would obtain a low $r_{\rm p}$. Put simply, the detection significance and $r_{\rm p}$ are independent.

\section{Comparison of CMASS and LOWZ Analytic Covariances}
\label{app:covar_compare}
\begin{figure}
    \centering
    \includegraphics[width=\textwidth]{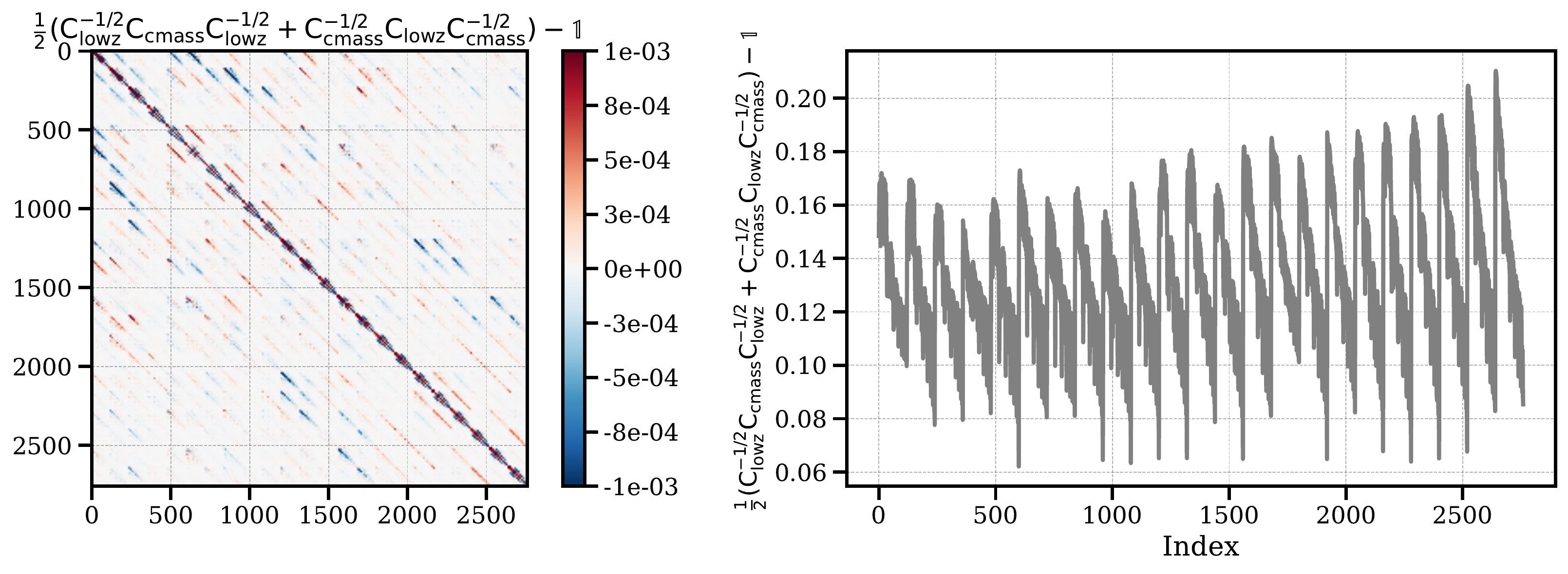}
    \caption{Comparison of the CMASS and LOWZ analytic covariance matrices, but with their volumes set to be equal. We do this because an overall scaling will not adversely impact a cross-correlation analysis, but non-trivial differences in the matrices do and it is these we want to display. We have mapped each set of channels and radial bins into 1D indices to make a 2D plot. The \textit{lefthand} panel shows a symmetrized test; notably, the diagonal is non-vanishing after subtracting the identity matrix.  The non-vanishing off-diagonal elements present in the lefthand panel indicate that the eigenbases of the two matrices also differ, as we might expect given that the shot noise impacts the covariance non-trivially and the two matrices are computed with different shot-noises (see Table \ref{tab:summary}). The \textit{righthand} panel shows just the diagonal of our test matrix; the residual is likely due to the slightly different number densities.} 
    \label{fig:half_inv_cmass_lowz}
\end{figure}

Here we compare the analytic covariances for the two samples (Fig. \ref{fig:half_inv_cmass_lowz}). This test shows that they differ non-trivially, thus precluding a cross-correlation analysis between the two samples if we wished to diagonalize as described in \S\ref{subsec:cross}.

\section{Redshift-Distribution Dependence on Imaging Depth}
\begin{figure}
     \begin{subfigure}[b]{0.45\textwidth}
         \centering
         \includegraphics[width=1\textwidth]{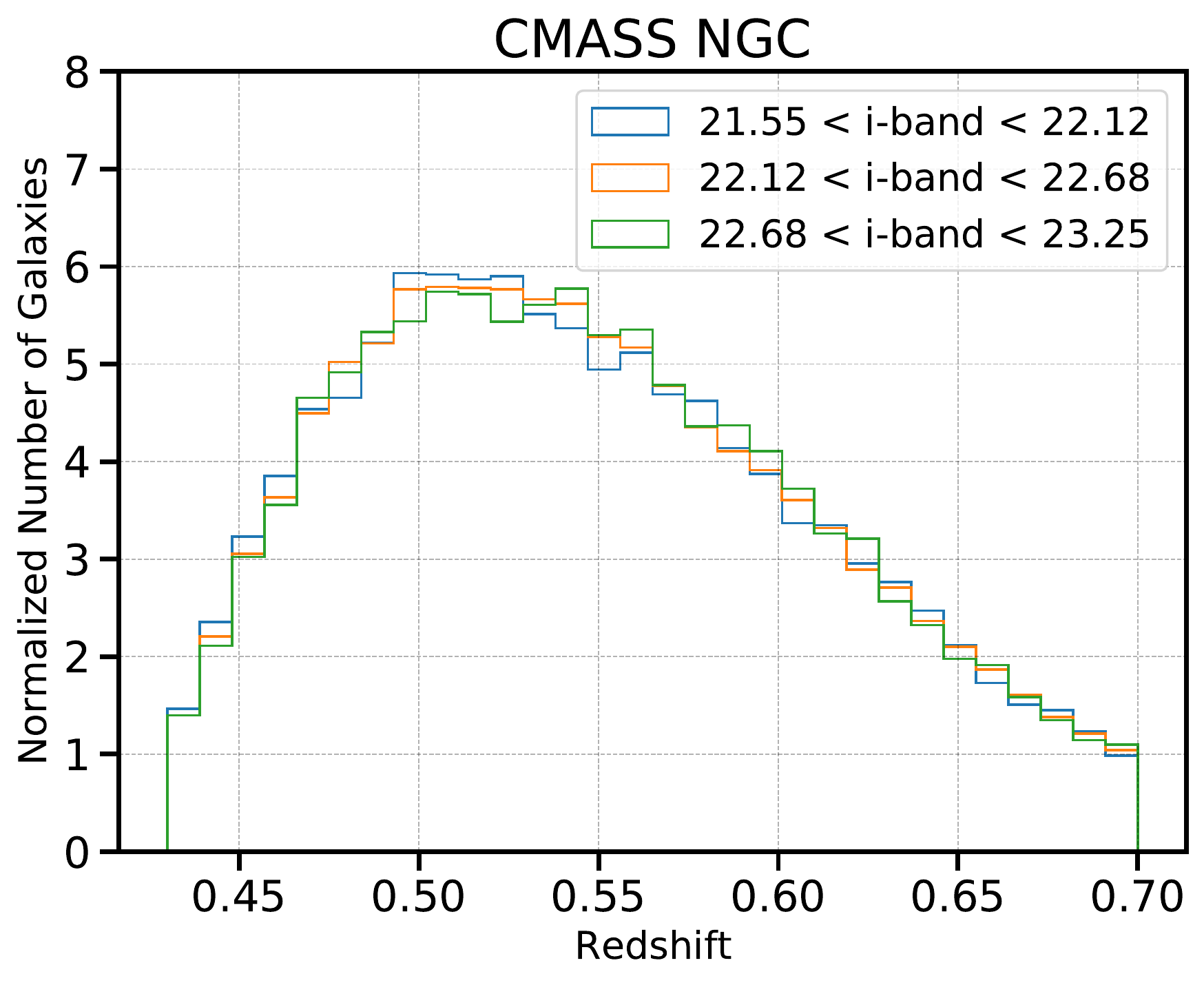}
     \end{subfigure}
     \begin{subfigure}[b]{0.45\textwidth}
         \centering
         \includegraphics[width=1\textwidth]{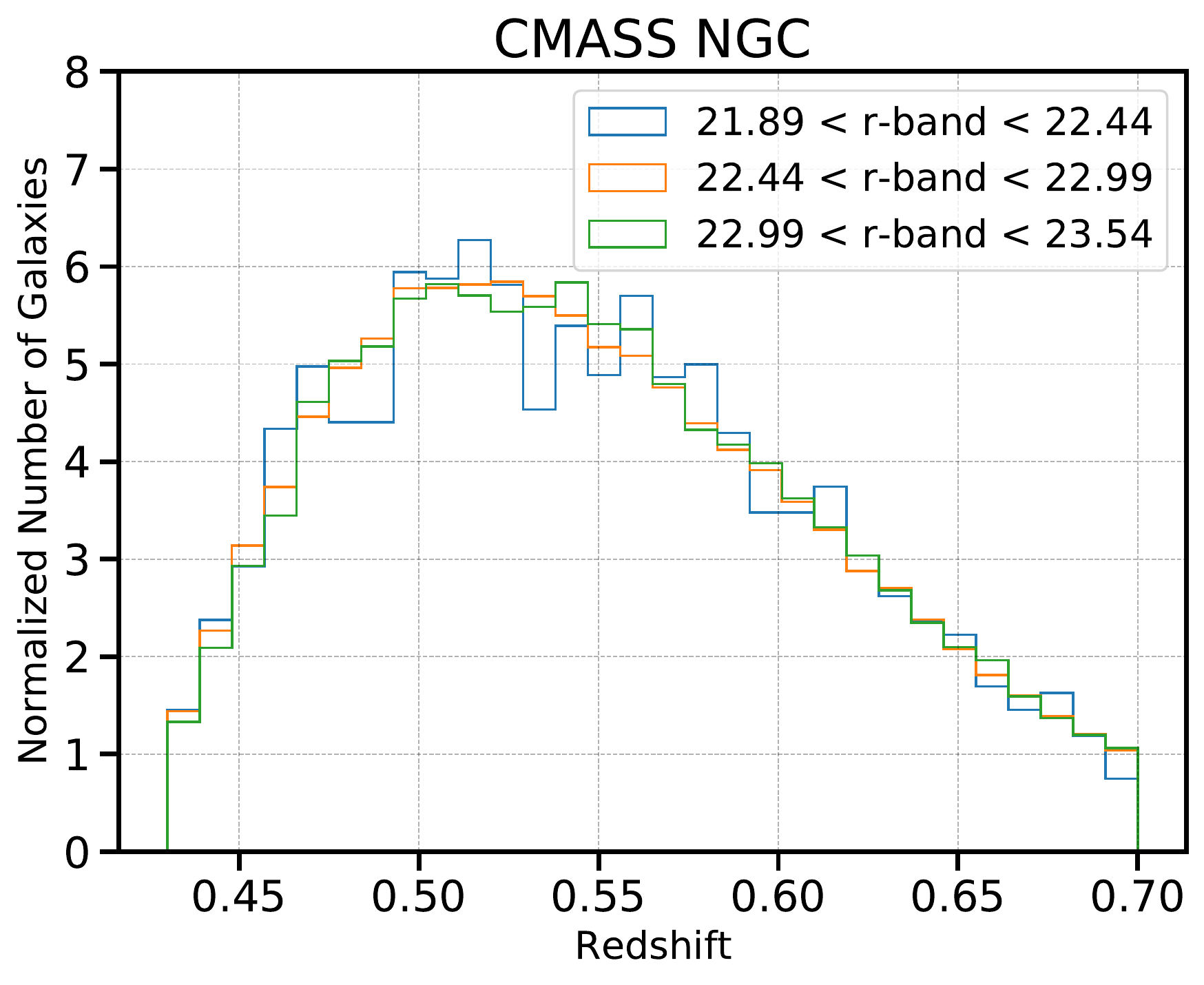}
     \end{subfigure}
\caption{Redshift-distribution dependence on imaging depth for CMASS NGC. {\it Left}: Normalized galaxy number counts as a function of redshift for $i$-band. We have split the sample into three bins in imaging depth. {\it Right}: Same as the left but for $r$-band. We see that the three bins in $i$- and $r$-band depths have very similar $n(z)$, implying that the imaging depth is unlikely to impact our analysis.}
\label{fig:image_depth_cmass_North}
\end{figure}

\begin{figure}
     \begin{subfigure}[b]{0.45\textwidth}
         \centering
         \includegraphics[width=1\textwidth]{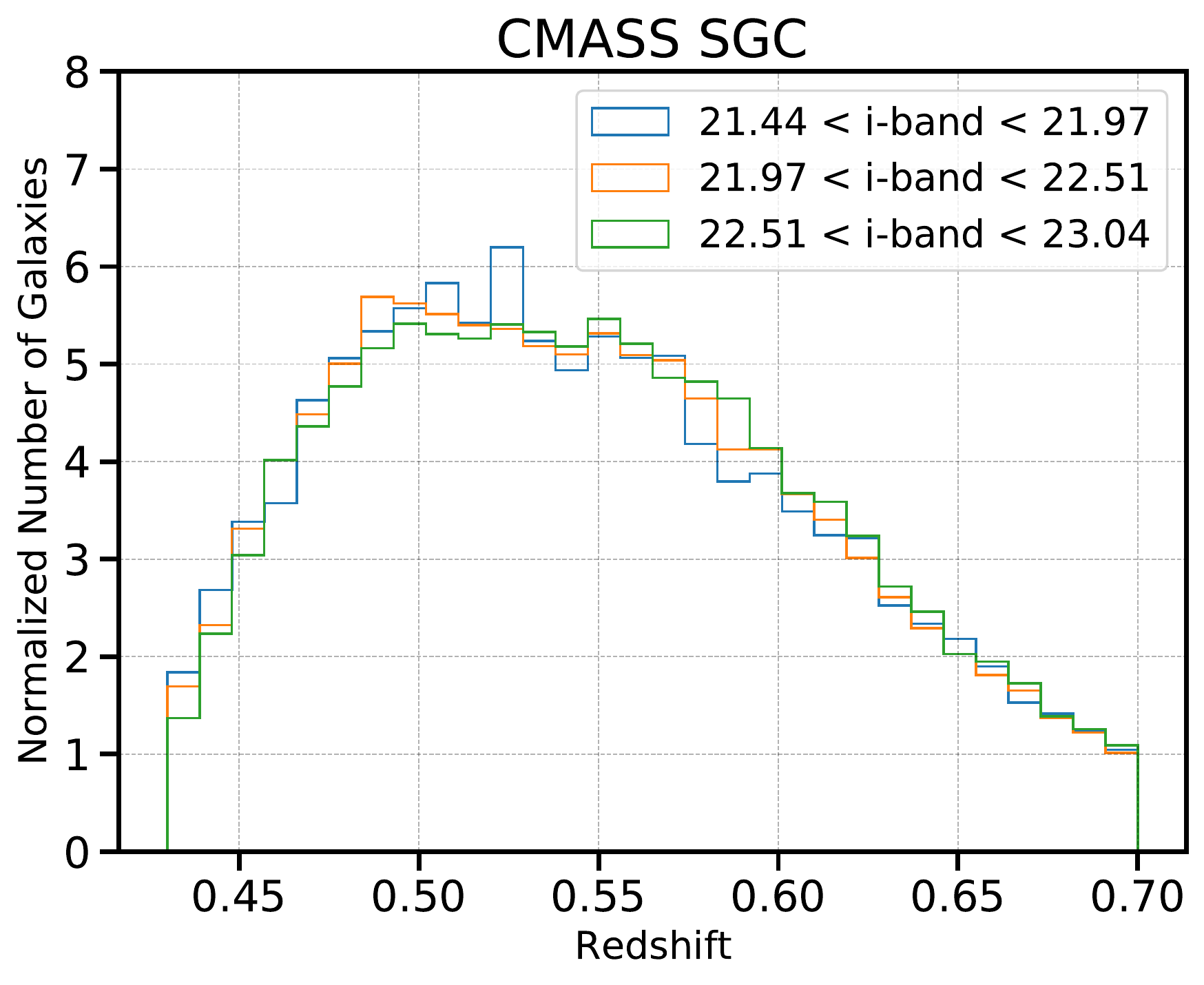}
     \end{subfigure}
     \begin{subfigure}[b]{0.45\textwidth}
         \centering
         \includegraphics[width=1\textwidth]{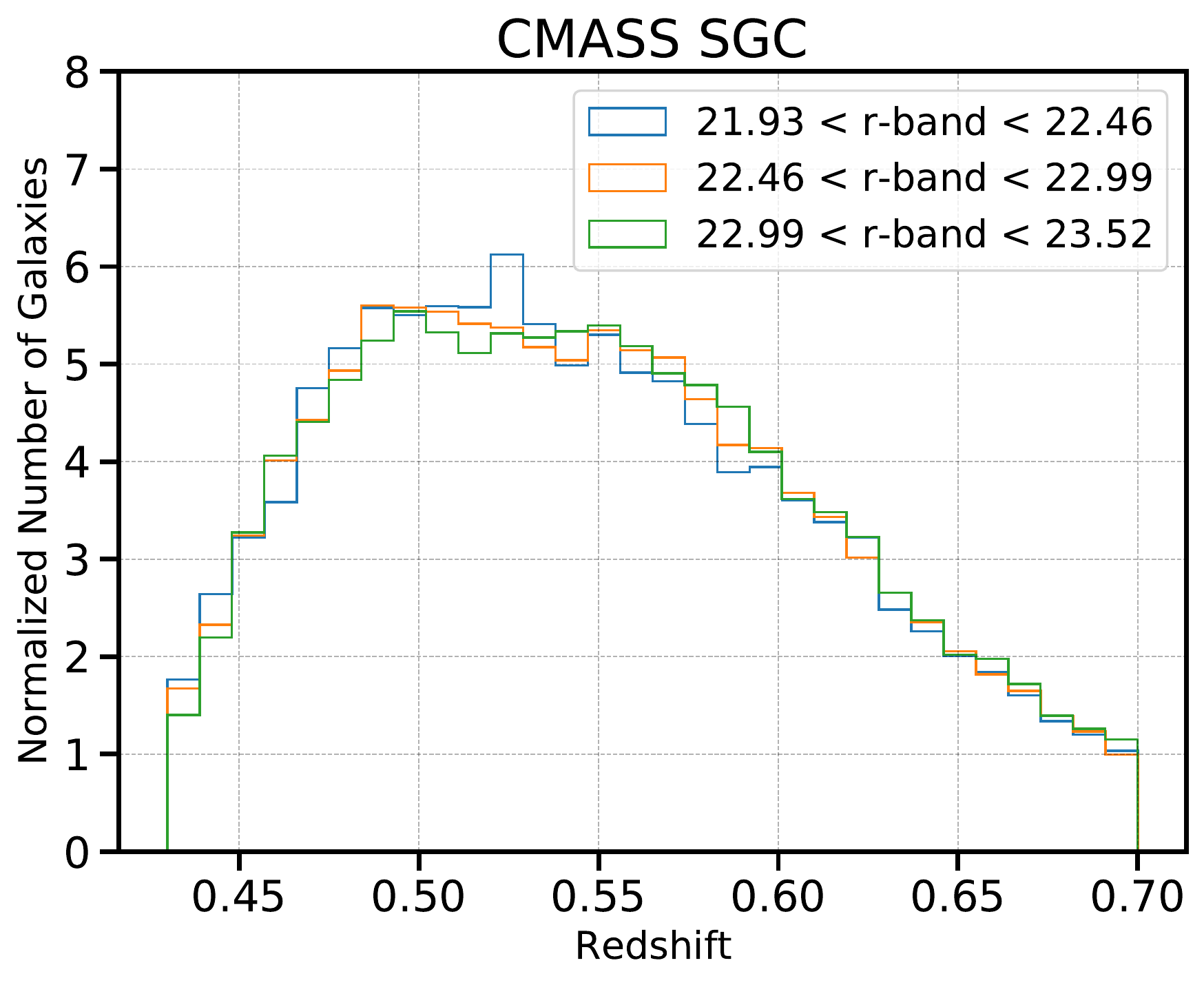}
     \end{subfigure}
     \caption{Same as Fig.~\ref{fig:image_depth_cmass_North} but for CMASS SGC. Again, the $i$- and $r$-bands have very similar $n(z)$ for the three bins in imaging depth, implying that imaging depth does not strongly impact our analysis for SGC.}
\label{fig:image_depth_cmass_South}
\end{figure}

We here assess how the imaging depth affects the number density as a function of redshift, focusing on CMASS. Figs.~\ref{fig:image_depth_cmass_North} and~\ref{fig:image_depth_cmass_South} illustrate the redshift distribution's dependence on the $r$-band and $i$-band imaging depths for respectively the NGC and the SGC. In each cap in both $r$-band and $i$-band, the redshift distribution is very similar across the three bins in imaging depth that we construct. This similarity implies that the imaging depth does not strongly impact the redshift distribution.

\section{Coarser Radial Binning}
\label{sec:coarse}

\begin{figure}
    \centering
    \includegraphics[width=.8\textwidth]{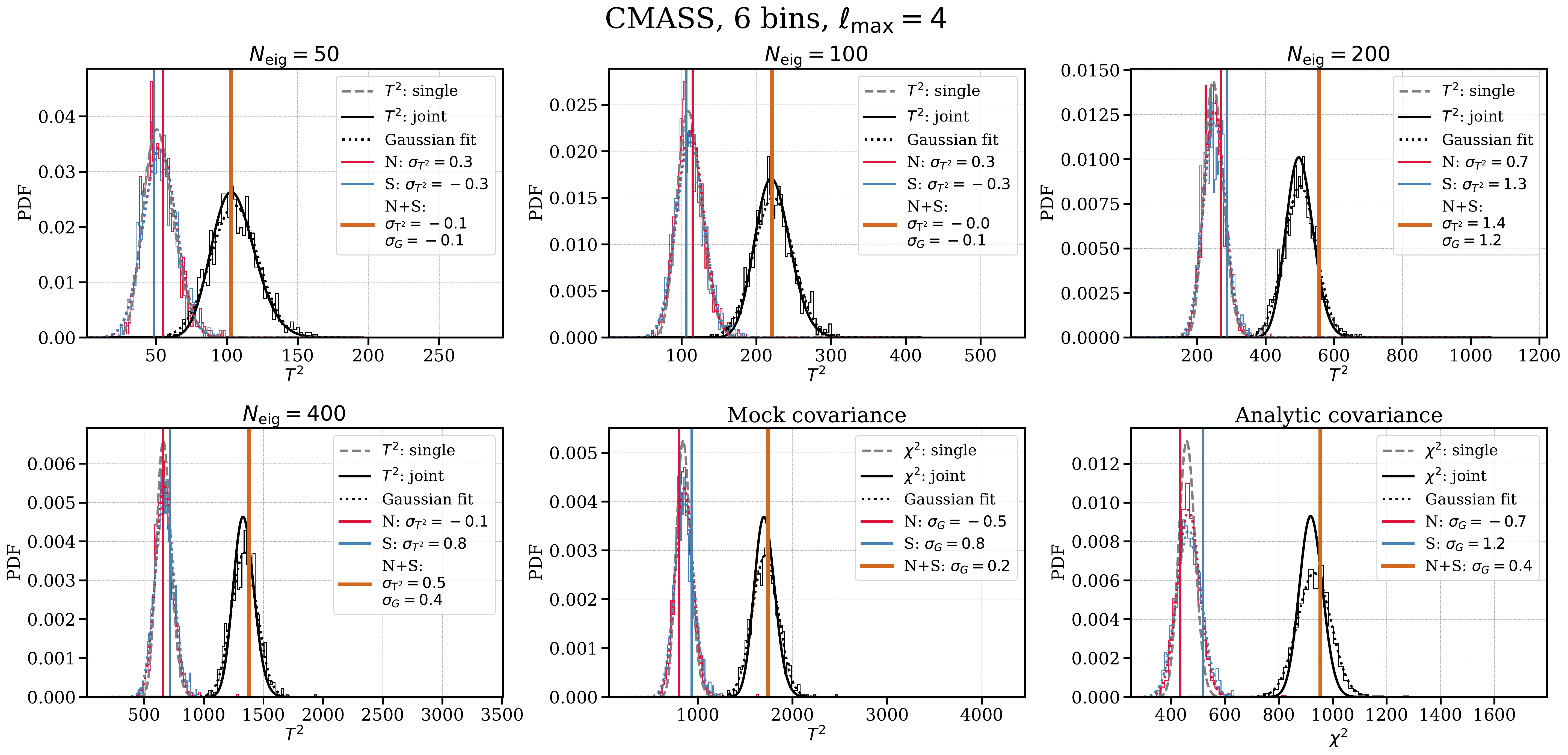}
    \caption{Here we show a compressed analysis (\S\ref{subsub:comp}), and direct approaches with both the mock covariance (lower middle panel) and the analytic covariance (lower right panel) for a 4PCF with just {\bf six} radial bins. This substantially reduces the number of degrees of freedom, permitting the use of the mock covariance. However, the detection significance is also degraded; we attribute this to much larger ``internal cancellation'' (\S\ref{subsec:int_canc}) than in our ten- and eighteen-bin analyses.}    
    \label{fig:chi2s_cmass_ns_6bin}
\end{figure}

Here we show the results of using a coarser radial binning by a factor of roughly two relative to our ten-bin case. We generally see lower detection significance, consistent with the theory that ``internal cancellation'' occurs to a greater extent in this coarser binning. Our results are displayed in Fig. \ref{fig:chi2s_cmass_ns_6bin}.

\section{Maximum scale cut and minimum bin separation}

To explore whether the use of small scales makes a difference in our analysis (as these are the scales on which the mocks may be most likely to imperfectly mirror the data, due to approximate treatment of non-linear structure formation), we test the detection significances by applying further restrictions to the radial bins. First, we force the minimum radial bin separation to be $\Delta r\geq 15\;\mpch$ such that the results are less sensitive to small scales. Second, we vary the maximum radial bin $r_{\rm max}=90\;\mpch$ and $r_{\rm max}=130\;\mpch$ such that we are less affected by any mismatch between mocks and data around the BAO position and towards larger scales (as was seen in some of the 2PCF measurements). 

\begin{figure}
    \centering
    \begin{subfigure}[b]{1\textwidth}
    \includegraphics[width=1\textwidth]{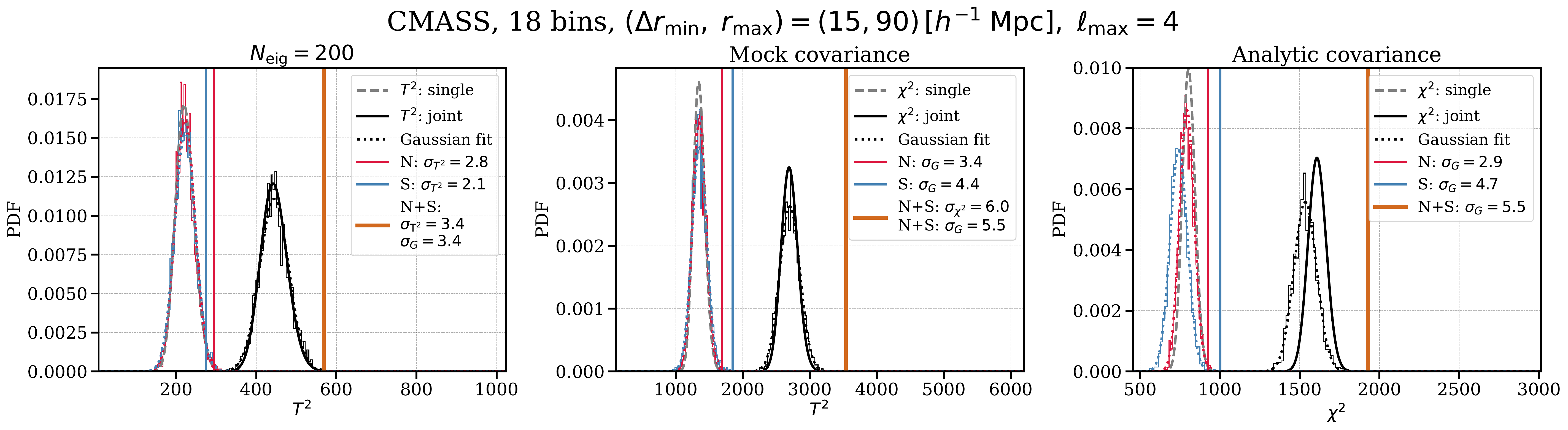}
    \end{subfigure}
    \begin{subfigure}[b]{1\textwidth}
    \includegraphics[width=1\textwidth]{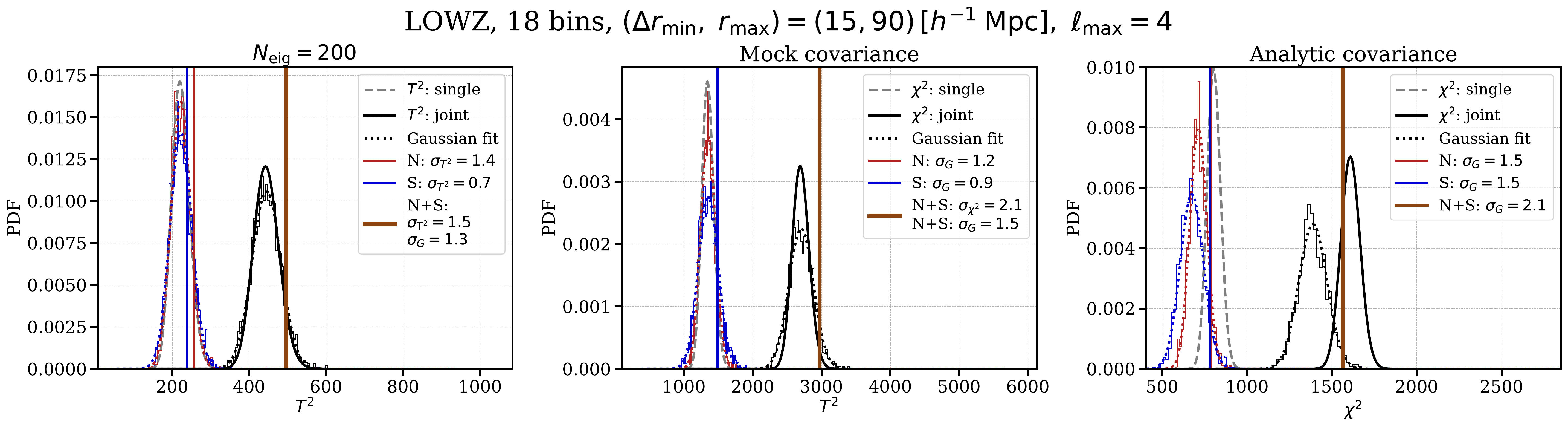}
    \end{subfigure}
    \caption{{\it Upper}: CMASS sample with minimum radial bin separation $\Delta r=15\; \mpch$ and maximum radial bin $r_{\rm max}=90\;\mpch$. The leftmost column uses the data compression method with $N_{\rm eig}=200$. The middle column directly uses the mock covariance. The rightmost column uses the analytic covariance.
    {\it Lower}: Same as the upper row but for LOWZ. In all these cases the detection significance is reduced relative to that in our fiducial analysis due to having fewer degrees of freedom.}
    \label{fig:chi2s_ns_nbins18_LMAX4_drCut15_rmax90}
\end{figure}

\begin{figure}
    \centering
    \begin{subfigure}[b]{1\textwidth}
    \centering
    \includegraphics[width=.9\textwidth]{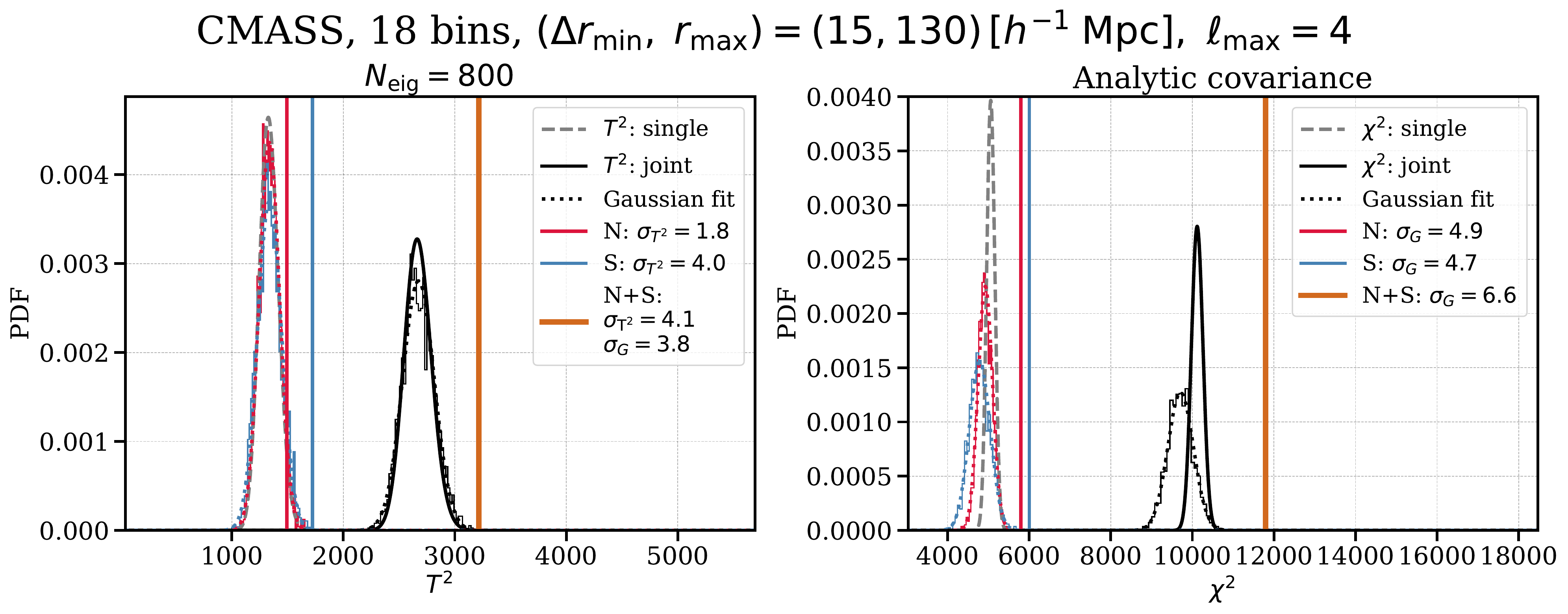}
    \end{subfigure}
    \begin{subfigure}[b]{1\textwidth}
    \centering
    \includegraphics[width=.9\textwidth]{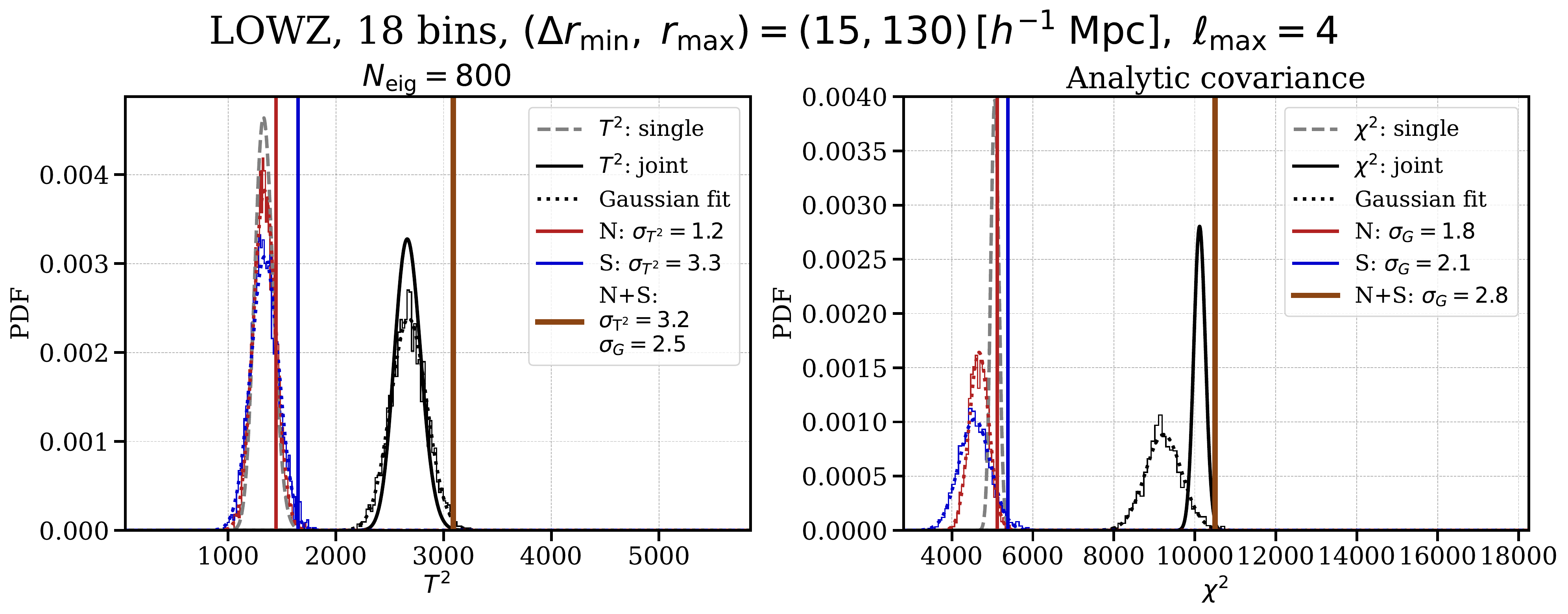}
    \end{subfigure}
    \caption{{\it Upper}: CMASS sample with minimum radial bin separation $\Delta r=15\; \mpch$ and maximum radial bin $r_{\rm max}=130\;\mpch$. The left column uses the data compression method with $N_{\rm eig}=200$. The right column uses the analytic covariance.
    {\it Lower}: Same as the upper row but for LOWZ. As in Fig.~\ref{fig:chi2s_ns_nbins18_LMAX4_drCut15_rmax90} the detection significance is reduced compared to that in our fiducial analysis.}
    \label{fig:chi2s_ns_nbins18_LMAX4_drCut15_rmax130}
\end{figure}

The results are shown in Figs.~\ref{fig:chi2s_ns_nbins18_LMAX4_drCut15_rmax90} and \ref{fig:chi2s_ns_nbins18_LMAX4_drCut15_rmax130}. In all these cases the detection significance reduces compared to the fiducial case (without minimum bin separation or maximum radial bin) given that the number of degrees of freedom is reduced. However, we still observe non-negligible detection significances for all cases. It is worth pointing out that there are in total only $35\times 23=805$ degrees of freedom when using $r_{\rm max}=90\;\mpch$, which allows us to use the mock covariance directly. Despite the mismatch between the analytic covariance matrix and the expected $\chi^2$ distribution, there is no substantial difference in the detection significances.

\section{All angular channels}
\label{app:data_all_angular_ell}

\begin{figure}
    \centering
    \includegraphics[width=0.8\textwidth]{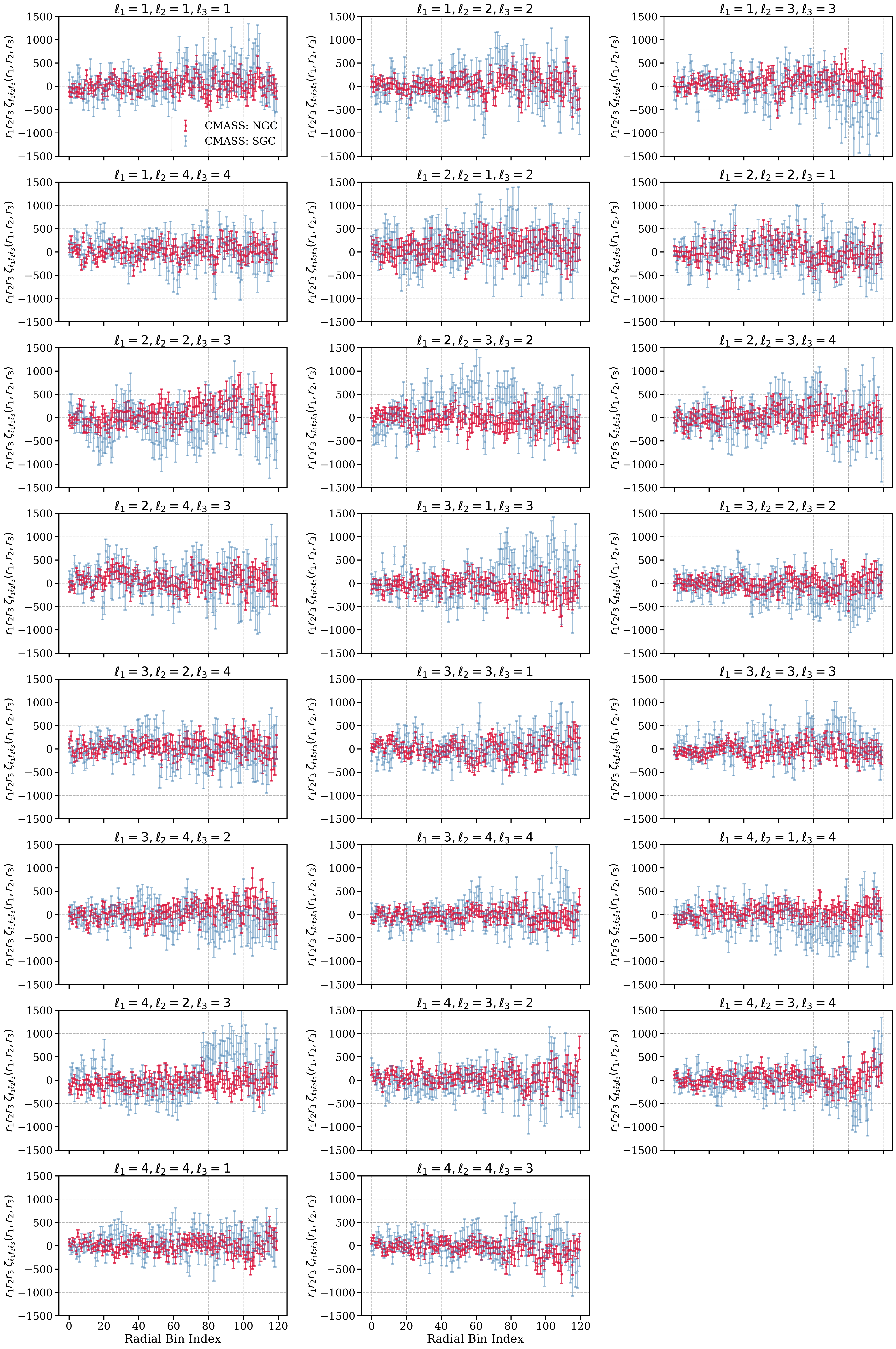}
    \caption{The parity-odd 4PCF for the BOSS CMASS data, with NGC in red and SGC in blue. The plot includes all the angular channels for {\bf ten} radial bins. The error bars are the rms of the \patchy mocks.}
\end{figure}

\begin{figure}
    \centering
    \includegraphics[width=0.8\textwidth]{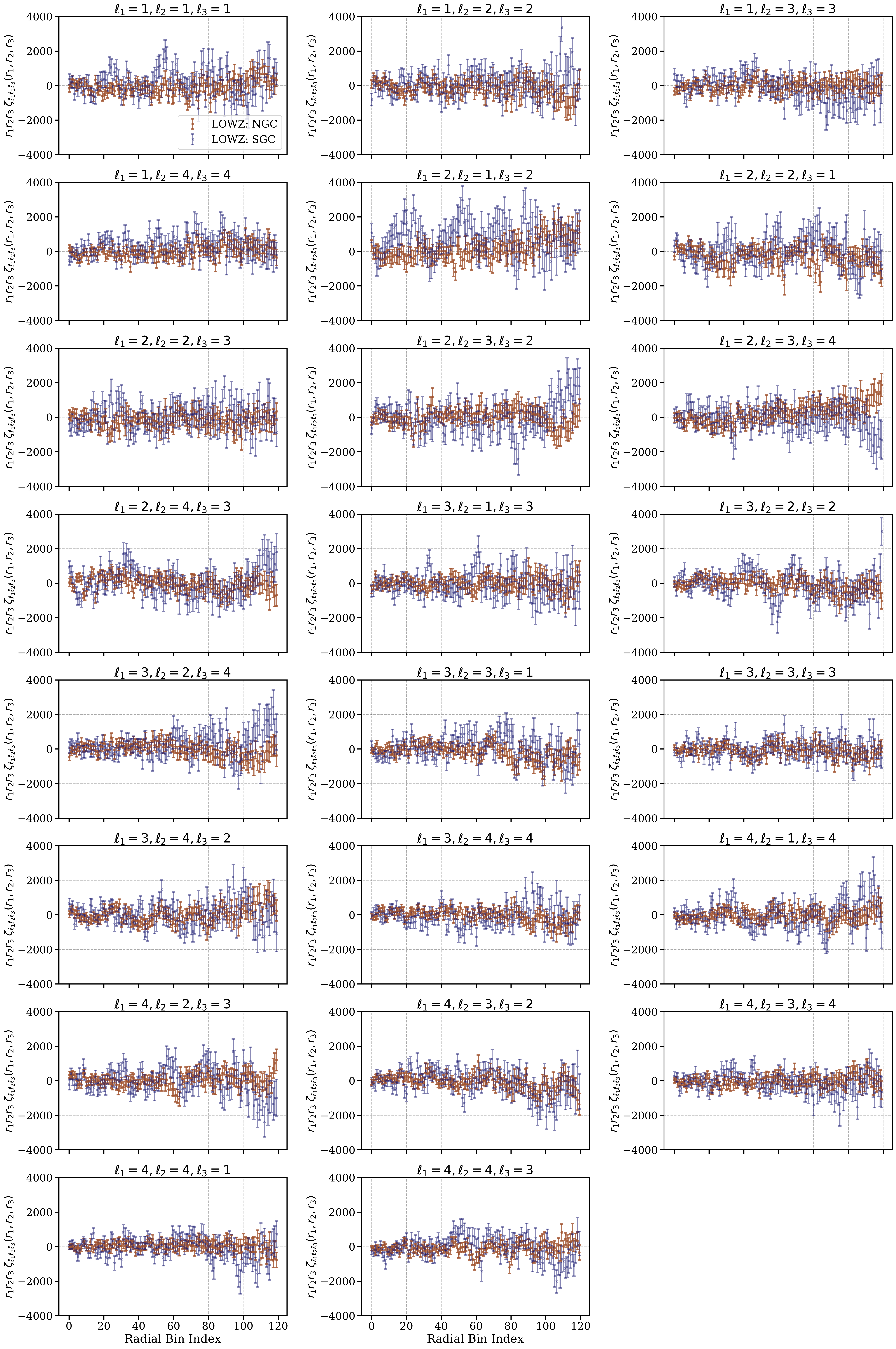}
    \caption{The parity-odd 4PCF for the BOSS LOWZ data including all the angular channels for {\bf ten} radial bins. NGC is in brown and SGC in blue; the error bars are the rms of the \patchy mocks.}
\end{figure}

\begin{figure}
    \centering
    \includegraphics[width=0.8\textwidth]{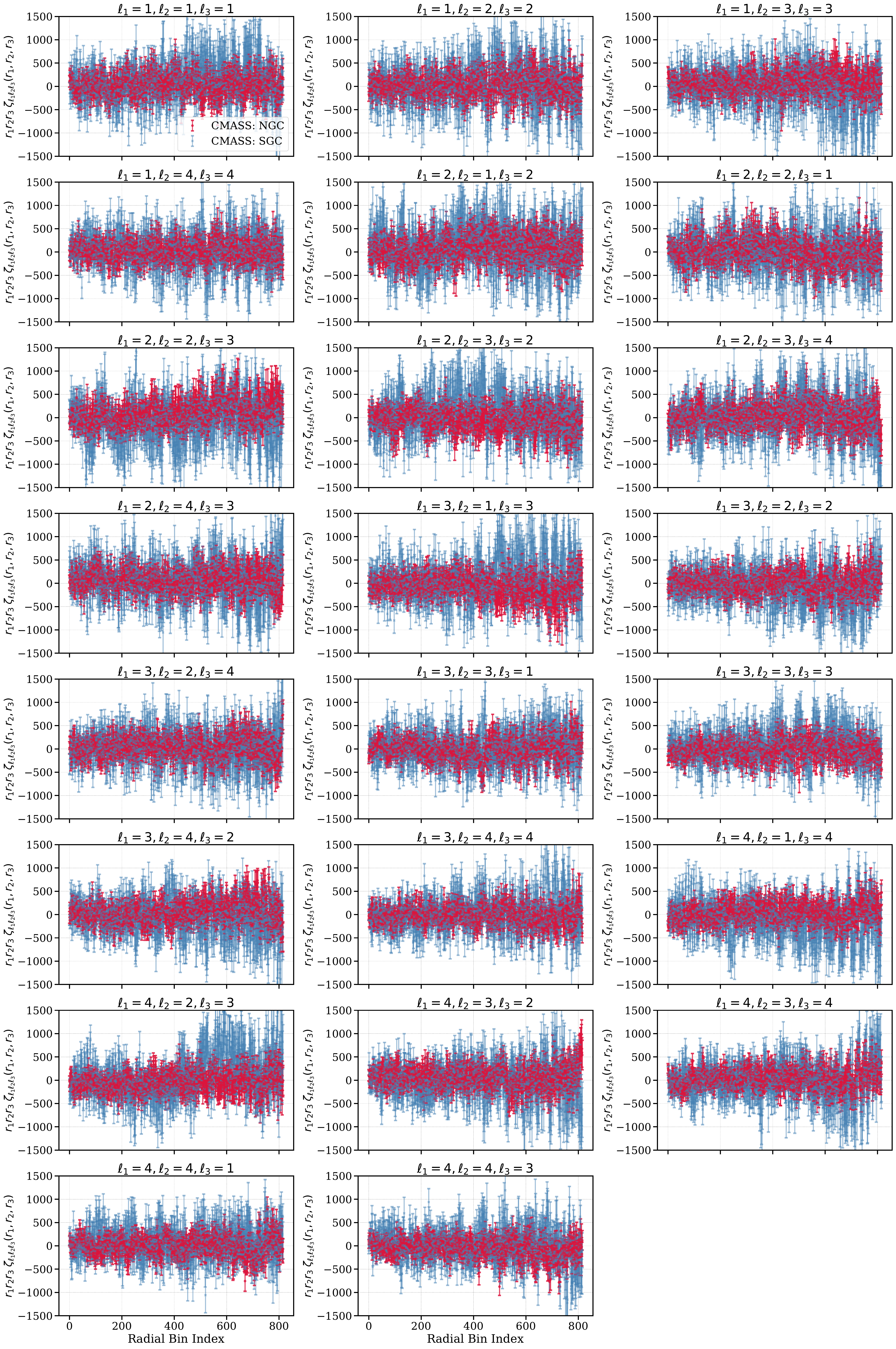}
    \caption{The parity-odd 4PCF for the BOSS CMASS data including all the angular channels for {\bf eighteen} radial bins. NGC is in red and SGC is in blue; the error bars are the rms of the \patchy mocks.}
\end{figure}

\begin{figure}
    \centering
    \includegraphics[width=0.8\textwidth]{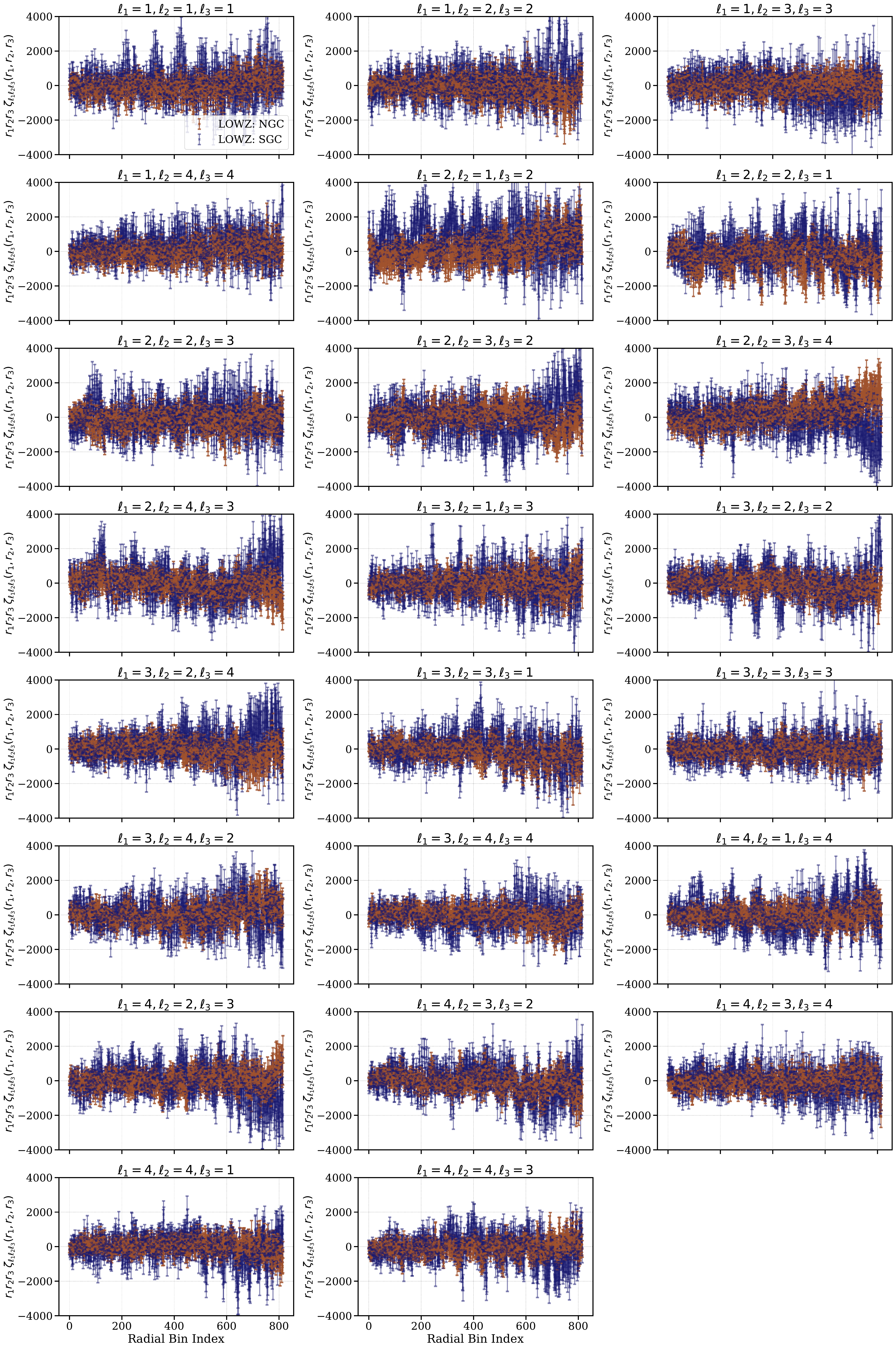}
    \caption{The parity-odd 4PCF for the BOSS LOWZ data including all the angular channels for {\bf eighteen} radial bins. NGC is in brown and SGC in blue; the error bars are the rms of the \patchy mocks.}
\end{figure}

\label{lastpage}
\end{document}